\documentclass[fleqn,usenatbib]{mnras}

\usepackage{newtxtext,newtxmath}
\usepackage[T1]{fontenc}

\DeclareRobustCommand{\VAN}[3]{#2}
\let\VANthebibliography\thebibliography
\def\thebibliography{\DeclareRobustCommand{\VAN}[3]{##3}\VANthebibliography}

\usepackage{graphicx}	% Including figure files
\usepackage{amsmath}	% Advanced maths commands

\usepackage{xcolor}

\usepackage[normalem]{ulem}

\newcommand{\imp}{\Rightarrow}
\newcommand{\1}{\left( }
\newcommand{\2}{\right) }
\newcommand{\3}{\left[ }
\newcommand{\4}{\right] }
\newcommand{\intd}{\, \textrm{d}}

\newcommand{\msun}{\mathrm{M}_{\odot}}

\newcommand{\mstarmax}{m_{\rm max}}
\newcommand{\mstarmin}{m_{\rm min}}
\newcommand{\mcl}{M_{\rm cl}}
\newcommand{\mmin}{M_{\rm cl}^{\rm min}}
\newcommand{\mmax}{M_{\rm cl}^{\rm max}}
\newcommand{\mlim}{m_{\star}^{\rm limit}}

\newcommand{\mhalo}{M_{\rm vir}}

\newcommand{\muv}{M_\mathrm{UV}}
\newcommand{\fstar}{f_\star}
\newcommand{\fmassive}{f_{\rm massive}}
\newcommand{\tsb}{t_{\rm sb}}
\newcommand{\tq}{t_{\rm q}}

\title[Cloud-scale star formation at Cosmic Dawn]{Bright Galaxies at Cosmic Dawn: A Cloud-Scale Star Formation Model Unifying Variable SFE, IMFs, and Stochasticity}

\author[Cueto \& Hutter]{
Elie Cueto,$^{1,2}$\thanks{elierc@mpa-garching.mpg.de}
Anne Hutter$^{2,3,4}$\thanks{anne.hutter@univie.ac.at}
\\
$^{1}$Max Planck Institut für Astrophysik, Karl Schwarzschild Straße 1, D-85741 Garching, Germany \\
$^{2}$Niels Bohr Institute, University of Copenhagen, Jagtvej 128, DK-2200, Copenhagen N, Denmark \\
$^{3}$Cosmic Dawn Center (DAWN) \\
$^{4}$Department of Astrophysics, University of Vienna, T\"urkenschanzstr. 17, 1180 Vienna, Austria
}

\date{Accepted XXX. Received YYY; in original form ZZZ}

\pubyear{\the\year{}}

\begin{document}
\label{firstpage}
\pagerange{\pageref{firstpage}--\pageref{lastpage}}
\maketitle

% Abstract of the paper
\begin{abstract}
To investigate the origins of the high abundance of UV-bright galaxies at $z>10$, we present a new semi-analytic model that bridges the gap between small-scale star formation physics and large-scale galaxy evolution by explicitly tracking discrete star-forming clouds within smoothly evolving dark matter potentials. Unlike conventional semi-analytic models, our approach naturally captures the stochasticity of star formation, allowing us to isolate how cloud properties, star formation efficiencies (SFEs), and stellar initial mass functions (IMFs) shape the star-formation burstiness of early galaxies. Clouds are drawn sequentially from a cloud mass distribution, assigned an SFE and IMF depending on cloud mass, metallicity, and redshift, and evolve under the influence of short-timescale stellar feedback.
We identify three distinct star formation regimes arising from the interplay between cloud masses, densities, and feedback timescales: a stochastic, feedback-limited regime in low-mass halos with long quiescent phases; a bursty regime regulated by the cloud mass distribution; and a smooth, continuous regime in massive halos. Our fiducial model adopts a dynamic IMF, an SFE linked to cloud properties and the IMF, and massive, moderately dense clouds. Varying assumptions reveals that top-heavy IMFs and enhanced SFEs in massive clouds amplify burstiness, while altering the upper cloud mass or density normalisation is secondary. Among model ingredients, the IMF most strongly impacts the UV luminosity function (LF), while switching to a constant SFE boosts the faint end, and reducing the maximum cloud mass decreases the bright end. These results demonstrate that cloud-scale physics critically shape early galaxy UV luminosity distributions.
\end{abstract}

% Select between one and six entries from the list of approved keywords.
% Don't make up new ones.

%%%%%%%%%%%%%%%%%%%%%%%%%%%%%%%%%%%%%%%%%%%%%%%%%%

%%%%%%%%%%%%%%%%% BODY OF PAPER %%%%%%%%%%%%%%%%%%

%%%%%%%%%%%%%%%%%%%%%%%%%%%%%%%%%%%%%%%%%%%%%%%%%%
\section{Introduction}
%%%%%%%%%%%%%%%%%%%%%%%%%%%%%%%%%%%%%%%%%%%%%%%%%%

The James Webb Space Telescope (JWST) has revealed a higher number density of ultraviolet (UV) bright galaxies at $z>10$ than had been expected by standard galaxy evolution models \citep[e.g.][]{ArrabalHaro2023, Bouwens2023b, Donnan2024, Finkelstein2023, Harikane2025, Labbe2023, McLeod2026, PerezGonzalez2023}. To reconcile this tension, various solutions have been proposed. These include bursty (highly fluctuating) star formation \citep{Mason2023, Shen2023, Sun2023, Gelli2024}, enhanced SFEs \citep{Dekel2023, Li2024, Ceverino2024, Andalman2025}, top-heavy IMFs \citep[][]{Trinca2024, cueto_astraeus_2024, hutter_astraeus_2025, Lu2025}, less attenuation by interstellar dust due to dust evacuated from star-forming sites \citep{Ferrara2023, Fiore2023, Ziparo2023, Toyouchi2025, Garaldi2026} or consisting of larger grains \citep{Toyouchi2025, Narayanan2026}, enhanced nebular emission \citep{Katz2025}, helium-enhanced stellar populations in massive star clusters \citep{Katz2024}, black holes accretion \citep{Pacucci2022}, and even dynamic dark energy variations \citep{Menci2024}. However, most of these physical processes have been explored in isolation, neglecting how they self-consistently interact with one another. 

High-resolution simulations of individual early galaxies, which form stars when a region exceeds a specific density threshold, and in some cases becomes Jeans unstable, indeed find elevated local SFE values \citep{Chen2026, Mayer2025}. Artificially raising this local SFE increases star formation variability and renders dust less concentrated at the galaxy centre \citep{Jeong2025, Andalman2025}, yet leaves the time-averaged galaxy-wide SFE largely unchanged \citep{Jeong2025, Voit2024a, Voit2024b}. Conversely, shifting to a more top-heavy IMF similarly enhances star formation variability but introduces more immediate stellar feedback that reduces both the local and global SFE \citep{Jeong2025, Menon2024, cueto_astraeus_2024}.  

While individual galaxy simulations capture local feedback loops depending on the flexibility of the included physical processes, they are too computationally expensive to model the large volumes required to predict the UV luminosity functions. Furthermore, even high-resolution simulations that tie star formation to local gas density generally do not allow the IMF of newly formed stars to vary dynamically with their birth environment \citep[e.g.][]{Andalman2025, Ceverino2024, Chen2026, Jeong2025}. Conversely, larger cosmological frameworks capable of probing these large volumes, such as semi-analytic models (SAMs), rely on galaxy-wide scaling relations \citep[e.g.][]{Somerville2025, Lu2025, hutter_astraeus_2021}. However, by neglecting local spatial and rapid temporal variations, SAMs treat parameters like the SFE and IMF as uniform constant across an entire galaxy and $\sim3-30$\,Myr timesteps. Consequently, there remains a gap for a framework that can self-consistently couple small-scale, burst-driving variations with large-scale cosmological volumes.

Observations of high-redshift star forming galaxies have increasingly revealed exceptionally dense, cluster-like star-forming regions across galaxies at $z\sim6-10$ \citep{Abdurrouf2025, Adamo2024, Bradac2025, Cullen2025, Mowla2024, Nakane2025, Vanzella2023a}. They suggest that early star formation is concentrated within short-lived, dense gas clouds, providing birthplaces for highly stochastic starbursts. This observational picture is supported by recent high-resolution simulations of individual galaxies at cosmic dawn \citep{Chen2026, Mayer2025}, which show that $\sim 90$\% of stars are born in dense clouds across redshifts $z>8$ and nears $100$\% at $z>15$. In these galaxies ($M_{\rm vir}\lesssim10^{11}\,\msun$), cloud masses can range from $10^5-10^8\,\msun$ and reach up to $5\times10^8\,\msun$, while $98$\% of the resulting star clusters are found to be single-population systems experiencing only a single starburst before being disrupted.
Further evidence for this localised, bursty mode of star formation comes from JWST detection of compact galaxies featuring enhanced nitrogen enrichment \citep[high N/O ratios; e.g.][]{Bunker2023, Cameron2024, Cameron2026, Isobe2023, Marques-Chaves2024, Senchyna2024, Topping2025, Watanabe2024}. The anomalous chemical profiles of these systems resemble local globular clusters, pointing to localised self-enrichment driven by extreme stellar densities and intense starbursts within individual clouds. 
Taken together, this evidence motivates a modelling approach that treats galactic star formation as a collection of discrete cloud-scale events, where the resulting star formation characteristics are set by local cloud properties such as mass, density, gas metallicity and initial turbulence. 

Detailed simulations of star-forming clouds show that the birth properties of a cloud drive its evolution by simultaneously regulating its SFE and IMF. On the one hand, higher cloud masses and densities shorten the gas free-fall time and increase the gravitational pressure, directly enhancing the SFE by converting a larger fraction of gas into stars \citep[cf.][]{Fukushima2020, Fukushima2021, Fukushima2023, Garcia2023, Geen2017, He2019, Kim2018, Reina-Campos2025}. On the other hand, the cold, metal-poor conditions of these same dense environments suppress gas fragmentation and enhance accretion onto individual protostars, shifting the resulting IMF toward a more top-heavy distribution \citep[cf.][]{Chon2022, Chon2024, Garcia2023, He2019, Menon2024, Tanvir2024}. 
While a fraction of the cloud's gas is immediately unbound by initial turbulence, its final SFE is set by the radiation feedback from massive stars, which determines when a starburst is shut off. In lower-density clouds photoevaporation dominates the unbinding of gas, whereas in denser clouds radiation pressure is required to overcome the deeper gravitational potential wells.
This creates a direct physical competition between gravitational collapse and stellar feedback. As \citet{Menon2024} demonstrated, a more top-heavy IMF injects more radiation into the birth environment, shutting down star formation earlier and lowering the cloud-scale SFE. However, this dampening effect can be heavily countered if the initial cloud density and mass are high enough to resist early disruption.

This competitive interplay establishes cloud-scale physics as the primary driver of galactic burstiness. If high-redshift clouds are denser and more massive, they should trigger a competition between rapid gravitational collapse (which increases the SFE) and a top-heavy stellar population (which accelerates feedback and dampens the SFE). How these opposing mechanisms balance across time, and what specific profile of star-formation burstiness they produce, remains an open question.

In this paper, we develop a novel, cloud-based semi-analytic framework that models galactic star formation as a sequence of discrete, self-gravitating cloud events. Inspired by radiation hydrodynamic simulations, our model couples a cloud's mass and density to both its IMF and transition between photoionisation and radiation pressure feedback to determine the SFE. Using this framework, we aim to answer the following questions:
How does star-formation burstiness, specifically the duration of bursts and subsequent quiescent phases, depend on the assumed cloud properties? Do these cloud characteristics dictate whether a galaxy shows bursty or continuous star formation? And how does this mass-and redshift-dependent burstiness affect the UV LFs?

This paper is organised as follows. In Section \ref{sec:model} we introduce our model, detailing the description for halo growth, baryon accretion, stellar feedback, metal enrichment, dynamic IMFs, and the formation of discrete star-forming clouds. Section \ref{sec:burstiness} presents our model variants and examines how cloud properties, assumed IMFs, and SFE values affect the burstiness of star formation. We specifically focus on the durations of starbursts, quiescent phases between bursts, and the resulting scatter in the UV luminosities of galaxies. Section \ref{sec:uvlf} discusses how the burstiness of star formation affects the UV luminosity functions, and we conclude in Section \ref{sec:conclusions}. 
Throughout this work, we adopt the AB magnitude system \citep{oke_secondary_1983}, and assume a standard $\Lambda$CDM cosmology with parameters from the \cite{planck_collaboration_planck_2016}: $\Omega_\Lambda=0.692885$, $\Omega_m=0.307115$, $\Omega_b=0.048206$, $H_0=100h=67.77 \rm km\, s^{-1}Mpc ^{-1}$.

%%%%%%%%%%%%%%%%%%%%%%%%%%%%%%%%%%%%%%%%%%%%%%%%%%%%%%%%%%
\section{The model}\label{sec:model}
%%%%%%%%%%%%%%%%%%%%%%%%%%%%%%%%%%%%%%%%%%%%%%%%%%%%%%%%%%

We introduce a cloud-based star-formation toy model built within a simplified galaxy-evolution framework. Our model follows gas accretion, IMF-dependent stellar feedback, and chemical enrichment from the \textsc{astraeus} framework \citep{cueto_astraeus_2024}. However, rather than tracking galaxy growth along numerical dark matter merger trees, our model runs on smooth, analytic halo mass assembly histories. 

In this model, star formation is quantised and restricted to dense, self-gravitating molecular clouds, mimicking the dense star-forming regions observed in high-redshift star-forming galaxies \citep{Abdurrouf2025, Adamo2024, Bradac2025, Mowla2024, Nakane2025, Vanzella2023a}. To resolve the short dynamical lifetimes of these dense clouds ($\sim$ a few Myr), we subdivide the snapshot intervals ($\sim 3$–$40$ Myr) of the underlying \textsc{vsmdpl} dark matter N-body simulation into fine sub-timesteps of $\Delta t_{\rm step}\sim 1$ Myr. The final sub-timestep of each snapshot is dynamically adjusted ($0.5$–$1.5$ Myr) to match the snapshot boundaries.

% ---------------------------------------------------------
\subsection{Halo mass assembly and gas accretion}\label{sec:halomassassembly}
% ---------------------------------------------------------

We model analytical halo growth trajectories following \cite{correa_accretion_2015}:
\begin{equation}\label{eq:halomassassembly}
    M_{\rm halo}(z) = M_{\rm halo,0}(1+z)^\alpha e^{\beta z},
\end{equation}
where $\alpha=0.25$ \citep{ferrara_super-early_2023}. We sample descendant masses at $z=0$ across $M_{\rm halo,0}=10^{10}$–$10^{14.5}\,\mathrm{M}_\odot$ and vary the assembly parameter $\beta$ from $-0.95$ to $-0.55$ (mean $-0.75$). This spans progenitor masses at $z=5$ from $1.35\times10^8$ to $3.16\times10^{13}\,\mathrm{M}_\odot$. To ensure cadence consistency, sub-timestep halo masses are linearly interpolated between the discrete \textsc{astraeus} snapshot boundaries, starting when a halo first exceeds $10^{7.4}\,\mathrm{M}_\odot$.

Assuming baryons trace dark matter on cosmological scales, the total gas mass accreted from the IGM over snapshot $i$ is:
\begin{equation}
    M_{\rm gas,acc}^i = \left(M_{\rm halo}^{i + 1} - M_{\rm halo}^i\right) \frac{\Omega_b}{\Omega_m}.
\end{equation}
This mass is distributed uniformly across the snapshot's sub-timesteps. The available diffuse gas at the start of sub-timestep $j$ is updated as:
\begin{equation}
    M_{\rm gas, ini}^j = M_{\rm gas}^{j-1} + M_{\rm gas,acc}^i\frac{\Delta t_{{\rm step},j}}{\Delta t_{{\rm snap},i}},
\end{equation}
where $M_{\rm gas}^{j-1}$ is the remaining diffuse gas from the previous sub-timestep, and $\Delta t_{{\rm step},j}$ and $\Delta t_{{\rm snap},i}$ are the sub-timestep and snapshot durations, respectively. 

We partition the interstellar medium (ISM) into two components: (i) a diffuse reservoir ($M_{\rm gas,ini}$) subject to accretion, reionisation filtering, and supernova ejection, and (ii) a dense, self-gravitating cloud component ($M_{\rm gas,cl}$). These clouds are assumed to be too dense to undergo external IGM photo-evaporation, shielding their gas until stellar feedback disrupts them.

Baryonic growth at the cosmic baryon fraction occurs only in the absence of radiative feedback processes. During reionisation, photo-heating raises the IGM temperature in ionised region, suppressing gas accretion and photo-evaporating gas in low-mass halos \citep{efstathiou_suppressing_1992, kravtsov_tumultuous_2004, gnedin_probing_1998,gnedin_cosmological_2000}. Following \citet{gnedin_cosmological_2000}, we model the effect of photo-heating with a filtering mass $M_{\rm F}$ that reduces the fraction of the cosmological-baryon gas a halo can retain to
\begin{equation}
    f_g(z,\mhalo) = \dfrac{1}{\11+0.26\frac{M_{\rm F}}{\mhalo}\2^3},
\end{equation}
and adopt the analytic expression for $M_\mathrm{F}$ from \citet{kravtsov_tumultuous_2004}. The maximum retained gas mass is then
\begin{equation}
    M_{\rm gas,max}=\mhalo\dfrac{\Omega_b}{\Omega_m}f_g
\end{equation}
and the total gas mass in a galaxy
\begin{equation}
    M_{\rm gas,total}=M_{\rm gas,ini}+M_{\rm gas,cl},
\end{equation}
where $M_{\rm gas,cl}$ depicts the summed gas mass of the galaxy's existing star-forming clouds.
At the beginning of each timestep, we adjust the (initial) diffuse component as
\begin{align}
    M_{\rm gas,ini}^{\rm adjust} &=
    \begin{cases}
        M_{\rm gas,ini}&M_{\rm gas,total}<M_{\rm gas,max},\\
        M_{\rm gas,max}-M_{\rm gas,cl}&M_{\rm gas,cl}\leq M_{\rm gas,max}\leq M_{\rm gas,total},\\
        0&M_{\rm gas,max}< M_{\rm gas,cl},
    \end{cases}
\end{align}
assuming existing star-forming clouds are too dense to be photo-evaporated. 
Suppression through a photo-heated IGM is strongest in low-mass halos ($\mhalo\lesssim10^{8.5}\msun$) at $z\lesssim8$ and negligible in galaxies with $\mhalo\gtrsim10^9\msun$.

% -----------------------------------------
\subsection{Star-forming cloud properties}\label{sec:cloud_properties}
% -----------------------------------------

At each sub-timestep, the newly adjusted diffuse gas reservoir ($M_{\rm gas,ini}^{\rm adjust}$) provides the mass budget for seeding new star-forming clouds. In contrast to the continuous star-formation recipes typical of global galaxy-scale models \citep[cf.][]{hutter_astraeus_2025}, we partition this available gas into discrete, individual cloud complexes. The unique mass and density of each sampled cloud then directly determine how efficiently it converts its gas into stars.

\subsubsection{Cloud mass function and boundary conditions}
Motivated by observations \cite[see e.g. reviews by][]{krumholz_big_2014, krumholz_star_2019, renaud_star_2018}, we draw cloud masses $M_{\rm cl}$ from a scale-free clouds mass function (CMF)
\begin{equation}
    \frac{\intd N}{\intd\ln M}=\dfrac{M_{\rm cl}^{\rm max,j} M_{\rm cl}^{\rm min}}{M_{\rm cl}^{\rm max,j} - M_{\rm cl}^{\rm min}} M^{\alpha}
\end{equation}
with $\alpha=2$ \citep[cf. high-resolution galaxy simulation from][]{Chen2026}, bounded between a lower resolution limit $M_{\rm cl}^{\rm min} = 10^4\,\mathrm{M}_\odot$ and a dynamically evolving upper mass bound $M_{\rm cl}^{\rm max,j}$. 
Because high gas accretion and stellar feedback keep the ISM turbulent, we adopt a turbulent Jeans-like scaling, assuming the turbulent velocity scales with the virial velocity as $v_{\rm turb}\propto \mhalo^{1/3}(1+z)^{1/2}$, which implies that the CMF upper bound scales as $M_{\rm cl}^{\rm max}(M_{\rm halo},z)\propto M_{\rm halo}(1+z)^{3/2}$. We implement
\begin{equation}\label{eq:cloudmasslimit}
    M_{\rm cl}^{\rm max,j}= \mathrm{min}\3\dfrac{\mhalo^j}{10^{4}}\1\dfrac{1+z_j}{1+9}\2^{3/2}f_{\rm max},M_{\rm cl}^{\rm max}\4,
\end{equation}
with calibration factor $f_{\rm max}=12$ and a hard global maximum cloud cap $M_{\rm cl}^{\rm max}$.

\subsubsection{Cloud densities}
Once a cloud mass is drawn, its density is mapped directly from the global properties of the host galaxy. Assuming uniform, spherical geometry, we set each cloud’s mean number density $n_0$ proportional to the ambient gas density within the virial radius $R_{\rm vir}$:
\begin{equation}
    n_0 = n_{s} \left(\frac{M_{\rm gas,ini}^{\rm adjust}}{\mu m_H}\right)\left(\frac{4\pi}{3} R_{\rm vir}^3\right)^{-1},
\end{equation}
where $\mu = 2.3$ is the mean molecular weight for molecular gas, $m_H$ is the hydrogen mass, and $n_s$ is a dimensionless calibration factor. To prevent uncalibrated extrapolations into unphysical regimes, $n_0$ is bounded between a minimum of $10^3\,\mathrm{cm}^{-3}$ and a maximum of $10^5\,\mathrm{cm}^{-3}$ (corresponding to surface densities of $\sim3\times10^3-7\times10^4\,\msun$pc$^{-2}$ for a $M_{\rm cl}=10^8\,\msun$ cloud). We adopt a calibration factor of $n_s=5\times 10^5$ to ensure that most cloud densities fall within these upper and lower limits. While strongly lensed high-redshift star-forming regions suggest localised densities can exceed these values (e.g. Cosmic Gems: \citealt{Adamo2024}; Sunrise: \citealt{Vanzella2023a}), this range safely captures the vast majority of standard star-forming environments in our baseline calibrations.

\subsubsection{Cloud surface densities and lifecycles}
For a cloud of drawn mass $M_{\rm cl}$ and density $n_0$, the mass density is $\rho = \mu m_H n_0$, corresponding to an initial radius of $R_{\rm cl} = (3 M_{\rm cl}/4\pi \rho)^{1/3}$. The surface density entering the SFE model is:
\begin{equation}
    \Sigma_0 = \frac{M_0}{\pi R_{\rm cl}^2} = \frac{M_{\rm cl}}{\pi R_{\rm cl}^2} \propto M_{\rm cl}^{1/3} n_0^{2/3}.
\end{equation}

This surface density dictates the time evolution of the cloud. We track two internal cloud timescales: an onset delay $t_{\rm sf,0}$ for core collapse and the active star-forming duration $t_{\rm sf}$. For a cloud formed at $t_{\rm form}$, the active star-forming window is $[t_{\rm form} + t_{\rm sf,0},\, t_{\rm form} + t_{\rm sf,0} + t_{\rm sf}]$. Parametrised from the radiation-hydrodynamic simulations of \citet{Kim2018} and \citet{Menon2024}, these timescales follow the empirical fits:
\begin{equation}
    \frac{t_{\rm sf}}{t_{\rm ff}} = \left(\frac{\Sigma_0}{100\,\mathrm{M}_\odot\,\mathrm{pc}^{-2}}\right)^{0.255} \mathrm{and} \quad t_{\rm sf,0} = 3.50\,\mathrm{Myr} \left(\frac{M_{\rm cl}/\mathrm{M}_\odot}{R_{\rm cl}/\mathrm{pc}}\right)^{-0.255},
\end{equation}
where $t_{\rm ff}= \sqrt{3\pi/(32 G \rho)}$ represents the baseline cloud free-fall time. 

% ---------------------------------------------------------------
\subsection{Cloud Sampling and Numerical Implementation}\label{sec:numerical_sampling}
% ---------------------------------------------------------------

\subsubsection{Discrete sampling (``bursty'' regime)}
To populate the galaxy with the molecular clouds characterised in Section~\ref{sec:cloud_properties}, we implement a sequential stochastic sampling routine at each sub-timestep $j$. We sequentially draw an individual cloud mass $M_{\rm cl}$ from the CMF, discarding and redrawing any sample where $M_{\rm cl} > M_{\rm cl}^{\rm max,j}$. For each accepted cloud, we compute its surface density $\Sigma_0$ and corresponding star-formation efficiency (SFE) $\epsilon_\star$ (Section~\ref{sec:fcl}). The resulting stellar mass, $\epsilon_\star M_{\rm cl}$, is added to the sub-timestep's cumulative stellar mass bucket, $M_{\rm tot}^j$. This process repeats, subtracting the consumed gas from the diffuse reservoir, until $M_{\rm tot}^j$ reaches the galaxy-level stellar mass cap $M_{\star, \rm max}^j$ (Section~\ref{sec:snfeedback}). Because clouds are processed sequentially, the properties of the final cloud are naturally constrained by the remaining available gas budget.

\subsubsection{CMF smoothing trigger (``continuous'' regime)}\label{sec:smoothing_trigger}
To maintain computational tractability at high gas masses, we switch from stochastic sampling to a deterministic ``smoothed star formation'' mode once the CMF is effectively fully sampled. We trigger this transition when the relative scatter in the number of clouds required to consume the diffuse gas reservoir falls below a threshold fraction, $T_{\rm smooth}$, of the expected number of clouds. Setting $T_{\rm smooth}=0.006$ ensures this switch occurs after a galaxy transitions out of its bursty phase. The corresponding diffuse gas mass threshold is defined as:
\begin{equation}\label{eq:Mgasthresh}
    M_{\rm gas,thresh} = \frac{\mathrm{Var}[M_{\rm cl}]}{\mathrm{E}[M_{\rm cl}]T_{\rm smooth}^2},
\end{equation}
where $\mathrm{E}[M_{\rm cl}]$ and $\mathrm{Var}[M_{\rm cl}]$ represent the mean and variance of the CMF (see Appendix~\ref{app:variance}). In this regime, we bypass the loop and adopt the CMF-averaged SFE, $\epsilon_\mu$, converting a total cloud gas mass of $M_{\star,\rm max}^j/\epsilon_\mu$ into stars.

\subsubsection{Time evolution}
During each sub-timestep, we advance all existing and newly formed clouds. A cloud converts its gas into stars with efficiency $\epsilon_\star$ over its active duration $t_{\rm sf}$, yielding a new stellar mass of
\begin{equation}\label{eq:newstarstep}
    M_{\star,{\rm new}} = M_{\rm cl} \epsilon_\star \frac{\Delta t_{\rm step,eff}}{t_{\rm sf}},
\end{equation}
where $\Delta t_{\rm step,eff}$ is the overlap between the current sub-timestep $\Delta t_{\rm step,j}$ and the cloud’s star-forming window $[t_{\rm form} + t_{\rm sf,0}, t_{\rm form} + t_{\rm sf,0} + t_{\rm sf}]$. 
Clouds are dissolved at the onset of the first core-collapse SNe. For a fully sampled IMF, this disruption occurs $\sim 3$\,Myr after star-formation onset; for cloud-mass-limited IMFs, it ranges from $3$ to $18$\,Myr depending on the maximum stellar mass present. Upon disruption, any remaining unconsumed cloud gas, metals, and dust are returned to the diffuse reservoir. Note that the active star-forming phase typically lasts only $\sim 1$--$2$ free-fall times ($t_{\rm sf} \sim 1$--$2$\,Myr), meaning clouds survive as inert structures for a brief period after star formation ceases before being fully disrupted by SNe. For bookkeeping, we add newly formed stars to age-metallicity bins to compute delayed chemical yields, SN feedback, and radiative outputs.

% ----------------------------------------------------------
\subsection{Stellar initial mass functions} \label{sec:IMFs}
% ----------------------------------------------------------

We consider four distinct stellar IMFs:
\begin{enumerate}
    \item Salpeter IMF \citep{salpeter_luminosity_1955}: A standard power-law, $\frac{\intd N}{\intd m}\propto m^{-2.35}$, defined over the mass range $0.1-100$~M$_\odot$.
    \item Evolving IMF \cite{cueto_astraeus_2024}: A Salpeter-like slope at the low-mass end transitioning to a log-flat high-mass tail. The IMF is parameterised such that a fraction $f_{\rm massive}$ of the total stellar mass lies in the log-flat part, given by $f_{\rm massive} = 1.07~[1 - 2^{X}] + 0.04 \cdot (2.67)^{X}~z$, where $X = 1 + \log_{10}(Z/Z_\odot)$, $Z$ is the gas metallicity, and $z$ is the redshift.
    \item[(iii)/(iv)] Cloud-mass-limited (CML) IMFs: We implement CML variants of the Salpeter and Evolving IMFs by imposing a cloud-dependent upper stellar mass. Because a finite cloud forms only a finite total stellar mass $M_\star^{\rm tot}$, it cannot sample arbitrarily high stellar masses. The maximum stellar mass is thus bounded by the maximum from random sampling, $m_{\max} = m_{\max}(M_\star^{\rm tot})$, derived explicitly in App. \ref{app:CMLIMFs}.
\end{enumerate}
To normalise the SFE across these variations, we define a fiducial ionising‑photon yield per unit stellar mass, $\Xi_0 = 5.05 \times 10^{46}\, {\rm s}^{-1} \msun^{-1}$, and express it via the dimensionless factor $\chi \equiv \Xi/\Xi_0$.

% -----------------------------------------------------
\subsection{Cloud-scale star formation efficiency}\label{sec:fcl}
% -----------------------------------------------------

The SFE of molecular clouds is uncertain and sensitive to many processes. Radiative–hydrodynamical (RHD) simulations spanning a range of cloud masses, surface densities, metallicities, and feedback prescriptions (e.g. photoheating/ionisation, UV/IR radiation pressure, and occasionally SNe) have mapped this dependence, exploring how clouds of different masses and densities evolve \citep[e.g.][]{Kim2018, menon_infrared_2022, menon_outflows_2023, Menon2024, menon_bursts_2024, Fukushima2020, Fukushima2021, fukushima_far_2022, He2019, krumholz_radiation_2010, tanvir_environmental_2022, Tanvir2024} or the effects of the metal and dust content in clouds \citep[e.g.][]{bate_statistical_2022, Chon2022, Fukushima2021, Fukushima2020, He2019, howard_universal_2018, menon_infrared_2022, menon_outflows_2023, Menon2024, menon_bursts_2024, Tanvir2024}.
Despite differing setups, two trends are robust: 
Firstly, the SFE of clouds increases with rising cloud surface density. While higher densities increase the significance of radiation pressure, photoionisation feedback becomes less impactful. This decline in photoionisation feedback dominates, leading to more efficient star formation.
Secondly, the impact of metallicity depends on the cloud's density regime. At low-to-moderate surface densities ($\Sigma_{\rm cl}\lesssim 1000$ M$_\odot$ pc$^{-2}$), higher metallicity boosts SFE because enhanced metal-line cooling weakens photoionisation feedback. Conversely, in very dense clouds, the trend reverses: increased metallicity and thus dust opacity traps infrared radiation, building intense internal radiation pressure that opposes gravity and suppresses SFE.
We synthesize most of these complementary findings into a cloud-scale SFE model based on the semi-analytic framework of \citet{Kim2018}. This framework describes the SFE as a function of cloud mass $M_{\rm cl}$ and surface density $\Sigma_0$, governed by two limiting regimes: photoionisation-limited and radiation-pressure limited.

\subsubsection{Photoionisation-limited model}
In this regime, UV photons from young stars create a pressure-driven H\,{\small II} region that photoevaporates the cloud. The SFE is set by a mass budget in which gravity (tracked by the initial column density $\Sigma_0$) competes with the time-integrated photoevaporation rate. 
Following \citet{Kim2018}, we define a characteristic surface density, 
\begin{equation}
    \Sigma_{\rm ion} = \mu_{\rm H} c_i \left(\frac{\Xi}{8G\alpha_B}\right)^{1/2} = 140 \left(\frac{\Xi}{\Xi_0}\right) \left(\frac{T_{\rm ion}}{8000\,\rm K}\right)^{0.85} \msun\,\rm pc^{-2},
\end{equation}
which is the gas column density that photoionisation heating can overcome or the amount of gas an H\,{\small II} region can plausibly evaporate. It increases with the ionising emissivity $\Xi$ and the ionised sound speed $c_i$ at temperature $T_{\rm ion}$; clouds with $\Sigma_0 \lesssim \Sigma_{\rm ion}$ are readily eroded (low SFE), whereas $\Sigma_0 \gg \Sigma_{\rm ion}$ are resistant (high SFE). $\mu_H$ is the mean mass per hydrogen nucleus, and $\alpha_B$ is the case-B recombination coefficient.
To account for departures from an idealised, smooth H\,{\small II} region (e.g. porous geometry (greater exposed surface area), clumping and self-shielding) we introduce a dimensionless factor $\phi_{\rm ion}$ that scales the instantaneous photoevaporation rate. Analogously, to account for source time variability and for how the H\,{\small II} region expands or is confined in a finite cloud, we introduce a dimensionless duration $\phi_t$ that sets how long strong photoevaporation operates. Together these give a time-integrated evaporated column density $(\phi_t \phi_{\rm ion}) \Sigma_{\rm ion}$. \citet{Kim2018} calibrated the product from RHD simulations and found a weak increase with the initial surface density, interpreted as denser clouds maintaining high-pressure, photoevaporating interfaces for longer:
\begin{equation}
    \phi_t \phi_{\rm ion} = -2.89 + 2.11\, \log_{10} \left(\frac{\Sigma_0}{M_\odot\mathrm{pc}^{-2}} + 25.3\right),
\end{equation}
a dimensionless quantity typically of order unity to a few across typical $\Sigma_0$.

Following \citet{Kim2018}, we partition mass conservation at cloud dispersal relative to the initial mass $M_0$:
\begin{align}
    1 &= \epsilon_\star + \epsilon_{\rm ion} + \epsilon_{\rm ej,turb},
\end{align}
where $\epsilon_\star \equiv M_\star/M_0$ and $\epsilon_{\rm ion} \equiv M_{\rm ion}/M_0$ normalise the total stellar mass formed ($M_\star$) and the photoevaporated gas mass ($M_{\rm ion}$), and $\epsilon_{\rm ej,turb}$ accounts for the mass fraction lost to initial turbulent clearing prior to stellar feedback. We assume this ejected gas promptly evacuates the core volume, leaving an active mass $M_{\rm active} = M_0(1-\epsilon_{\rm ej,turb})$ subject to photoionisation. Physically, this clearing carves low-density channels through the cloud structure, allowing radiation to penetrate the remaining mass without being wasted on the evacuated gas. Consequently, the active surface density scales linearly with this mass reduction, $\Sigma_{\rm active}=\Sigma_0(1-\epsilon_{\rm ej,turb})$. Unless stated otherwise, we adopt $\epsilon_{\rm ej,turb} = 0.13$, consistent with simulations indicating $10-15\%$ prompt turbulent losses prior to radiative feedback for an initial virial parameter of $\alpha_{\rm vir}=2$ \citep{Raskutti2016}.
To model the remaining active cloud self-similarly, we scale the local photoevaporation efficiency relative to the post-turbulent mass. The photoevaporated mass fraction scales then as $\epsilon_{\rm ion, active} = \phi_t \phi_{\rm ion} ( \Sigma_{\rm ion}/\Sigma_{\rm active} ) \epsilon_{\star, \rm active}^{1/2}$, where $\epsilon_{\star, \rm active} \equiv M_\star/M_{\rm active} = \epsilon_\star / (1-\epsilon_{\rm ej,turb})$. Substituting our expressions for $M_{\rm active}$ and $\Sigma_{\rm active}$ to translate this scaling relation back relative to $M_0$ yields
\begin{align}
    \epsilon_{\rm ion} &= \phi_t \phi_{\rm ion} \left( \frac{\Sigma_{\rm ion}}{\Sigma_0\sqrt{1-\epsilon_{\rm ej,turb}}} \right) \epsilon_\star^{1/2}.
\end{align}
Introducing the pristine baseline parameter $\xi \equiv \frac{\Sigma_0}{(\phi_t \phi_{\rm ion}) \Sigma_{\rm ion}}$, the global mass balance reduces to the quadratic equation $\xi\sqrt{1-\epsilon_{\rm ej,turb}}\,y^2 + y - \xi(1-\epsilon_{\rm ej,turb})^{1.5} = 0$, whose positive root provides the photoionisation-limited SFE:
\begin{equation}
    \epsilon_{\star} = (1 - \epsilon_{\rm ej,turb}) \left( \frac{2\xi(1 - \epsilon_{\rm ej,turb})}{1 + \sqrt{1 + 4\xi^2(1 - \epsilon_{\rm ej,turb})^2}} \right)^2.
\end{equation}
This expression limits to $\epsilon_{\star}\simeq (1 - \epsilon_{\rm ej,turb})\xi^2$ for $\xi\ll 1$ and $\epsilon_{\star}\to (1 - \epsilon_{\rm ej,turb})$ for $\xi\gg 1$.  Thus, while efficient photoevaporation ($\xi\ll 1$) yields a low SFE, clouds whose surface density greatly exceeds their photoevaporative capacity ($\xi\gg 1$) exhibit SFE saturating to $(1 - \epsilon_{\rm ej,turb})$. Because $\xi \propto \Sigma_0$, photoevaporation becomes increasingly ineffective at halting star formation at high surface density. 

\subsubsection{Radiation pressure-limited model}
In the high–surface-density regime, cloud disruption is set by momentum balance. We assume (i) a fixed turbulent ejection fraction $\epsilon_{\rm ej,turb}$ prior to feedback (the same $\epsilon_{\rm ej,turb}=0.13$ we used above), (ii) a single characteristic mass-weighted ejecta speed $v_{\rm ej}$, and (iii) that the cumulative radial momentum injected by feedback scales with stellar mass as $(p_\star/m_\star) M_\star$. Balancing injected and required ejecta momentum at dispersal,
\begin{equation}
    \left(\frac{p_\star}{m_\star}\right) M_\star \;=\; v_{\rm ej}\,\Big[(1-\epsilon_{\rm ej,turb})M_0 - M_\star\Big],
\end{equation}
yields the radiation–pressure–limited SFE
\begin{equation}
    \label{eq:sfe_rp0}
    \epsilon_{\star {\rm RP},0} \;=\; \frac{1-\epsilon_{\rm ej,turb}}{1 + (p_\star/m_\star)/v_{\rm ej}}.
\end{equation}
Calibration from the RHD simulations of \citet{Kim2018} gives the total radial momentum ejected per unit stellar mass formed,
\begin{align}
    \frac{p_\star}{m_\star} &= 135~{\rm km\,s^{-1}}\left(\frac{\Sigma_0}{10^2~M_\odot~{\rm pc^{-2}}}\right)^{-0.74}. \label{eq:p_over_m}
\end{align}
The mass-weighted ejecta velocity is fit as
\begin{align}
    v_{\rm ej} &= 4.15\left(\frac{M_0}{M_\odot}\right)^{0.143}~{\rm km\,s^{-1}}, \label{eq:vej}
\end{align}
a power-law interpolation of the values measured by \cite{Kim2018} in their three cloud mass simulations ($v_{\rm ej} \simeq 15$, $23$, and $29$~km~s$^{-1}$ for $M_{\rm cl}=10^4,10^5,10^6\,\msun$, respectively). 

Both the radial momentum (Eqn.~\ref{eq:p_over_m}) and ejecta velocity (Eqn.~\ref{eq:vej}) are calibrated for a fiducial IMF with an ionising-photon yield $\Xi_0$. 
A more top‑heavy IMF produces a larger UV luminosity per stellar mass ($\Psi_{\rm UV}$) and hence a larger ionising yield $\Xi$, which increases the momentum injected per unit stellar mass and lowers the SFE. To quantify this dependence, we adopt the local slope measured by \citet{Menon2024} at $M_0=10^6\,\msun$, $Z=10^{-2}Z_\odot$, and $\Sigma_0=3\times10^3\,\msun{\rm pc^{-2}}$:
\begin{equation}
    \frac{{\rm d}\epsilon_\star}{{\rm d}\Psi_{\rm UV}} \simeq -\frac{1}{30},
\end{equation}
and assume $\Xi\propto\Psi_{\rm UV}$. We express the UV output in units of its fiducial value by defining
\begin{equation}
    \chi \equiv \frac{\Xi}{\Xi_0} \approx \frac{\Psi_{\rm UV}}{\Psi_{{\rm UV},0}},
\end{equation}
so that $\chi=1$ for the fiducial IMF and the \cite{Menon2024} sensitivity applies per unit change in $\chi$ (i.e. in these normalised units).
We then define the SFE ($\epsilon$) relative to the post‑turbulent gas mass $M_0=(1-\epsilon_{\rm ej,turb})M_{\rm cl}$ and approximate $\epsilon$ by a first‑order Taylor expansion about $\chi=1$:
\begin{align}
    \epsilon(\chi) &\equiv \frac{M_\star}{M_0} \simeq \epsilon_0 + \left.\frac{{\rm d}\epsilon}{{\rm d}\chi}\right|_{\chi=1}(\chi-1) =\epsilon_0 \left[1 - \frac{\chi-1}{30}\right] \nonumber \\
    &= \epsilon_0 \frac{31-\chi}{30}, 
    \quad \epsilon_0 \equiv \epsilon(\chi{=}1).
\end{align}
Physically, as the IMF becomes more top‑heavy ($\chi$ increases), the fraction of the available gas that turns into stars declines, and the fraction expelled by feedback rises.
In the radiation‑pressure–limited model, momentum balance implies that the ejected‑to‑stellar mass ratio equals the momentum ratio:
\begin{equation}
    \frac{M_{\rm ej}}{M_\star} = \frac{p_\star/m_\star}{v_{\rm ej}}
    \quad\Rightarrow\quad
    1+\frac{p_\star/m_\star}{v_{\rm ej}} = \frac{1}{\epsilon}.
\end{equation}
Thus the Kim‑baseline denominator rescales with $\chi$ as
\begin{equation}
    \Bigg[1+\frac{p_\star/m_\star}{v_{\rm ej}}\Bigg]_{\chi}
    = \frac{1}{\epsilon(\chi)}
    = \frac{30}{31-\chi}\frac{1}{\epsilon_0}
    = \frac{30}{31-\chi}\Bigg[1+\frac{p_\star/m_\star}{v_{\rm ej}}\Bigg]_{\chi=1}.
\end{equation}
Substituting into Eqn.~\ref{eq:sfe_rp0} gives the IMF‑corrected radiation-pressure-limited SFE:
\begin{align}
    \epsilon_{\star {\rm RP}}(\chi)
    = \frac{31-\chi}{30} \frac{1-\epsilon_{\rm ej,turb}}{\Big[1+(p_\star/m_\star)/v_{\rm ej}\Big]_{\chi=1}}
    = \frac{31-\chi}{30} \epsilon_{\star {\rm RP},0}.
\end{align}
Here $\epsilon_{\star,{\rm RP},0}$ is the Kim‑baseline radiation-pressure-limited SFE (fiducial IMF), and $(p_\star/m_\star)(\Sigma_0)$ and $v_{\rm ej}(M_0)$ are given by Eqns.~\ref{eq:p_over_m}–\ref{eq:vej}. This formulation makes explicit that a more top‑heavy IMF ($\chi>1$) linearly suppresses the radiation-pressure‑limited SFE relative to the fiducial case. 
We note the linear correction is calibrated locally (for the stated $M_0$, $Z$, $\Sigma_0$) and is most reliable for modest deviations from $\chi=1$. At very high $\Sigma_0$ and $Z$, indirect (IR) radiation pressure on dust may further reduce $\epsilon_\star$, which we do not include.

\subsubsection{Transition between regimes}
There is no clear consensus on where exactly photoionisation ceases to regulate star formation and radiation pressure takes over \citep{krumholz_star_2019}, but the transition appears to occur when clouds escape velocities reach around $10-20\rm km\,s^{-1}$. We model the transition as occurring when the escape velocity equals the ionised gas sound speed, $v_{\rm esc}\sim c_i$, which typically lies in this range. The ionised sound speed is $c_i = \sqrt{\gamma k_B T_e/(\mu_H m_H)}$ with adiabatic index $\gamma$, electron temperature $T_e$, mean particle mass $\mu_i$ (for fully ionised gas), and hydrogen mass $m_H$. The electron temperature depends on the gas metallicity of the cloud as metal lines dominate cooling in H\,{\small II} regions. Using the photoionisation code \textsc{cloudy}, \cite{balser_metallicity-electron_2024} find that $T_e$ correlates with the gas-phase metallicity in a manner well fit by the \citet{shaver_galactic_1983} parameterisation:
\begin{align}
    \log_{10}{\rm O/H} + 12 &= 9.82 - 1.49 \frac{T_e}{10^4\rm K}
\end{align}
They report slight variation with cloud density ($10^1-10^4\rm cm^{-3}$) and only weak sensitivity to the ionising source temperature ($T_{\rm eff}=44616\mathrm{K},\,38151\mathrm{K},$ and $31524\mathrm{K}$ for an O3, O6, and O9 star, respectively), consistent with \cite{Menon2024}, who argue that the spectral shape in the UV, and hence the electron temperature, is only weakly affected by the IMF top-heaviness.
For a uniform, spherical cloud of mass $M$ and radius $R$, the escape speed is $v_{\rm esc} \sim \sqrt{GM/R}$, and the surface density is $\Sigma_{\rm cl} = M/(\pi R^2)$ where $R=(3M/4\pi\rho)^{1/3}$, and $\rho$ is the mean volume density, 
and we take $M=(1-\epsilon_{\rm ej,turb})M_{\rm cl}$.
We combine the photoionisation‑regulated and radiation-pressure‑regulated SFEs with a smooth transition controlled by $v_{\rm esc}/c_i$:
\begin{align}
    \epsilon_\star &= \epsilon_{\star \rm PH} e^{-(v_{\rm esc}/c_i)^a} + \epsilon_{\star \rm RP} \1 1-e^{-(v_{\rm esc}/c_i)^a}\2
\end{align}
where $a$ sets the sharpness of the transition (we adopt $a=2.3$). The ionising emissivity $\Xi$ entering the radiation pressure and photoionisation terms is defined at $t=0$ (Section~\ref{sec:LUV}).

We note that the assumption that the ionised gas temperature is IMF independent may not be entirely valid when $\mlim$ becomes small, as we are then missing the hot, massive end of the IMF entirely, and not simply, as \cite{Menon2024} assumes, changing the slope of the IMF. It is not expected that this will have a great impact on any results, however.

\subsubsection{SFE dependency on cloud mass and IMF}
\begin{figure*}
    \centering
    \includegraphics[width=\linewidth]{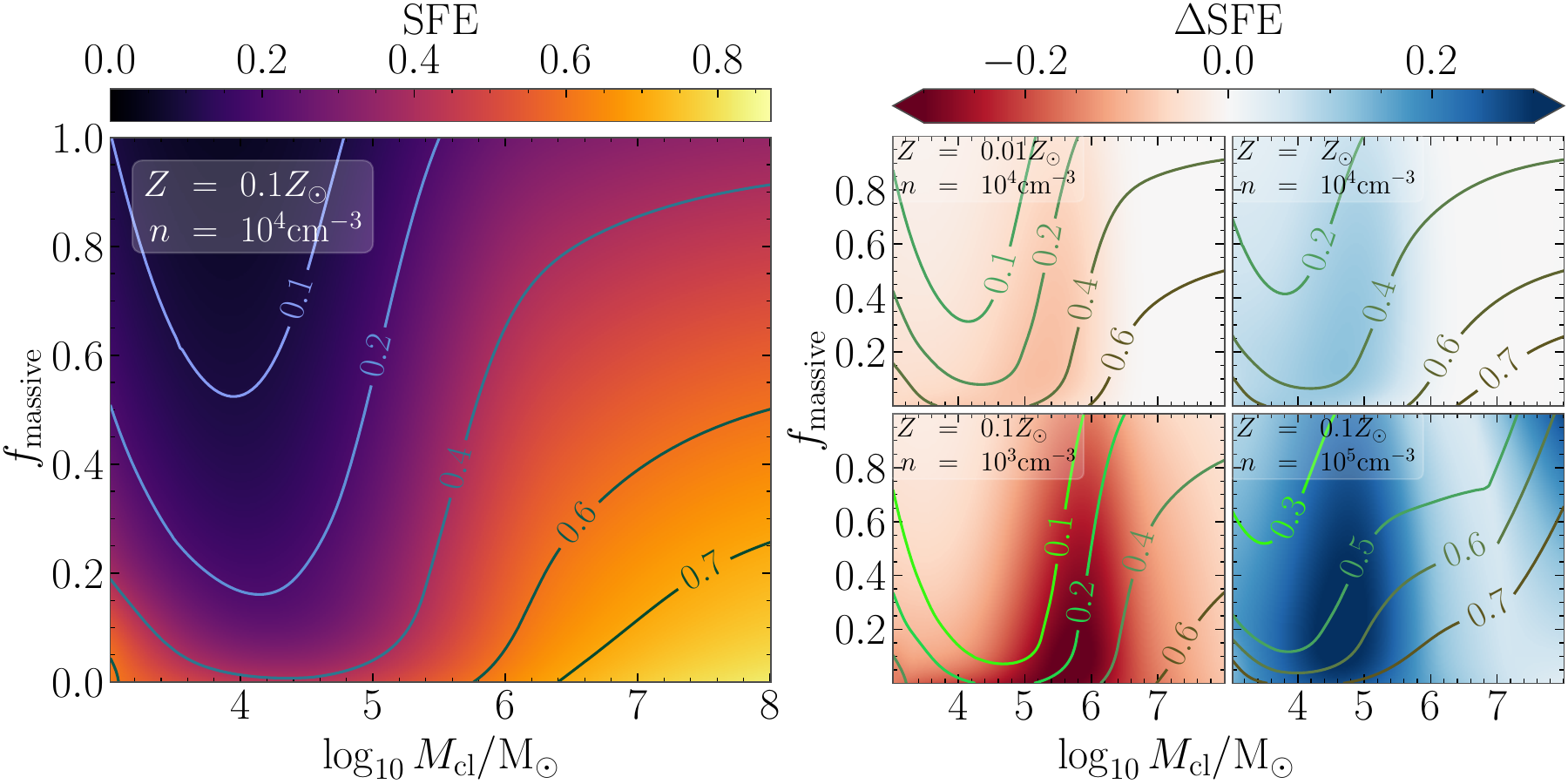}
    \caption{Star formation efficiency as a function of $\mcl$ and $\fmassive$ with the Cloud Mass Limited evolving IMF. \textbf{Left:} $\epsilon_\star$ with $Z=0.1Z_\odot$ and $n_0=10^4\rm cm^{-3}$. Contours show lines of constant SFE. \textit{Right:} Difference in SFE from the left panel for smaller and larger (\textit{left and right columns}) metallicity and density (\textit{top and bottom rows}). Contours show the SFE for these values of $Z$ and $n_0$. The bottom of each panel, with $\fmassive=0$ is equivalent to the Cloud Mass Limited Salpeter IMF.
    }
    \label{fig:SFE}
\end{figure*}

In Fig.~\ref{fig:SFE} we show how the SFE in clouds depends on $\fmassive$ and cloud mass (left panel), and how the SFE changes with metallicity, and cloud density (right panels), for the cloud mass limited evolving IMF. In lower mass clouds ($\mcl\lesssim10^4\msun$) we see the SFE decreases with increasing mass. This is due to $\mlim$ increasing with cloud mass in this regime, which in turn increases $\Xi$. In the same way, the SFE decreases with increasing $\fmassive$, as a more top-heavy IMF also increases $\Xi$. This results in a difference of about $0.4$ in $\epsilon_\star$ from $\fmassive=1$ to $0$. In more massive clouds ($\mcl\gtrsim 10^4\msun$) where the IMF is fully sampled, the SFE increases with cloud mass, providing a difference in SFE of about $0.4$ between $\mcl=10^4\msun$ and $10^8\msun$. 
On the top right panels in Fig.~\ref{fig:SFE} we see how cloud SFE increases with increasing metallicity in the photoionisation-regulated regime, where $\mcl\lesssim 10^6 \msun$, as more metal-rich clouds cool more efficiently and thus can form stars more efficiently, leading to a difference in $\epsilon_\star$ of up to about $0.2$ between $Z=0.01Z_\odot - Z_\odot$. In the radiation pressure-regulated regime ($\mcl\gtrsim10^6\msun$), metallicity does not meaningfully affect the SFE per our modelling, a dependency we aim to include in future work. The lower right panels of Fig.~\ref{fig:SFE} finally show how cloud SFE increases  with increasing cloud density. We see in particular how the higher escape velocities of denser clouds shift the transition between the photoionisation-limited and radiation pressure-limited regimes to lower cloud masses, which increases the SFE of lower mass clouds, leading to a an SFE difference peaking at about $0.3$ in clouds with $10^{5.3}\msun\lesssim\mcl\lesssim10^{6.3}\msun$ between $n=10^3-10^4\rm cm^{-3}$, and a similar difference peaking in clouds with $10^4\msun\lesssim\mcl\lesssim10^{5}\msun$ between $n=10^4-10^5\rm cm^{-3}$.

% ----------------------------------------------------
\subsection{Galaxy-wide star formation and supernova feedback}\label{sec:snfeedback}
% ----------------------------------------------------

Core-collapse supernovae (SNe) from massive stars heat and expel interstellar gas, reducing the cold-gas reservoir and thereby suppressing star formation, especially in low-mass halos with shallow potential wells. We adopt the delayed-feedback model used in \textsc{astraeus} \citep{hutter_astraeus_2025}. Type II SNe events occur as massive stars ($8~\msun\leq m\leq50~\msun$) die according to the stellar mass-lifetime relation \citep{padovani_stellar_1993},
\begin{equation}\label{eq:sntime}
    t_{\rm SN} = \31.2\times10^3\1\dfrac{M}{\msun}\2^{-1.85}+3\4\rm Myr,
\end{equation}
releasing an energy of $E_{51}= 10^{51}\rm erg$ per explosion. So $8~\msun$ and $50~\msun$ stars explode after $28.6~\rm Myr$ and $3.86~\rm Myr$, respectively. 
This SN feedback shapes the galaxy-wide star formation efficiency. As in \citet{cueto_astraeus_2024}, we model it as the minimum of an ejection-limited efficiency ($f_\star^{\rm ej}$ dominant in low-mass halos) and a baseline, dynamical-time–scaled efficiency ($f_\star^{\rm max}$, relevant for higher-mass halos):
\begin{align}\label{eq:fstareff}
    f_\star^{\rm eff} &= \mathrm{min}(f_\star^{\rm max},f_\star^{\rm ej}).
\end{align}
Equating the SN energy coupled to winds with the diffuse-gas binding energy yields the ejection-limited efficiency at timestep $j$, i.e. the star-formation fraction required to unbind the remaining gas:
\begin{align}\label{eq:fej}
    f_\star^{\rm ej} &= 1 - \dfrac{f_w,E_{51},\sum_{k=0}^{j-1}\nu_{kj},M_{\star,k}^{\rm new}}{M_{\rm gas,ini}^j,v_{c,j}^2},,
\end{align}
where $v_{c,j}=\sqrt{G M_{\rm vir}/R_{\rm vir}}$ is the circular velocity of the galaxy at timestep $j$, $M_{\star,k}^{\rm new}$ is the stellar mass formed in timestep $k$, $\nu_{kj}$ is the IMF-dependent fraction of stellar mass formed in timestep $k$ and exploding as SNII in timestep $j$, and $f_w$ is the fraction of SN energy coupling to the gas and driving winds. We cap $f_\star^{\rm ej}$ to the range $[0,1]$; by construction, $f_\star^{\rm ej}$ is the maximum fraction of the gas reservoir convertible into stars before SNe unbind the rest (if $f_\star^{\rm ej}\le 0$, star formation is fully suppressed in that step).
This follows \cite{hutter_astraeus_2023}, except we omit SNe from stars formed in the current timestep because all our timesteps are shorter than the $\sim3~$Myr lifetime of the most massive stars. The derivation of $\nu_{kj}$ is given in Appendix B of \cite{hutter_astraeus_2023}.

The baseline (dynamical-time–scaled) efficiency is
\begin{align}\label{eq:fstar}
    f_\star^{\rm max} &= \frac{\Delta t}{20\ \mathrm{Myr}}\ \fstar \ \1\frac{9+1}{z+1}\2^{-3/2},
\end{align}
where $f_\star$ is a free normalisation parameter. The factor $\1\frac{9+1}{z+1}\2^{-3/2}=\frac{\tau_{\rm dyn}(z=9)}{\tau_{\rm dyn}(z)}$ with $\tau_{\rm dyn}(z)=\sqrt{R_{\rm vir}^3/(G M_{\rm vir})}\propto(1+z)^{-3/2}$, ensures scaling with the halo dynamical time at redshift $z$.
To account for quiescent intervals between starbursts, we define $\Delta t$ as the sum of the current timestep duration, $\Delta t_{\rm step}$, and the time elapsed since the end of the last star-forming episode. If quiescence is caused by SN-driven removal of all gas, the elapsed time is measured from when the galaxy resumes gas accretion after stripping. We reset this clock at the onset of each new star-forming episode; if no episode has yet occurred, we measure from the start of the simulation (or the halo’s first appearance).

The maximum stellar mass that can be formed in the current step is then,
\begin{align}
    M_{\star,\max}^j &= f_\star^{\rm eff}\, M_{\rm gas,ini}^j.
\end{align}

% ----------------------------------------------------
\subsection{Stellar yields}\label{sec:yields}
% ----------------------------------------------------

We compute time-delayed stellar metal enrichment following the \textsc{astraeus} scheme \citep[see][for details]{ucci_astraeus_2022}. We precompute lookup tables from \cite{kobayashi_origin_2020} as functions of initial stellar mass $m$ and metallicity $Z$, including gas yields and the effective metal ejecta mass per star, $m_y(m,Z) = y(m,Z) + Z\,(m-M_{\rm R}(m,Z))$, where $M_{\rm R}$ is the remnant mass and $y(m,Z)$ denotes the mass of newly synthesised metals ejected by the star. This quantity represents the total mass of metals returned to the ISM by a star of mass $m$ and metallicity $Z$, accounting for both newly synthesised and recycled pre-existing metals, while excluding material locked into the stellar remnant.
We adopt a mass–normalised IMF, $\phi(m)$, such that $\int_{m_{\min}}^{m_{\max}} m\,\phi(m)\,{\rm d}m = 1$ and $\phi$ has units of $\msun^{-1}$. In discrete mass bins of width $\Delta m = 0.1\,\msun$, this implies $\sum_i m_i\,\phi(m_i)\,\Delta m = 1$ for each IMF variant (including cloud–mass–limited upper–mass cuts).

At timestep $j$, we sum the ejecta from all past star–formation cohorts formed at timestep $k$ with metallicity bin $Z_\theta$ and IMF variant $v$ (standard or CML). For each cohort, we include only those initial–mass bins $D_{k\to j}$ whose lifetimes place their deaths within timestep $j$, i.e. $t_k + \tau(m_i, Z_\theta) \in [t_j,\, t_j+\Delta t_j]$. Using the metal ejecta mass $m_y(m_i,Z_\theta)$ for a star of initial mass $m_i$, the total metal mass ejected into the ISM during timestep $j$ is
\begin{equation}
    \Delta M_Z^{\,j} \;=\; \sum_{k=0}^{j-1}\sum_{\theta}\sum_{v} M_{\star}^{k\theta v}
    \sum_{i\in D_{k\to j}} \phi_v(m_i;\,m_{\max,v})\, m_y(m_i, Z_\theta)\,\Delta m,
\end{equation}
where $M_{\star}^{k\theta v}$ is the stellar mass formed at timestep $k$ with metallicity $Z_\theta$ under IMF variant $v$, and $m_{\max,v}$ is the corresponding IMF upper limit (for CML, cloud–dependent; see Appendix \ref{app:CMLIMFs}). For the standard IMF, the sum over $v$ reduces to a single term.

For dust, we use a fixed yield $y_{\rm SNII}=0.45\msun$ per SNII event and adopt the AGB dust yield from \citet{kobayashi_origin_2020}. Accounting also for dust grain growth, dust destruction by SN shock waves, and astration, we follow the dust evolution model of \citet{hutter_astraeus_2023} and \citet{dayal_alma_2022}.

% ----------------------------------------------------
\subsection{UV luminosities and ionising emissivities}\label{sec:LUV}
% ----------------------------------------------------

We compute the time‑dependent ionising emissivity and UV specific luminosity of each single‑age stellar population (SSP) formed in the current timestep and store these quantities for all subsequent timesteps. We generate SSP spectra with the stellar population synthesis (SPS) code \textsc{Starburst99} \citep{leitherer_starburst99_1999} on a grid of IMF parameters. The IMF consists of a Salpeter slope between $0.1\,\msun$ and $m_c$, followed by a log-flat slope between $m_c$ and $100\,\msun$. We sample a range of transition masses $m_c$ and maximum stellar masses $m_{\rm max}=10$--$100\,\msun$.
For each grid point, we fit a set of smooth piecewise power‑law functions to the SPS outputs, yielding analytic expressions for the UV specific luminosity and ionising photon production rate per unit initial stellar mass as functions of SSP age. Explicit expressions are provided in Appendix~\ref{app:SSP_lum_fits}.

% ---------------------------------------
\subsection{Dust attenuation in galaxies}\label{sec:dustattenuation}
% ---------------------------------------

Having computed the intrinsic UV luminosity, we model dust attenuation following the implementation adopted in \textsc{astraeus} \citep{hutter_astraeus_2023}.
We assume that dust and gas are uniformly mixed throughout each galaxy. The dust distribution radius is parameterised as
\begin{equation}
    r_{\rm dust} = 4.5\lambda R_{\rm vir} \left(\frac{1+z}{6}\right)^{1.8},
    \label{eq}
\end{equation}
where $R_{\rm vir}$ is the halo virial radius and $\lambda$ is the halo spin parameter. We adopt $\lambda=0.035$, corresponding to the characteristic value of the spin-parameter distribution in \textsc{Astraeus}. This yields a dust surface density $\Sigma_{\rm dust} = M_{\rm dust}/(\pi r_{\rm dust}^2)$, where $M_{\rm dust}$ is the total dust mass. Assuming graphite/carbonaceous dust grains with radius $a=0.03\,\mu{\rm m}$ and internal density $s=2.25\,{\rm g\,cm^{-3}}$ \citep{todini_dust_2001}, the optical depth to UV continuum photons is $\tau_{\rm UV} = 3\Sigma_{\rm dust}/(4as)$. Assuming a slab geometry, the observed UV luminosity is then given by
\begin{equation}
    L_{\rm UV}^{\rm obs} = f_{\rm esc} L_{\rm UV}^{\rm intr} ~=~ \left[ \frac{1-\exp(-\tau_{\rm UV})}{\tau_{\rm UV}} \right]\, L_{\rm UV}^{\rm intr},
    \label{eq:L_UV_obs}
\end{equation}
where $L_{\rm UV}^{\rm intr}$ is the intrinsic luminosity defined in Section~\ref{sec:LUV} and $f_{\rm esc}$ the UV escape fraction. Throughout the remainder of this work, $L_{\rm UV}$ and $M_{\rm UV}$ denote dust-attenuated (observed) luminosities and magnitudes unless stated otherwise.

% -----------------------------------------------------------
\subsection{Simulations and weighting by the halo mass function}\label{sec:halomassfunction}
% -----------------------------------------------------------

For each set of simulation parameters, we generate 12 realisations. We sample nine values of $\beta$ uniformly in the range $-0.55$ to $-0.95$ (step 0.04) and 300 values of $\mhalo(z=0)$ spanning $10^{10}$--$10^{14.5},\msun$ (0.015 dex spacing), yielding a uniform sampling in halo mass.
The halo mass distribution in the Universe is non-uniform, with high-mass halos being progressively rarer as described by the halo mass function (HMF). To account for this, we weight each simulated galaxy by the HMF evaluated at its $z=4.5$ halo mass. The HMF is computed using the \textsc{HaloMod} calculator \citep{murray_hmfcalc_2013, murray_thehalomod_2021} assuming a Planck13 cosmology. This ensures that derived statistics are representative of a cosmological halo population rather than the uniform halo-mass sampling of the simulations.

%%%%%%%%%%%%%%%%%%%%%%%%%%%%%%%%%%%%%%%%%%%%%%%%%%%%%%%%%%%%%
\section{Sensitivity of star-formation burstiness to physical parameters}\label{sec:burstiness}

Bursty star formation emerges from the competition between (i) feedback-driven gas evacuation/heating (setting post-burst quenching depth and duration) and (ii) IGM gas accretion and cloud assembly (setting the cloud formation rate). 
We explore how four parameters modulate this balance: the maximum cloud mass ($M_{\rm cl}^{\max}$), the stellar IMF, the cloud star‑formation efficiency (SFE), and the characteristic cloud density. Our \texttt{Fiducial} model adopts $M_{\rm cl}^{\max}=10^8\,{\rm M_\odot}$ \citep[cf.][]{Chen2026}, a cloud-mass-limited evolving (top‑heavy) IMF (Section~\ref{sec:IMFs}), a mass‑dependent SFE, and $n_s=5\times 10^5$. 
To isolate the impact of each physical mechanism, we compare this baseline against four variant models where only one parameter is altered: 
(i) \texttt{Cloud} model ($M_{\rm cl}^{\max}=10^7\,{\rm M_\odot}$), 
(ii) \texttt{IMF} model (cloud‑mass–limited Salpeter IMF), 
(iii) \texttt{SFE} model (constant $\epsilon_\star=0.45$), and 
(iv) \texttt{Density} model ($n_s=1.5\times 10^6$).

\begin{figure*}
    \centering
    \includegraphics[width=1\linewidth]{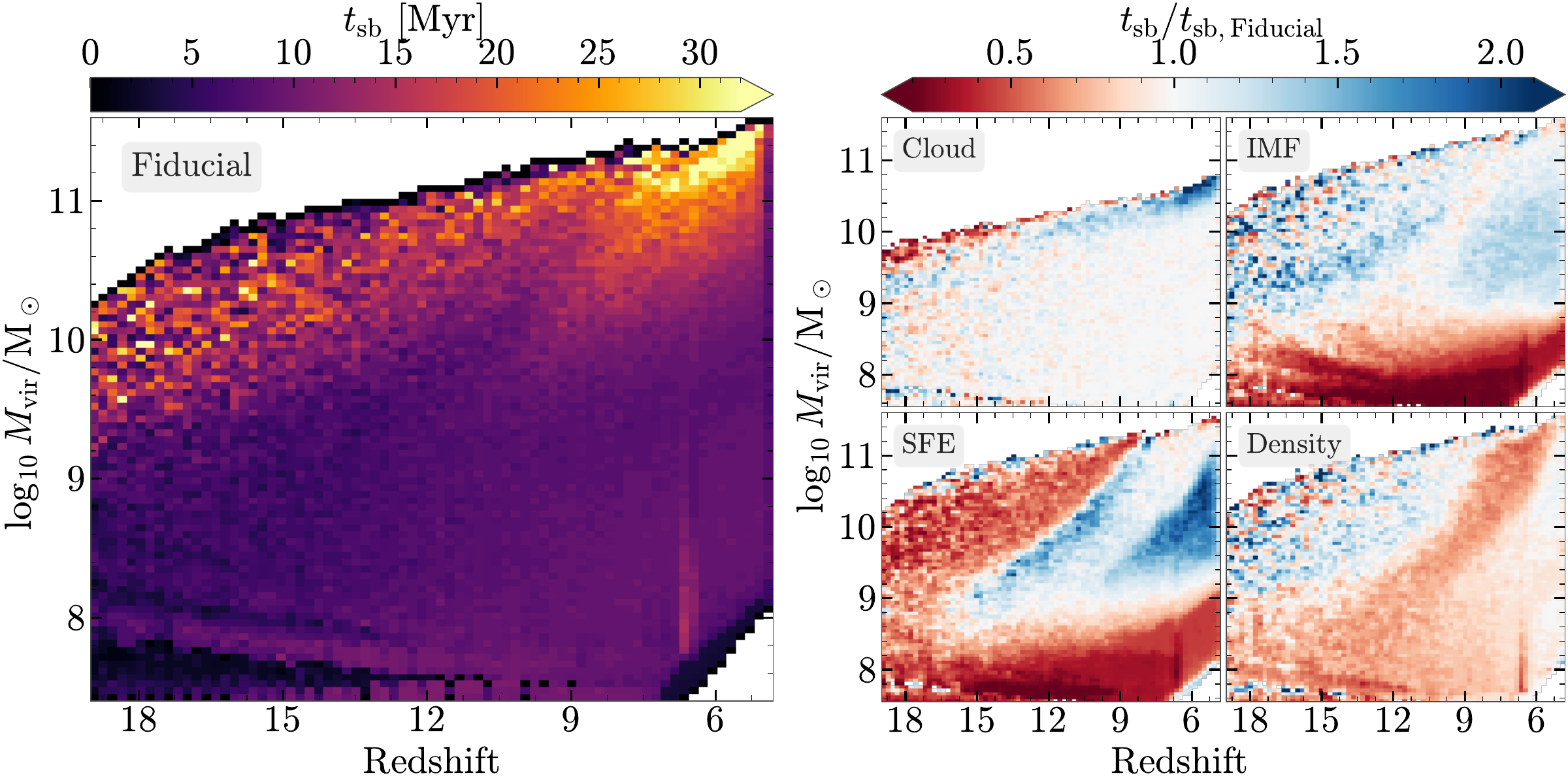}
    \caption{\textit{Left:} Median duration of starbursts as a function of halo mass and redshift for the \texttt{Fiducial} model. \textit{Right:} Fractional change in median starburst duration between the \texttt{Fiducial} model and the four variation models, $t_{\rm sb}/t_{\rm sb,Fiducial}$.}
    \label{fig:sb}
\end{figure*}
\begin{figure*}
    \centering
    \includegraphics[width=1\linewidth]{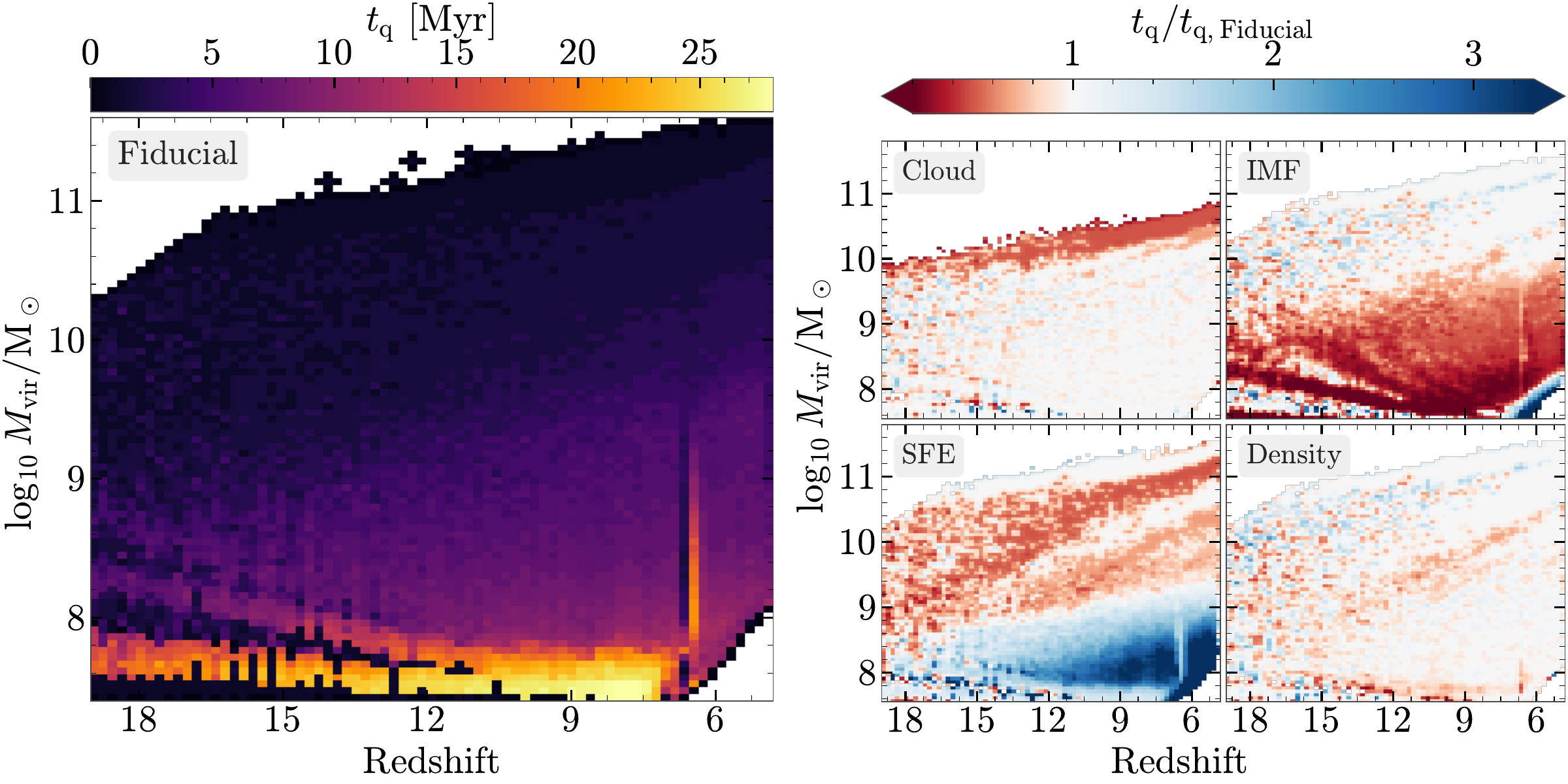}
    \caption{\textit{Left:} Median duration of quiescent periods as a function of halo mass and redshift for the \texttt{Fiducial} model. \textit{Right:} Fractional change in median quiescent duration between the \texttt{Fiducial} model and the four variation models, $t_{\rm q}/t_{\rm q,Fiducial}$.}
    \label{fig:q}
\end{figure*}

\subsection{Fiducial picture: feedback-dominated, cloud-mass-limited, and continuous star formation regimes}
\label{sec:fiducial-regimes}

Fig.~\ref{fig:sb} and \ref{fig:q} (left panels) show the median burst ($t_{\rm sb}$; contiguous intervals with $\mathrm{SFR} > 0$) and quiescent ($t_{\rm q}$; contiguous intervals with $\mathrm{SFR} = 0$) durations as a function of halo mass and redshift for the \texttt{Fiducial} model. We define the bursty phase as the 
evolutionary period prior to sustained star formation, when the system has not 
yet reached a regime of uninterrupted star formation.
The global burst-quench cycle is primarily governed by the competition between the gas accumulation timescale ($t_{\rm form} \simeq M_{\rm cl,max}/\dot{M}_{\rm gas}$), representing the time required to assemble the most massive cloud, and the effective SN feedback timescale ($t_{\rm SN}$). When $t_{\rm form} \gtrsim t_{\rm SN}$, feedback disrupts the gas reservoir before replenishment, driving bursty star formation with extended quiescent phases. Conversely, when $t_{\rm form} \lesssim t_{\rm SN}$, gas accumulates faster than feedback can suppress star formation, resulting in shorter quiescent phases and eventually continuous star formation.
Alongside this global timescale balance, the internal evolution of individual clouds introduces an additional modulation through the cloud SFE. Depending on cloud surface density, the SFE is set either by photoionisation ($\Sigma_0 < \Sigma_{\rm crit}$) or radiation pressure ($\Sigma_0 > \Sigma_{\rm crit}$), with the transition occurring at $\Sigma_0 = \Sigma_{\rm crit}$. Because $\Sigma_0 \propto M_{\rm cl}^{1/3} n_0^{2/3} \propto M_h^{1/3} (1+z)^{5/2}$, the transition shifts to higher halo masses toward lower redshift.
Driven by these mechanisms, we identify three distinct evolutionary regimes across the halo mass - redshift space:

\noindent (1) \emph{Feedback-dominated regime} ($\mhalo \lesssim 10^{8.5}\,\msun$): 
Because $t_{\rm form} > t_{\rm SN}$, individual starbursts inject sufficient energy to efficiently evacuate the gas reservoir, suppressing star formation until fresh gas can be accreted. Consequently, these halos exhibit long quiescent periods ($\tq \sim 30\,\rm Myr$ vs. $\sim 10\,\rm Myr$ in massive systems). Once the gas reservoir replenishes, a new cloud rapidly forms and the cycle resumes. 
Starbursts are also extended in this regime ($\tsb \lesssim 12\,\rm Myr$). This reflects the fact that most clouds in low-mass halos are insufficiently massive to fully sample the IMF and exhibit lower SFEs due to photoionisation regulation. This combination of inefficient gas-to-star conversion and a deficit of massive stars postpones the onset of SN feedback, allowing star formation to proceed for longer before it can be suppressed by SN feedback.

\noindent (2) \emph{Cloud-mass-limited regime} ($10^{8.5} \lesssim \mhalo \lesssim 10^{10.5}\,(10^{9.5})\,\msun$ at $z=5\,(15)$): 
Here, $t_{\rm form} \sim t_{\rm SN}$, meaning starbursts are terminated by the finite ability of the system to continuously assemble clouds rather than by feedback-driven  gas evacuation. This regime exhibits two distinct trends in $t_{\rm sb}$, reflecting the changing balance between cloud-scale feedback and gas reservoir growth; at high redshifts and in more massive halos, the burst duration is short and approximately constant ($\tsb \sim 6$--$8\,\rm Myr$), whereas at lower redshifts and in less massive halos, $t_{\rm sb}$ first decreases and then increases with halo mass.
In the high-redshift, massive-halo branch, the SFE is regulated by radiation pressure and is high, rapidly converting accreted gas into stars. The burst duration is thus controlled by the cloud assembly rate and the maximum cloud mass. Since both $M_{\rm cl}^{\max}$ and the gas accretion rate increase proportionally with halo mass, their ratio (setting $t_{\rm form}$) remains approximately constant, leaving $t_{\rm sb}$ invariant. 
Conversely, in less massive halos at lower redshifts, the SFE is lower under photoionisation regulation, allowing gas to accumulate between bursts. The initial decrease of $t_{\rm sb}$ up to $M_h \lesssim 10^{9.5}\,\msun$ occurs because increasingly massive clouds sample the high-mass IMF more completely, triggering earlier and more effective SN feedback. At higher masses, the trend reverses and $t_{\rm sb}$ increases again, because the higher gas accretion rate allows the reservoir to sustain massive cloud formation more continuously.
Meanwhile, the quiescent duration ($t_{\rm q}$) decreases continuously from $\sim 10\,\rm Myr$ to zero with increasing mass and redshift, showing no such division. This suggests that $t_{\rm q}$ is primarily governed by the recovery time of the gas reservoir rather than internal cloud physics. Increasing halo masses deepen the gravitational potential well and reduce gas ejection efficiency while elevating gas accretion rates, which shortens the time required to replenish the star-forming reservoir. At higher redshifts, the enhanced gas accretion rates further accelerate this recovery, progressively reducing the duration of quiescent phases toward zero.

\noindent (3) \emph{Continuous star formation} ($\mhalo \gtrsim 10^{10.5}\,(10^{9.5})\,\msun$ at $z=5\,(15)$): 
Gas is replenished and assembled into star-forming clouds more rapidly than SN feedback can regulate the star formation cycle ($t_{\rm form} \ll t_{\rm SN}$), leading to a breakdown of the burst–quench structure. Galaxies then continuously form the most massive clouds allowed by the model’s upper mass limit, resulting in effectively uninterrupted star formation. 
In this regime, the measured $\tsb$ reflects the duration of these uninterrupted star-forming episodes rather than isolated bursts, while quiescent phases vanish ($\tq \rightarrow 0$) because feedback cannot halt ongoing gas inflow and star formation. 
The transition to this regime is characterized by a sharp increase in $\tsb$ to $\sim 20$--$30\,\rm Myr$, tracking the sustained cloud formation at the maximum cloud mass scale. At higher redshifts ($z \gtrsim 15$), elevated gas accretion rates further reduce $t_{\rm form}$, allowing some galaxies to reach this continuous star-forming state before fully saturating the cloud-mass limit. This premature transition produces the steeper gradients in $\tsb$ and $\tq$ seen at the high-mass end of the colormaps in Figs.~\ref{fig:sb} and \ref{fig:q}.

In summary, our framework suggests that star formation in early galaxies progresses through distinct physical phases: transitioning from feedback-regulated burstiness in low-mass halos to continuous star formation in massive ones, driven by a systematic reduction in quiescent timescales and a shift from short, cloud-regulated bursts to sustained star formation across the intermediate-to-high mass regime.

\subsection{Variations between models}\label{sec:model_variations}

Modifying core parameters alters these evolutionary phases by shifting the underlying balance between feedback and gas accumulation.

\begin{itemize}
    \item \textbf{The \texttt{Cloud} model}, assuming an order-of-magnitude lower maximum cloud mass ($M_{\rm cl}^{\rm max}=10^7\,\msun$), closely follows the \texttt{Fiducial} model for $\mhalo \lesssim 10^{10}\,(10^{9})\,\msun$ at $z=5\,(15)$. Above this mass, galaxies reach the imposed upper cloud-mass limit sooner, roughly $1$~dex lower in halo mass than in the \texttt{Fiducial} model, which shortens $t_{\rm form}$ and forces systems into the continuous star-forming regime ($t_{\rm form} \ll t_{\rm SN}$) at lower masses. Consequently, the bursty phase terminates earlier, leading to a decrease in $t_{\rm q,Cloud}/t_{\rm q,Fiducial}$ and a corresponding increase in $t_{\rm sb,Cloud}/t_{\rm sb,Fiducial}$.\footnote{For massive halos ($\mhalo\gtrsim10^{9.5}\,\msun$) at $z\gtrsim12$, the decrease in $t_{\rm sb,Cloud}/t_{\rm sb,Fiducial}$, may be driven by low-number statistics.}

    \item \textbf{The \texttt{IMF} model} assumes a cloud-mass-limited Salpeter IMF. As shown in the top-right panels of Figs.~\ref{fig:sb} and \ref{fig:q}, the impact of this change depends strongly on halo mass. 
    These variations arise because modifying the IMF alters two coupled mechanisms: (i) the fraction of stellar mass allocated to high-mass stars per unit $M_\star$, setting the SN energy and ionising luminosity per cloud, and (ii) the total stellar mass ($M_\star = \epsilon_\star M_{\rm cl}$) assembled before feedback halts star formation via the IMF–SFE coupling. Together, these factors govern the stochastic sampling of the IMF within each cloud, determining the realised maximum stellar mass ($m_{\max}(M_\star)$) that triggers the first SN and dissolves the cloud.
    
    In low-mass halos ($M_h \lesssim 10^9\,\msun$), both timescales are reduced relative to the \texttt{Fiducial} model ($t_{\rm sb,IMF}/t_{\rm sb,Fiducial} \simeq 0.3$--$0.5$ and $t_{\rm q,IMF}/t_{\rm q,Fiducial} \simeq 0.3$--$0.5$). 
    Typical cloud masses ($M_{\rm cl}$) here are small, making IMF sampling incomplete and $m_{\max}(M_\star)$ highly sensitive to the total stellar mass budget. Moving from the evolving \texttt{Fiducial} IMF to a Salpeter IMF reduces the high-mass fraction per unit stellar mass, which weakens the feedback per unit stellar mass and boosts the cloud SFE ($\epsilon_\star$) via IMF--SFE coupling, yielding a higher $M_\star$ per cloud.
    The resulting higher $M_\star$ per cloud raises the higher effective $m_{\max}(M_\star)$ at fixed $M_{\rm cl}$, accelerating the arrival of the first SN event. The resulting premature cloud disruption shortens $\tsb$. The reduction in $\tq$ occurs because weaker individual SN energy injections lower the efficiency of global gas ejection, accelerating the re-formation of star-forming clouds.
    
    In higher-mass halos ($M_h \gtrsim 10^9\,\msun$), starbursts become longer ($t_{\rm sb,IMF}/t_{\rm sb,Fiducial} \simeq 1.2$--$1.5$), while quiescent intervals $\tq$ remain virtually unchanged. Here, clouds are massive enough to fully sample the IMF, so $m_{\max}$ is no longer limited by stochastic effects. Instead, the behavior is dominated by the difference in feedback strength per unit stellar mass, propagated through the IMF–SFE equilibrium: the Salpeter IMF produces fewer massive stars than the top-heavy phases of the evolving \texttt{Fiducial} model, resulting in weaker and more extended feedback-driven cloud clearing. This delays cloud disruption, even though SN onset is no longer sampling-limited, and prolongs $\tsb$. Meanwhile, $\tq$ remains invariant because the inter-burst timescale is governed by large-scale gas supply and gas reservoir replenishment rather than the IMF-dependent variation in cloud disruption physics.

    \item \textbf{The \texttt{SFE} model} assumes a constant cloud-scale SFE of $\epsilon_\star=0.45$. This creates a clear transition relative to the \texttt{Fiducial} model, visible in Figs.~\ref{fig:sb} and \ref{fig:q}: it forces higher cloud SFE values in low-mass halos ($\mhalo \lesssim 10^{9.4}\,(10^9)\,\msun$ at $z=5\,(15)$) but lower SFE values in more massive halos. The resulting impacts divide into three distinct regimes:
    
    In low-mass halos, starbursts shorten ($t_{\rm sb, SFE}/t_{\rm sb, Fiducial} \simeq 0.3$--$1$) while quiescent phases lengthen ($t_{\rm q, SFE}/t_{\rm q, Fiducial} \simeq 1$--$4$). The higher cloud SFE leads to higher gas-to-star conversion and a higher stellar mass per cloud, triggering an earlier onset of the first SNe. The resulting stronger SN feedback expels more gas, yielding shorter, more intense bursts followed by extended quiescent phases, during which gas can re-accrete.
    
    In massive halos at lower redshifts ($z \lesssim 9$), the trend reverses: quiescent durations decrease ($t_{\rm q, SFE}/t_{\rm q, Fiducial} \simeq 0.5$) while burst durations generally increase ($t_{\rm sb, SFE}/t_{\rm sb, Fiducial} \simeq 1$--$2$). Here, the lower SFE relative to the \texttt{Fiducial} model reduces stellar mass production per cloud and delays SN feedback, allowing star formation to persist longer once sufficient gas has accumulated.

    At the highest halo masses and highest redshifts, starburst phases also shorten ($t_{\rm sb, SFE}/t_{\rm sb, Fiducial} \simeq 0.3$--$0.7$). In this radiation-pressure-regulated regime, star formation becomes strictly supply-limited. Because the cloud SFE is lower than in the \texttt{Fiducial} model, the galaxy requires substantially more gas inflow into the dense cloud reservoir to reach the same stellar mass target. This accelerated conversion into stellar mass rapidly depletes the surrounding diffuse gas reservoir ($M_{\rm gas, ini}$), leaving insufficient fuel to sustain subsequent star formation and truncating the burst.

    \item \textbf{The \texttt{Density} model} assumes a higher baseline cloud density ($n_s = 1.5 \times 10^6$), which increases local volume and surface densities ($\Sigma_0 \propto M_{\rm cl}^{1/3} n_0^{2/3}$), shifting cloud SFEs to higher values. The impacts on $\tsb$ and $\tq$ depend on whether clouds form in the photoionisation-regulated or radiation-pressure-regulated regime:
    
    In the photoionisation-regulated regime, both timescales decrease slightly. Quiescent durations ($\tq$) shorten because higher gas densities accelerate gravitational contraction, compressing the free-fall time ($t_{\rm ff} \propto n_0^{-1/2}$) and therefore the onset delay between cloud formation and the start of star formation ($t_{\rm sf,0} \propto n_0^{-0.085}$). Concurrently, $\tsb$ shortens because the active star-forming window of individual clouds scales inversely with density ($t_{\rm sf} \propto n_0^{-0.33}$), meaning clouds rapidly form stars and dissolve sooner.
    
    In the radiation-pressure-regulated regime, the trend reverses, and higher densities lengthen $\tsb$. Here, higher gas densities deepen the local gravitational potential well so effectively that radiation-pressure clearing is suppressed. This elevates the cloud SFE, resulting in the galaxy requiring less diffuse gas to achieve its target stellar mass per time step ($M_{\rm cloud} = M_{\star, \max} / \epsilon_\star$). This conserves the diffuse gas reservoir ($M_{\rm gas, ini}$) instead of depleting it, allowing a sustained supply of gas to feed sequential cloud formation over consecutive sub-timesteps and lengthening $\tsb$.
\end{itemize}
These variations show that star formation timescales are controlled by the balance between gas supply rates and feedback timing.
Altering the maximum cloud mass or baseline density modifies the gas supply rate, either accelerating the transition to continuous star formation or conserving the diffuse gas reservoir to prolong bursts. Conversely, modifying the IMF or SFE shifts the feedback timescales, either shortening timescales through stochastic sampling at low masses or delaying cloud disruption to extend starbursts in massive systems.

\begin{figure*}
    \centering
    \includegraphics[width=\linewidth]{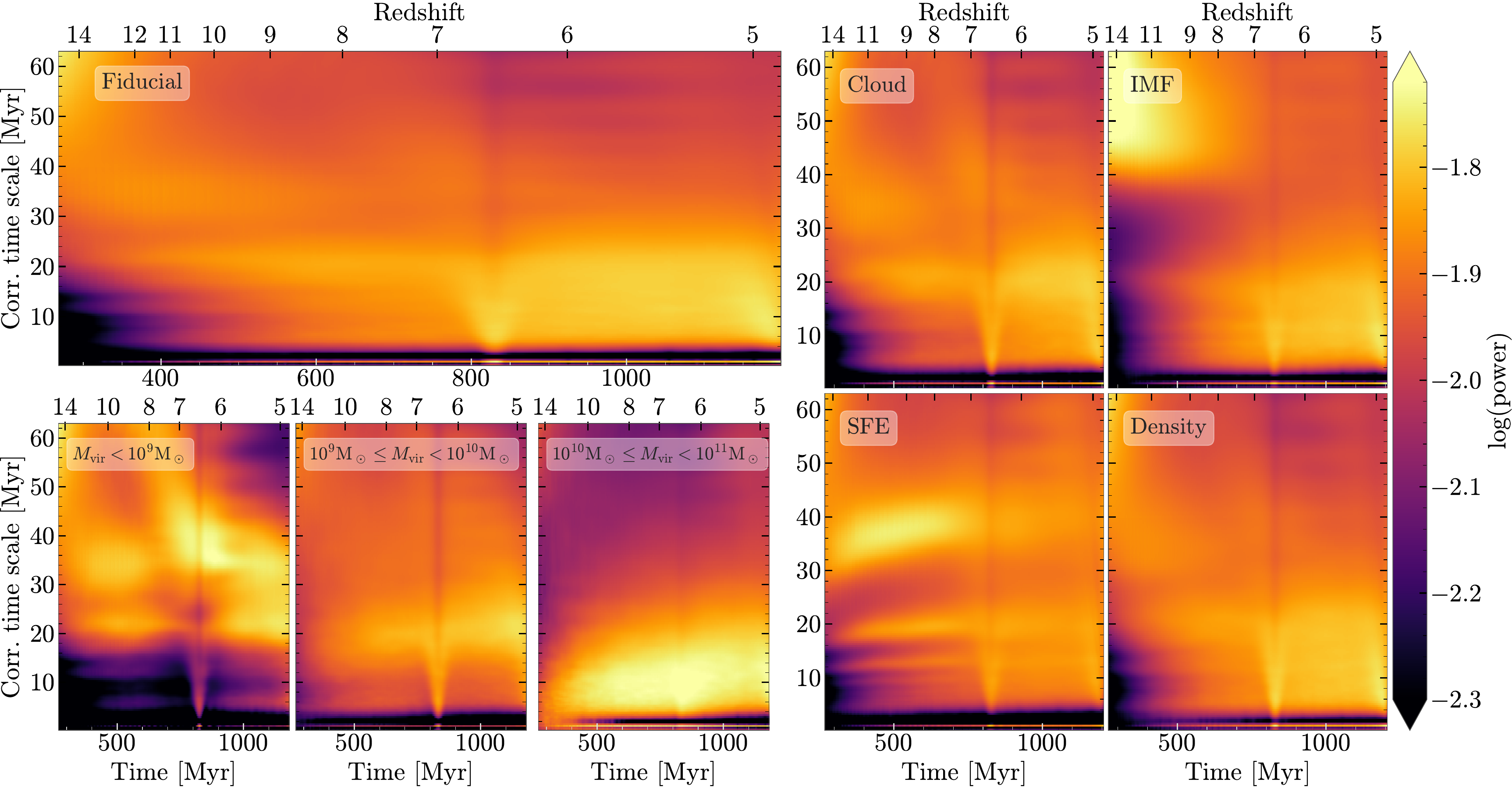}
    \caption{Wavelet power spectra of SFR, summed over all galaxies in the models, weighted by the halo mass function. \textit{Left:} The fiducial model, for all halos (\textit{top}), and for halos during time steps when their halo masses are $M_{\rm vir}<10^9\msun$ (\textit{bottom left}), $10^9\msun\leq M_{\rm vir} <10^{10}\msun$ (\textit{bottom center}), and $10^{10}\msun\leq\mhalo\leq 10^{11}\msun$ (\textit{bottom right}). \textit{Right:} All galaxies in each of the variation models. Note that the vertical feature around $z=6.5$ in all the spectra is a numerical artefact, and not a meaningful feature.}
    \label{fig:cwt}
\end{figure*}

\subsection{Wavelet analysis of star formation rate variability}
\label{sec:wavelet_analysis}

To identify the dominant frequencies of star-formation burstiness, we compute continuous wavelet power spectra of the total star formation rate (SFR) using Morlet wavelets \citep[following][adopting the method for our 1-dimensional SFR histories]{cheng__new_2020}. Fig.~\ref{fig:cwt} shows these spectra summed over the simulated populations and weighted by the halo mass function (HMF) at $z=4.5$. Elevated power highlights the characteristic timescales of the star-formation cycle at an epoch. Because the HMF drops sharply towards massive halos, these global spectra are dominated by the more abundant lower-mass halo population.

The \texttt{Fiducial} model (top-left) shows a broadly distributed wavelet power, whereas its peak shifts towards shorter timescales over cosmic time. At early  epochs ($t \lesssim 500\,\rm Myr$, $z \gtrsim 9$), power concentrates at long timescales ($\sim 30$--$50\,\rm Myr$), driven by high accretion rates that trigger intense starbursts and SN feedback clearings that prolong the formation of star-forming clouds. As time progresses, this power shifts below $\lesssim 15\,\rm Myr$. 
The mass-decomposed spectra (bottom-left panels) illuminate the physics behind this frequency shift. 
In low-mass halos ($M_{\rm vir} < 10^9\,\rm M_\odot$), the power is confined to long-period fluctuations ($\sim 30$--$50\,\rm Myr$), confirming the \emph{feedback-dominated regime} where SN feedback-driven gas evacuation enforces long quiescent phases that dominate the SFH. 
In intermediate-mass halos ($10^9\,\rm M_\odot \le M_{\rm vir} < 10^{10}\,\rm M_\odot$), the power shifts to intermediate timescales ($\sim 20\,\mathrm{Myr}$), reflecting a shortening of the burst–quench cycle. Deeper potential wells accelerate gas accretion and recovery, shortening quiescent phases while extending the previously short burst phases.
In massive halos ($10^{10}\,\rm M_\odot \le M_{\rm vir} < 10^{11}\,\rm M_\odot$), the power concentrates at the shortest resolved scales ($\sim 3$--$5\,\rm Myr$). This high-frequency signature marks the transition to the \emph{continuous star formation regime}, where the burst-quench cycle breaks down ($t_q \to 0$), leaving only the rapid, stochastic flickering of individual cloud assembly.

Varying the four model parameters displays distinct frequency shifts that map directly to their altered cloud-scale physics (Fig.~\ref{fig:cwt}, right panels):
\begin{itemize}
    \item \textbf{The \texttt{Cloud} model} displays a slight reduction in power at larger timescales ($\sim 30-50\,\rm Myr$) relative to the \texttt{Fiducial} model. Its lower cloud-mass ceiling causes intermediate-to-massive halos to reach the maximum allowed cloud mass and enter the continuous star-formation regime earlier in their growth histories.

    \item \textbf{The \texttt{IMF} model} significantly enhances power at early cosmic times ($t < 400\,\rm Myr$, $z \gtrsim 11$), concentrated at long timescales ($\sim45 -60\,\rm Myr$). At these high redshifts, star formation is dominated by low-mass clouds that cannot fully sample the high-mass end of the IMF. Switching from the dynamically evolving IMF of the \texttt{Fiducial} model to a Salpeter IMF reduces the massive stellar fraction, resulting in weaker feedback per unit stellar mass. Via the IMF--SFE coupling, this weaker feedback increases the cloud SFE, producing intense starbursts that are truncated early by SN events, followed by accelerated gas recovery phases. While the fiducial evolving IMF varies with environment and distributes power broadly, the Salpeter IMF confines the early low-mass galaxy population to a more uniform, rapid duty cycle.

    \item \textbf{The \texttt{SFE} model} shows horizontal bands of enhanced power at discrete timescales ($\sim9$, $12$, $20$, and $30$--$40$\,Myr) persisting throughout cosmic dawn ($t \lesssim 1000\,\rm Myr$). This quantisation occurs because a constant cloud SFE eliminates the cloud-mass-dependent SFE variations of the \texttt{Fiducial} model: gas clouds of different masses undergo uniform gas-to-star conversion, triggering similar starburst and quiescent timescales across the population. The high-frequency bands ($\sim 9-20$\,Myr) map the prompt feedback destruction of individual or consecutive cloud generations, while the long-period band ($\sim 30-40$\,Myr) reflects the similar timescales of halo gas replenishment.

    \item \textbf{The \texttt{Density} model} suppresses power at long timescales ($\gtrsim 30\,\rm Myr$) across all epochs, shifting the bulk of the variability down to $\lesssim 30\,\rm Myr$. Compressing the free-fall time ($t_{\rm ff} \propto n_0^{-1/2}$) shortens both the cloud onset delay ($t_{\rm sf,0}$) and the active star-forming window ($t_{\rm sf}$), which accelerates the entire star-formation cycle.
\end{itemize}
The variability of the SFH directly reflects the physical ``clock'' governing cloud assembly, gas consumption, and feedback disruption. While the \texttt{Fiducial} model features a smooth gradient of timescales that dynamically shorten as galaxies grow, fixing the underlying cloud parameters (\texttt{IMF}, \texttt{SFE}) strips away this diversity, resulting in uniform, quantised frequencies. Conversely, compressing the cloud free-fall time (\texttt{Density}) accelerates the entire star-formation engine across all epochs.

\begin{figure}
    \centering
    \includegraphics[width=\linewidth]{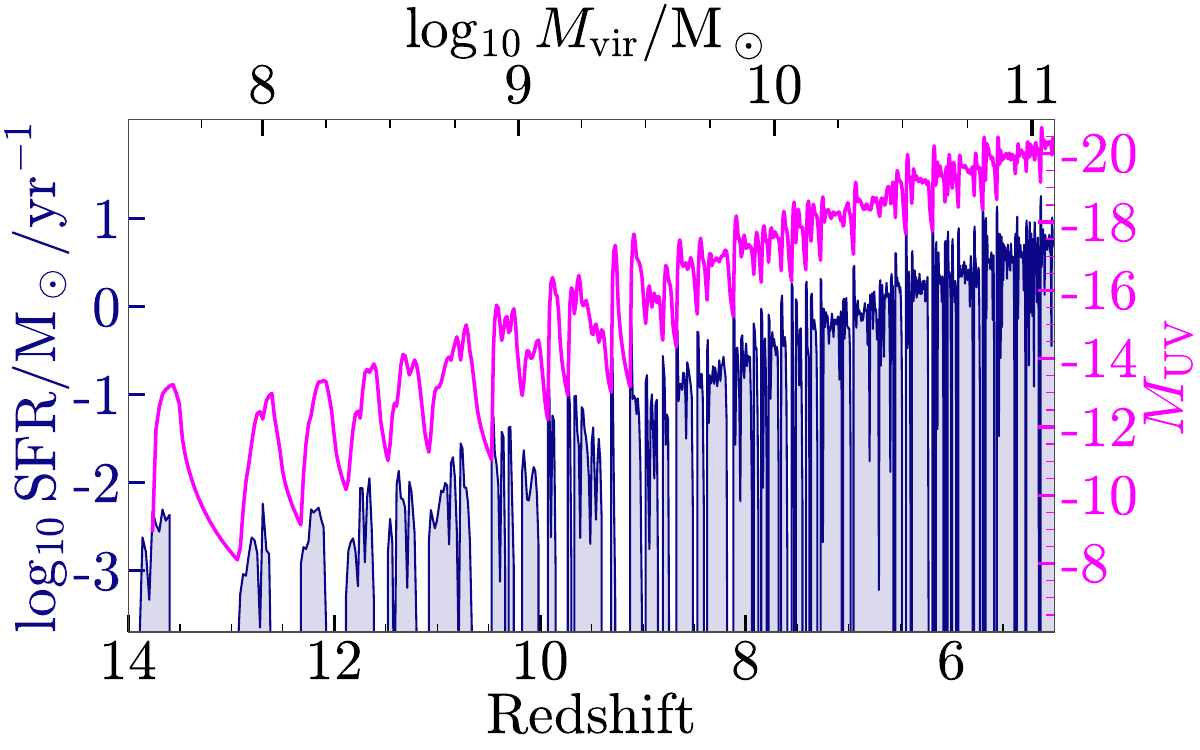}
    \caption{The star formation history and UV luminosity of a halo in the fiducial model.}
    \label{fig:sfhex}
\end{figure}

\subsection{Observational imprints: scatter in $\muv$}

\begin{figure*}
    \centering
    \includegraphics[width=1\linewidth]{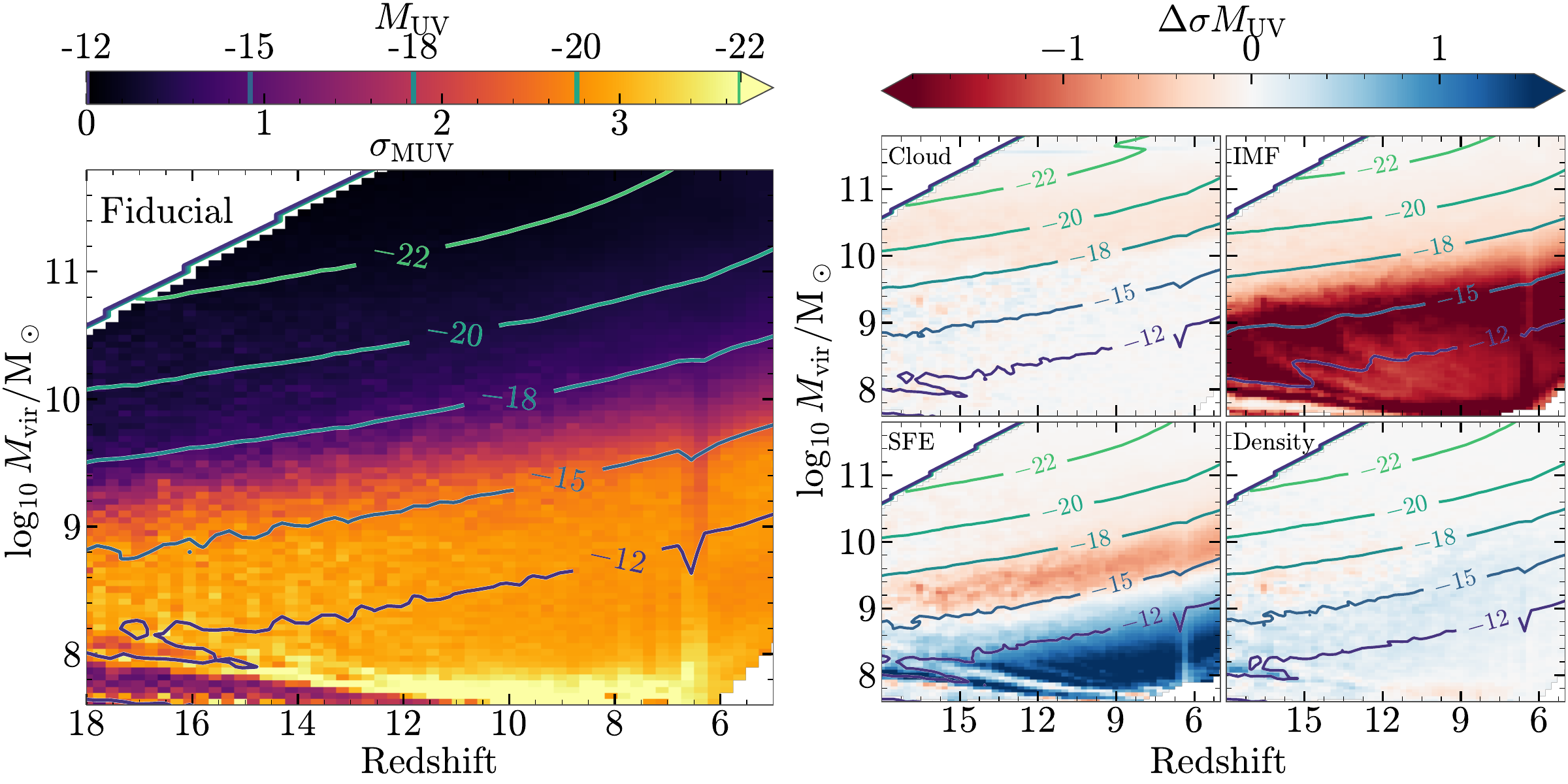}
    \caption{ \textit{Left:} $1\sigma$ scatter in luminosity as a function of redshift and halo mass. Contours show median luminosities. \textit{Right:} Difference in luminosity scatter between the variation models and the \texttt{Fiducial} model, so blue signifies the model has greater scatter in luminosity than the \texttt{Fiducial} model, and vice versa for red.}
    \label{fig:MvirMuv}
\end{figure*}

UV emission, dominated by massive short-lived stars, traces short-term SFR variability (see Fig.~\ref{fig:sfhex}), while its scatter at fixed halo mass and redshift, $\sigma_{\muv}$, encodes the star-formation duty cycle and burst contrast. Fig.~\ref{fig:MvirMuv} shows the corresponding $16$–$84$th percentile scatter (left) for the \texttt{Fiducial} model and the differences of the model variations relative to \texttt{Fiducial} (right).

In the \texttt{Fiducial} model, the largest scatter ($\sigma_{\muv}\simeq 2\text{--}3\,\rm mag$) occurs in low‑mass, SN feedback‑ and cloud-dominated halos ($\mhalo\lesssim 10^{9}\,\msun$). In this regime, long quiescent periods ($t_{\rm q} \gtrsim 5$\,Myr) allow galaxies to fade by several orders of magnitude between bursts as their massive stars die, followed by rapid rebrightening once sufficient gas is accreted and cooled to trigger the next starburst. 
With increasing halo mass, $\sigma_{\muv}$ decreases. Up to halo masses of about $10^{9.5}\,(10^{9})\,\msun$ at $z=5\,(15)$, where $\sigma_{\muv}\gtrsim2\rm\, mag$, this decline is driven primarily by shorter quiescent times $\tq$. Above this threshold - consistent with the increase in $\tsb$ in Fig.~\ref{fig:sb} that indicates less bursty star formation - $\sigma_{\muv}$ tapers toward zero as the bursty regime ends. Shorter quiescent phases leave insufficient time for massive stellar populations to die off completely between bursts, heavily reducing the luminosity contrast and flattening the population-wide scatter.
This behaviour is also evident in Fig.~\ref{fig:sfhex}, which shows the SFR and $\muv$ evolution of a single galaxy as a function of $z$ and $M_{\rm halo}$. At $z\gtrsim10$ ($M_{\rm halo}\lesssim10^{9},M_\odot$), luminosity variations reach $\sim4,\mathrm{mag}$. As this individual halo accretes mass over time, its burst intervals shorten; this drives a steady decline in its individual luminosity variability, directly mirroring the global population trend where higher-mass systems transition into the low-scatter, continuous star-formation regime.

Fig.~\ref{fig:MvirMuv} (right panels) also shows how modifying core parameters changes the UV scatter ($\Delta\sigma_{\muv}$) relative to the \texttt{Fiducial} model, highlighting how cloud-scale physics dictates macro-scale observational variability:
\begin{itemize}
    \item \textbf{The \texttt{Cloud} model} suppresses scatter ($\Delta\sigma_{\muv} \sim -0.2\,\rm mag$) in intermediate-to-high mass halos ($\mhalo \sim 10^{9.5}$--$10^{11}\,\msun$). Lowering the cloud mass ceiling forces growing systems to hit their maximum allowed cloud size earlier, accelerating their transition into the continuous star-formation regime. This truncates large-scale variations and smooths the aggregate UV output.

    \item \textbf{The \texttt{IMF} model} severely reduces scatter ($\Delta\sigma_{\muv} \sim -2\,\rm mag$) in low-mass halos across all epochs. Because small clouds at early times cannot fully sample a Salpeter IMF, the high-mass stellar fraction drops, yielding weaker SN feedback per unit stellar mass. Via the IMF–SFE coupling, this weaker feedback drives rapid, efficient starbursts with short duty cycles (cf. shorter $\tsb$ and $\tq$ values in Fig.~\ref{fig:sb} and \ref{fig:q}). This fast-paced cycle prevents galaxies from spending extended periods in deeply faded, quiescent states, truncating the low-luminosity tail and narrowing the overall UV scatter.

    \item \textbf{The \texttt{SFE} model} suppresses scatter at intermediate masses, but enhances it ($\Delta\sigma_{\muv} \sim 2\,\rm mag$) in the lowest-mass halos ($\mhalo \lesssim 10^{8.5}\,\msun$). This enhancement is the direct observational signature of the quantised timescales identified in Sec.~\ref{sec:wavelet_analysis}. Forcing a high, constant SFE causes these small halos to exhaust their gas, forcing them into prolonged quiescent recovery states ($t_{\rm q, SFE}/t_{\rm q, Fiducial} \sim 2$--$4$). These galaxies operate in an extreme binary ``blinking'' mode - intensely bright during a burst or entirely dark for tens of millions of years - driving up the UV scatter to its absolute maximum. Conversely, at intermediate masses, the fixed SFE is lower than the values in the \texttt{Fiducial} model, shortening quiescent phases and reducing UV variability.

    \item \textbf{The \texttt{Density} model} increases scatter ($\Delta\sigma_{\muv} \sim 0.3\,\rm mag$) in low-to-intermediate mass halos at early times. While compressing the free-fall time accelerates cloud lifetimes and shortens the active starburst windows, preserving mass-dependent SFE and IMF scaling maintains high internal variety. This accelerated clock, paired with strong cloud-scale mass variance, introduces rapid, chaotic stochastic spikes in the SFR history, that boost high-frequency scatter across the growing halo population.
\end{itemize}

%%% ---------------------------------------------------
\section{Impact on observed UV luminosity functions}\label{sec:uvlf}
%%% ---------------------------------------------------

\begin{figure*}
    \centering
    \includegraphics[width=\linewidth]{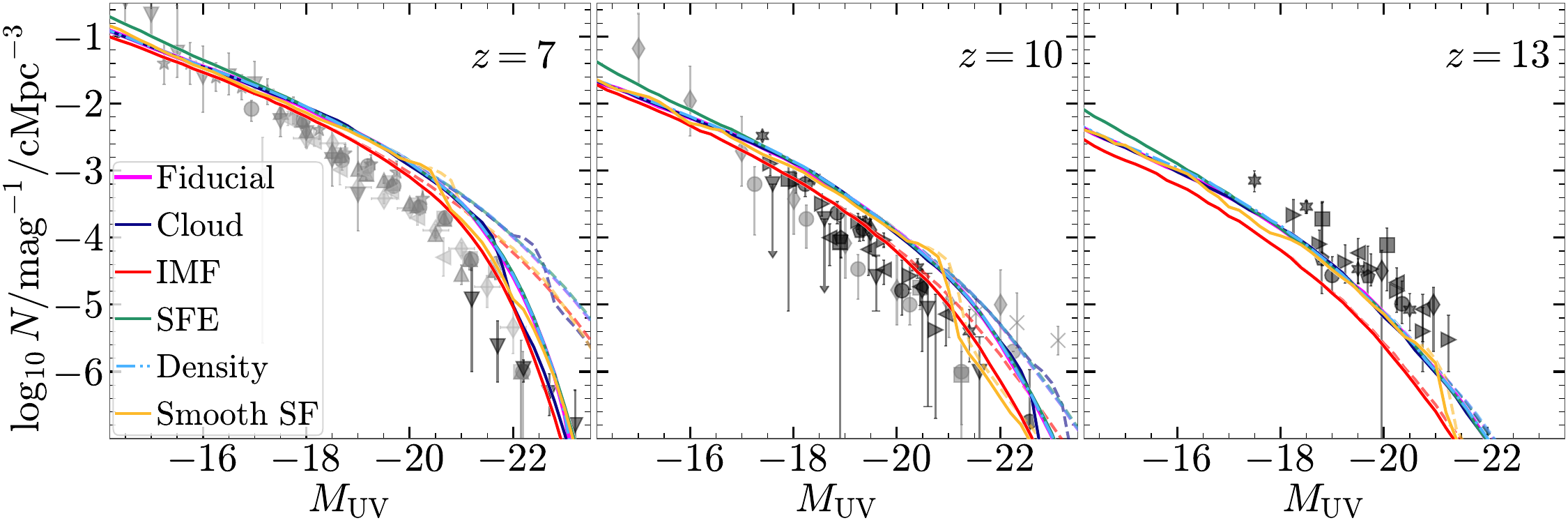}
    \caption{UV luminosity functions at $z=7$, $10$, and $13$, for our five model variations as well as a \texttt{Smooth SF} model. Dashed lines show intrinsic UV LFs, while solid and dash-dotted lines show dust attenuated UVLFs. Light gray points show pre-JWST observational results from \citep{atek_new_2015,bouwens_uv_2015, bouwens_bright_2016,  bouwens_new_2021, calvi_bright_2016, finkelstein_evolution_2015, ishigaki_full-data_2018, Livermore2017, mclure_new_2013, schenker_uv_2013}, while dark gray points show results from JWST observations \citep{adams_epochs_2024, bouwens_evolution_2023, bouwens_uv_2023,donnan_evolution_2023, Donnan2024, harikane_pure_2024, Whitler2025, willott_steep_2024}.}
    \label{fig:uvlf}
\end{figure*}

Finally, we assess how different assumptions on cloud-scale star formation affect the observed cosmic UV Luminosity Functions (LFs) at $z=7$, $10$, and $13$ (Fig.~\ref{fig:uvlf}). We further compare to a \texttt{Smooth SF} case, in which we adopt the smoothed star formation mode (Section \ref{sec:smoothing_trigger}) everywhere regardless of gas mass, to completely remove the effects of stochastic cloud formation. 

The LF shifts across our five main model variations are governed by scatter, where lower-mass systems caught during peak starburst spikes temporarily scatter upward and dominate the exponentially rare bright-end bins. 
In the \texttt{IMF} model, the severe collapse of UV scatter ($\Delta\sigma_{\muv} \sim -2\,\rm mag$) eliminates these upward excursions. By truncating these spikes, the model systematically produces less bright galaxies across all epochs, falling below the observational data at $z \gtrsim 10$. 
Conversely, the \texttt{SFE} model steepens the LF at early times ($z \gtrsim 10$). Its high, constant cloud SFE in lower-mass halos triggers immediate, intense bursts that boost the faint end ($M_{\rm UV} \gtrsim -16$).
Meanwhile, the \texttt{Cloud} model drops off towards the bright end ($M_{\rm UV} \lesssim -21$) at lower redshift ($z\lesssim 7$) where the earlier end to the bursty star formation phase in galaxies reduces the bright scatter in massive galaxies.
Finally, the \texttt{Density} model and the fainter end of the \texttt{Cloud} model ($\muv\gtrsim -21$) closely track the \texttt{Fiducial} baseline, indicating that the global distribution remains robust as long as environmental mass-dependent scaling and variety are preserved.
This is further highlighted by the \texttt{Smooth SF} comparison: removing burstiness entirely cuts off these upward luminosity excursions, causing the bright-end tail to drop prematurely. This demonstrates that short-term stochastic variability is required to match the observed abundance of bright galaxies at Cosmic Dawn.

%%%%%%%%%%%%%%%%%%%%%%%%%%%%%%%%%%%%%%%%%%%%%%%%%%%%%%%%%%%%%
\section{Conclusions}\label{sec:conclusions}

We have developed a novel semi-analytic, cloud-based framework for star formation, wherein star formation events are restricted to dense, self-gravitating clouds sequentially drawn as galaxies accumulate cold gas. The underlying cloud population follows a scale-free mass distribution, with an upper mass boundary that scales with the host halo's Jeans mass up to a globally enforced ceiling. Within these clouds, the SFE is regulated by two distinct feedback regimes, transitioning from photoionisation-limited at low-to-moderate cloud densities to radiation pressure-limited in high-density environments, and is coupled to a dynamic IMF. 
The fiducial IMF becomes increasingly top-heavy at lower gas metallicities and higher redshifts, while random sampling within the parent cloud determines the mass of the most massive stars formed. 
The coupled treatment of the SFE, the IMF and underlying cloud properties presents the primary novelty of this framework, directly motivated by radiation hydrodynamic cloud simulations.

To establish how critical parameters, namely cloud masses, densities, and SFE/IMF prescriptions, affect the star formation burstiness, we integrated this framework into a simple galaxy model, accounting for gas accretion, star formation, SN feedback, and metal and dust enrichment. The baseline behaviour of star-forming galaxies within this framework is characterised by: 
\begin{enumerate}
    \item Stellar feedback, gas accretion and cloud formation regulate star formation burstiness. A galaxy transitions through three distinct regimes as it accumulates mass. 
    \begin{itemize}
        \item \emph{Feedback-limited regime:} In low-mass halos ($M_{\rm vir} \lesssim 10^{8.5}\, \msun$), the gas accumulation timescale exceeds the SN feedback timescale, resulting in short bursts followed by extended quiescent periods. Due to stochastic undersampling of the high-mass IMF in small clouds, the onset of SN feedback is delayed to $\sim 10\,$Myrs. Because the quiescent phases are long enough for bursts to completely fade, the cycle depends strongly on gas replenishment, yielding the highest UV luminosity scatter.
        \item \emph{Cloud-mass-limited regime:} In intermediate-mass halos ($10^{8.5}\, \msun \lesssim M_{\rm vir} \lesssim 10^{10.5}\, \msun$), gas accumulation and SN feedback timescales become comparable. Starbursts are limited by how quickly the system can assemble clouds from the available gas. As the burst and quiescent phases shorten, subsequent bursts overlap before the UV emission fully fades, decreasing the UV luminosity scatter.
        \item \emph{Continuous star formation:} In massive halos ($M_{\rm vir} \gtrsim 10^{10.5}\, \msun$), gas is replenished and assembled into clouds faster than SN feedback can regulate the star formation cycle, erasing quiescent phases. The UV luminosity scatter reaches its minimum, with fluctuations driven primarily by variations in the gas accretion histories.
    \end{itemize}

    \item The feedback mechanisms that set the SFE govern the nature of the star formation cycle. At high redshifts and in massive halos, the SFE is regulated by radiation pressure and remains high. Because accreted gas is rapidly converted into stars, the burst duration is controlled by the maximum cloud mass and the cloud assembly rate. Conversely, at lower redshifts and in less-massive halos, the SFE is lower and regulated by photoionisation, allowing gas to accumulate between subsequent bursts and leading to longer burst durations.
\end{enumerate}
By varying one parameter at a time from our fiducial configuration, we quantified how each ingredient sets the duration of star-forming and quiescent phases, the star formation duty cycle, the UV luminosity scatter, and the UV LFs at $z\simeq5-13$. Our key findings are:
\begin{enumerate}
    \item \textbf{Lowering the maximum cloud mass terminates the bursty phase earlier.} This reduces the UV luminosity scatter in intermediate-mass halos and lowers UV luminosities in the most massive halos, causing a noticeable drop at the bright end of the UV LF.

    \item \textbf{An IMF that becomes increasingly top-heavy at lower metallicities and higher redshifts (relative to a Salpeter-like IMF) shortens burst durations in massive halos, while extending both the burst and quiescent phases in lower-mass halos.} This systematic shift amplifies the UV luminosity scatter, with the most pronounced increase occurring in lower-mass systems. Simultaneously, the lower mass-to-light ratio shifts the UV LF toward higher luminosities at higher redshifts.

    \item \textbf{Transitioning from a constant SFE to our variable model, where the SFE scales with cloud mass and density but drops with an increasingly top-heavy IMF, decreases (increases) the burst-to-quiescent duration ratio in massive (lower-mass) galaxies.} Consequently, the UV luminosity scatter increases for intermediate-mass halos but decreases for low-mass halos, with the latter driving a noticeable flattening at the faint end of the UV LF.

    \item \textbf{Raising the cloud density increases the cloud-scale SFE but shortens the cloud formation and lifetimes}, resulting in shorter, more frequent bursts. This increased burst frequency translates to a larger UV luminosity scatter in lower-mass halos, while leaving the UV LFs largely unchanged.
\end{enumerate}
From these parameter dependencies we conclude that an evolving IMF and larger maximum cloud mass primarily boost the bright end of the UV LF. Conversely, a cloud-property-dependent SFE and higher cloud densities govern the frequency of the burst-quiescent cycle, minimally impacting the UV LFs but potentially the spatial clustering of UV-bright galaxies.

While our cloud-based star formation framework shows that more top-heavy IMFs and enhanced SFE in more massive clouds enhance star-formation burstiness and UV luminosities of intermediate to massive galaxies, it makes a number of simplified assumptions that highlight exciting pathways for future refinement. 
First, cloud densities and the maximum cloud mass ($M_{\rm cl}^{\rm max}$) may be even higher in the very first galaxies. Our current values reflect behaviours up to surface densities of $\Sigma_0\simeq10^5\,\msun \rm pc^{-2}$, whereas observations of young star clusters in $z\gtrsim6$ galaxies \citep[e.g.][]{Abdurrouf2025, Adamo2024, Bradac2025, Mowla2024, Vanzella2023a} reveal stellar surface densities of $\Sigma_0\sim10^4-10^6\,\msun\rm pc^{-2}$, implying higher  initial gas surface densities. Although star formation in such massive, very dense clouds remains barely numerically explored, our trends suggest these massive clusters would form very rapidly and efficiently.

Secondly, such extreme, highly dense environments on cloud scales are strongly linked to galaxy-wide properties, yet our model currently limits star formation in massive halos via a uniform, maximum galaxy-wide SFE ($f_\star$). This hard limit serves as a simple sub-grid proxy for gas cooling, heating, and turbulence injection and dissipation, which are not explicitly traced. Physically, in halos approaching $M_{\rm vir} \sim 10^{11}\,\msun$, shock-heating lengthens cooling timescales, while accretion and SN feedback drive severe turbulence that suppresses efficient star formation. However, this proxy neglects temporal variations. If the global SFE cap were temporarily elevated -- e.g. in dense, metal-enriched, highly accreting galaxies -- our model would form more massive clouds during the same starburst in the radiation-pressure-limited regime. This would produce shorter, more intense starburst phases, potentially reversing the relative burst duration between our \texttt{SFE} and \texttt{Fiducial} models. Paired with longer quiescent phases, these compressed starbursts would amplify the UV luminosity scatter for more massive galaxies and raise the bright end of the UV LF.

Thirdly, our model assumes all clouds are born in virial equilibrium with a fixed virial parameter of $\alpha_{\rm vir}=2$. However, the ambient ISM pressure in dense, high-redshift galaxies is significantly higher than in local environments, which likely drives initial cloud states to lower $\alpha_{\rm vir}$ values likely. This weaker initial turbulent support would reduce the fraction of gas that becomes unbound by internal velocity dispersions, thereby raising the cloud-scale SFEs. In this sense, our current model provides a conservative lower limit on the potential burstiness and SFE of these early, dense star-forming environments.

Fourthly, our framework assumes smooth gas accretion, omitting the violent dynamical effects of galaxy mergers. At high redshifts, frequent major mergers act as catalysts for starbursts. The resulting tidal forces channel gas into galactic centers, naturally creating the ultra-dense environments and temporal variations in the SFE cap ($f_\star$) discussed above. Replacing smooth accretion with stochastic merger tracking will allow us to capture these episodic events that likely dominate early galaxy duty cycles.

In conclusion, our could-based star formation model demonstrates that top-heavy IMFs and enhanced SFE in massive clouds amplify star-formation burstiness, boosting the bright end of the high-redshift UV LFs. By linking parsec-scale star formation processes to the cosmological evolution of high-redshift galaxies, our framework synthesises compelling explanations for the extreme UV luminosities observed in the early Universe.
%In conclusion, our cloud-based star formation framework links parsec-scale star formation processes to the cosmological evolution of high-redshift galaxies. By demonstrating that top-heavy IMFs and enhanced SFE in massive clouds amplify star-formation burstiness, this framework synthesises compelling explanations for the extreme UV luminosities observed in the early Universe. 
While future refinements incorporating dynamic cloud virial parameters and time-dependent SFE caps will detail this picture, our current framework provides a first self-consistent, physical model. 
Ultimately, we intend to couple our cloud-scale star formation framework with a more realistic gas assembly model and apply it directly to cosmological merger trees. This will allow us to explicitly test whether incorporating galaxy-scale gas dynamics and hierarchical merging further enhances the UV luminosity variance and boost to the bright end of the UV LF.

\section*{Acknowledgements}
We thank Viola Gelli for useful discussions and comments, and Charlotte Mason for comments and support.
EC acknowledges support from the ERC synergy grant 101166930-RECAP. AH acknowledges support by the VILLUM FONDEN under grant 37459. The Cosmic Dawn Center (DAWN) is funded by the Danish National Research Foundation under grant DNRF140. 
This work has made use of the publicly available software packages \texttt{matplotlib} \citep{hunter_matplotlib_2007} and \texttt{numpy} \citep{van_der_walt_numpy_2011}.

%%%%%%%%%%%%%%%%%%%%%%%%%%%%%%%%%%%%%%%%%%%%%%%%%%
\section*{Data Availability}

The source code of the framework is publicly available at \url{https://github.com/E-Cueto/SimpleGal}. The simulation results and analysis scripts used in this work will be made available upon reasonable request to the authors.

%%%%%%%%%%%%%%%%%%%% REFERENCES %%%%%%%%%%%%%%%%%%

% The best way to enter references is to use BibTeX:

\bibliographystyle{mnras}
\bibliography{MScPapers, CGM, clouds, dust, imf, massive_stars, models, observations, SFE_IMF, uvlf, example} % if your bibtex file is called example.bib

@ARTICLE{Voit2024a,
       author = {{Voit}, G. Mark and {Pandya}, Viraj and {Fielding}, Drummond B. and {Bryan}, Greg L. and {Carr}, Christopher and {Donahue}, Megan and {Oppenheimer}, Benjamin D. and {Somerville}, Rachel S.},
        title = "{Equilibrium States of Galactic Atmospheres. I. The Flip Side of Mass Loading}",
      journal = {\apj},
         year = 2024,
        month = dec,
       volume = {976},
       number = {2},
          eid = {150},
        pages = {150},
          doi = {10.3847/1538-4357/ad81d6},
archivePrefix = {arXiv},
       eprint = {2406.07631},
 primaryClass = {astro-ph.GA},
       adsurl = {https://ui.adsabs.harvard.edu/abs/2024ApJ...976..150V}
}

@ARTICLE{Voit2024b,
       author = {{Voit}, G. Mark and {Carr}, Christopher and {Fielding}, Drummond B. and {Pandya}, Viraj and {Bryan}, Greg L. and {Donahue}, Megan and {Oppenheimer}, Benjamin D. and {Somerville}, Rachel S.},
        title = "{Equilibrium States of Galactic Atmospheres. II. Interpretation and Implications}",
      journal = {\apj},
         year = 2024,
        month = dec,
       volume = {976},
       number = {2},
          eid = {151},
        pages = {151},
          doi = {10.3847/1538-4357/ad81d5},
archivePrefix = {arXiv},
       eprint = {2406.07632},
 primaryClass = {astro-ph.GA},
       adsurl = {https://ui.adsabs.harvard.edu/abs/2024ApJ...976..151V}
}

@article{correa_accretion_2015,
	title = {The accretion history of dark matter haloes – {II}. {The} connections with the mass power spectrum and the density profile},
	volume = {450},
	issn = {0035-8711},
	url = {https://doi.org/10.1093/mnras/stv697},
	doi = {10.1093/mnras/stv697},
	number = {2},
	urldate = {2025-03-26},
	journal = {MNRAS},
	author = {Correa, Camila A. and Wyithe, J. Stuart B. and Schaye, Joop and Duffy, Alan R.},
	month = jun,
	year = {2015},
	pages = {1521--1537},
}

@article{leitherer_starburst99_1999,
	title = {Starburst99: {Synthesis} {Models} for {Galaxies} with {Active} {Star} {Formation}},
	volume = {123},
	issn = {0067-0049},
	shorttitle = {Starburst99},
	url = {https://iopscience.iop.org/article/10.1086/313233/meta},
	doi = {10.1086/313233},
	language = {en},
	number = {1},
	urldate = {2025-03-27},
	journal = {ApJS},
	author = {Leitherer, Claus and Schaerer, Daniel and Goldader, Jeffrey D. and Delgado, Rosa M. González and Robert, Carmelle and Kune, Denis Foo and Mello, Duília F. de and Devost, Daniel and Heckman, Timothy M.},
	month = jul,
	year = {1999},
	note = {Publisher: IOP Publishing},
	pages = {3},
}

@article{krumholz_radiation_2010,
	title = {{RADIATION} {FEEDBACK}, {FRAGMENTATION}, {AND} {THE} {ENVIRONMENTAL} {DEPENDENCE} {OF} {THE} {INITIAL} {MASS} {FUNCTION}},
	volume = {713},
	issn = {0004-637X, 1538-4357},
	url = {https://iopscience.iop.org/article/10.1088/0004-637X/713/2/1120},
	doi = {10.1088/0004-637X/713/2/1120},
	language = {en},
	number = {2},
	urldate = {2024-11-11},
	journal = {ApJ},
	author = {Krumholz, Mark R. and Cunningham, Andrew J. and Klein, Richard I. and McKee, Christopher F.},
	month = apr,
	year = {2010},
	pages = {1120--1133},
}

@article{howard_universal_2018,
	title = {A universal route for the formation of massive star clusters in giant molecular clouds},
	volume = {2},
	issn = {2397-3366},
	url = {https://www.nature.com/articles/s41550-018-0506-0},
	doi = {10.1038/s41550-018-0506-0},
	language = {en},
	number = {9},
	urldate = {2024-11-11},
	journal = {Nature Astronomy},
	author = {Howard, Corey S. and Pudritz, Ralph E. and Harris, William E.},
	month = jun,
	year = {2018},
	pages = {725--730},
}

@article{tanvir_environmental_2022,
	title = {Environmental variation of the low-mass {IMF}},
	volume = {516},
	copyright = {https://academic.oup.com/journals/pages/open\_access/funder\_policies/chorus/standard\_publication\_model},
	issn = {0035-8711, 1365-2966},
	url = {https://academic.oup.com/mnras/article/516/4/5712/6702431},
	doi = {10.1093/mnras/stac2642},
	language = {en},
	number = {4},
	urldate = {2024-11-11},
	journal = {MNRAS},
	author = {Tanvir, Tabassum S and Krumholz, Mark R and Federrath, Christoph},
	month = sep,
	year = {2022},
	pages = {5712--5725},
}

@article{krumholz_star_2019,
	title = {Star {Clusters} {Across} {Cosmic} {Time}},
	volume = {57},
	issn = {0066-4146},
	url = {https://ui.adsabs.harvard.edu/abs/2019ARA&A..57..227K},
	doi = {10.1146/annurev-astro-091918-104430},
	urldate = {2025-01-13},
	journal = {ARA\&A},
	author = {Krumholz, Mark R. and McKee, Christopher F. and Bland-Hawthorn, Joss},
	month = aug,
	year = {2019},
	note = {ADS Bibcode: 2019ARA\&A..57..227K},
	pages = {227--303},
}

@article{fukushima_far_2022,
	title = {Far and extreme {UV} radiation feedback in molecular clouds and its influence on the mass and size of star clusters},
	volume = {511},
	issn = {0035-8711},
	url = {https://ui.adsabs.harvard.edu/abs/2022MNRAS.511.3346F},
	doi = {10.1093/mnras/stac244},
	urldate = {2025-01-13},
	journal = {MNRAS},
	author = {Fukushima, Hajime and Yajima, Hidenobu},
	month = apr,
	year = {2022},
	note = {Publisher: OUP
ADS Bibcode: 2022MNRAS.511.3346F},
	pages = {3346--3364},
}

@ARTICLE{planck_collaboration_planck_2016,
       author = {{Planck Collaboration} and {Aghanim}, N. and {Ashdown}, M. and {Aumont}, J. and {Baccigalupi}, C. and {Ballardini}, M. and {Banday}, A.~J. and {Barreiro}, R.~B. and {Bartolo}, N. and {Basak}, S. and {Battye}, R. and {Benabed}, K. and {Bernard}, J.-P. and {Bersanelli}, M. and {Bielewicz}, P. and {Bock}, J.~J. and {Bonaldi}, A. and {Bonavera}, L. and {Bond}, J.~R. and {Borrill}, J. and {Bouchet}, F.~R. and {Boulanger}, F. and {Bucher}, M. and {Burigana}, C. and {Butler}, R.~C. and {Calabrese}, E. and {Cardoso}, J.-F. and {Carron}, J. and {Challinor}, A. and {Chiang}, H.~C. and {Colombo}, L.~P.~L. and {Combet}, C. and {Comis}, B. and {Coulais}, A. and {Crill}, B.~P. and {Curto}, A. and {Cuttaia}, F. and {Davis}, R.~J. and {de Bernardis}, P. and {de Rosa}, A. and {de Zotti}, G. and {Delabrouille}, J. and {Delouis}, J.-M. and {Di Valentino}, E. and {Dickinson}, C. and {Diego}, J.~M. and {Dor{\'e}}, O. and {Douspis}, M. and {Ducout}, A. and {Dupac}, X. and {Efstathiou}, G. and {Elsner}, F. and {En{\ss}lin}, T.~A. and {Eriksen}, H.~K. and {Falgarone}, E. and {Fantaye}, Y. and {Finelli}, F. and {Forastieri}, F. and {Frailis}, M. and {Fraisse}, A.~A. and {Franceschi}, E. and {Frolov}, A. and {Galeotta}, S. and {Galli}, S. and {Ganga}, K. and {G{\'e}nova-Santos}, R.~T. and {Gerbino}, M. and {Ghosh}, T. and {Gonz{\'a}lez-Nuevo}, J. and {G{\'o}rski}, K.~M. and {Gratton}, S. and {Gruppuso}, A. and {Gudmundsson}, J.~E. and {Hansen}, F.~K. and {Helou}, G. and {Henrot-Versill{\'e}}, S. and {Herranz}, D. and {Hivon}, E. and {Huang}, Z. and {Ili{\'c}}, S. and {Jaffe}, A.~H. and {Jones}, W.~C. and {Keih{\"a}nen}, E. and {Keskitalo}, R. and {Kisner}, T.~S. and {Knox}, L. and {Krachmalnicoff}, N. and {Kunz}, M. and {Kurki-Suonio}, H. and {Lagache}, G. and {Lamarre}, J.-M. and {Langer}, M. and {Lasenby}, A. and {Lattanzi}, M. and {Lawrence}, C.~R. and {Le Jeune}, M. and {Leahy}, J.~P. and {Levrier}, F. and {Liguori}, M. and {Lilje}, P.~B. and {L{\'o}pez-Caniego}, M. and {Ma}, Y.-Z. and {Mac{\'\i}as-P{\'e}rez}, J.~F. and {Maggio}, G. and {Mangilli}, A. and {Maris}, M. and {Martin}, P.~G. and {Mart{\'\i}nez-Gonz{\'a}lez}, E. and {Matarrese}, S. and {Mauri}, N. and {McEwen}, J.~D. and {Meinhold}, P.~R. and {Melchiorri}, A. and {Mennella}, A. and {Migliaccio}, M. and {Miville-Desch{\^e}nes}, M.-A. and {Molinari}, D. and {Moneti}, A. and {Montier}, L. and {Morgante}, G. and {Moss}, A. and {Mottet}, S. and {Naselsky}, P. and {Natoli}, P. and {Oxborrow}, C.~A. and {Pagano}, L. and {Paoletti}, D. and {Partridge}, B. and {Patanchon}, G. and {Patrizii}, L. and {Perdereau}, O. and {Perotto}, L. and {Pettorino}, V. and {Piacentini}, F. and {Plaszczynski}, S. and {Polastri}, L. and {Polenta}, G. and {Puget}, J.-L. and {Rachen}, J.~P. and {Racine}, B. and {Reinecke}, M. and {Remazeilles}, M. and {Renzi}, A. and {Rocha}, G. and {Rossetti}, M. and {Roudier}, G. and {Rubi{\~n}o-Mart{\'\i}n}, J.~A. and {Ruiz-Granados}, B. and {Salvati}, L. and {Sandri}, M. and {Savelainen}, M. and {Scott}, D. and {Sirri}, G. and {Sunyaev}, R. and {Suur-Uski}, A.-S. and {Tauber}, J.~A. and {Tenti}, M. and {Toffolatti}, L. and {Tomasi}, M. and {Tristram}, M. and {Trombetti}, T. and {Valiviita}, J. and {Van Tent}, F. and {Vibert}, L. and {Vielva}, P. and {Villa}, F. and {Vittorio}, N. and {Wandelt}, B.~D. and {Watson}, R. and {Wehus}, I.~K. and {White}, M. and {Zacchei}, A. and {Zonca}, A.},
        title = "{Planck intermediate results. XLVI. Reduction of large-scale systematic effects in HFI polarization maps and estimation of the reionization optical depth}",
      journal = {\aap},
         year = 2016,
        month = dec,
       volume = {596},
          eid = {A107},
        pages = {A107},
          doi = {10.1051/0004-6361/201628890},
archivePrefix = {arXiv},
       eprint = {1605.02985},
 primaryClass = {astro-ph.CO},
       adsurl = {https://ui.adsabs.harvard.edu/abs/2016A&A...596A.107P}
}

@ARTICLE{oke_secondary_1983,
       author = {{Oke}, J.~B. and {Gunn}, J.~E.},
        title = "{Secondary standard stars for absolute spectrophotometry.}",
      journal = {\apj},
         year = 1983,
        month = mar,
       volume = {266},
        pages = {713-717},
          doi = {10.1086/160817},
       adsurl = {https://ui.adsabs.harvard.edu/abs/1983ApJ...266..713O}
}

@article{cueto_astraeus_2024,
	title = {{ASTRAEUS}: {IX}. {Impact} of an evolving stellar initial mass function on early galaxies and reionisation},
	volume = {686},
	copyright = {https://creativecommons.org/licenses/by/4.0},
	issn = {0004-6361, 1432-0746},
	shorttitle = {{ASTRAEUS}},
	url = {https://www.aanda.org/10.1051/0004-6361/202349017},
	doi = {10.1051/0004-6361/202349017},
	language = {en},
	urldate = {2025-02-26},
	journal = {A\&A},
	author = {Cueto, Elie R. and Hutter, Anne and Dayal, Pratika and Gottlöber, Stefan and Heintz, Kasper E. and Mason, Charlotte and Trebitsch, Maxime and Yepes, Gustavo},
	month = jun,
	year = {2024},
	pages = {A138},
}

@article{cheng__new_2020,
	title = {A new approach to observational cosmology using the scattering transform},
	volume = {499},
	copyright = {https://academic.oup.com/journals/pages/open\_access/funder\_policies/chorus/standard\_publication\_model},
	issn = {0035-8711, 1365-2966},
	url = {https://academic.oup.com/mnras/article/499/4/5902/5924461},
	doi = {10.1093/mnras/staa3165},
	language = {en},
	number = {4},
	urldate = {2026-07-13},
	journal = {MNRAS},
	author = {Cheng, Sihao and Ting, Yuan-Sen and Ménard, Brice and Bruna, Joan},
	month = nov,
	year = {2020},
	pages = {5902--5914},
}

@article{hutter_astraeus_2025,
	title = {{ASTRAEUS}: {X}. {Indications} of a top-heavy initial mass function in highly star-forming galaxies from {JWST} observations at \textit{z} {\textgreater} 10},
	volume = {694},
	copyright = {https://creativecommons.org/licenses/by/4.0},
	issn = {0004-6361, 1432-0746},
	shorttitle = {{ASTRAEUS}},
	url = {https://www.aanda.org/10.1051/0004-6361/202452460},
	doi = {10.1051/0004-6361/202452460},
	language = {en},
	urldate = {2025-02-26},
	journal = {A\&A},
	author = {Hutter, Anne and Cueto, Elie R. and Dayal, Pratika and Gottlöber, Stefan and Trebitsch, Maxime and Yepes, Gustavo},
	month = feb,
	year = {2025},
	pages = {A254},
}

@article{ucci_astraeus_2022,
	title = {Astraeus {V}: {The} emergence and evolution of metallicity scaling relations during the {Epoch} of {Reionization}},
	volume = {518},
	issn = {0035-8711, 1365-2966},
	shorttitle = {Astraeus {V}},
	url = {http://arxiv.org/abs/2112.02115},
	doi = {10.1093/mnras/stac2654},
	language = {en},
	number = {3},
	urldate = {2025-03-26},
	journal = {MNRAS},
	author = {Ucci, Graziano and Dayal, Pratika and Hutter, Anne and Kobayashi, Chiaki and Gottloeber, Stefan and Yepes, Gustavo and Hunt, Leslie and Legrand, Laurent and Tortora, Crescenzo},
	month = nov,
	year = {2022},
	note = {arXiv:2112.02115 [astro-ph]},
	pages = {3557--3575},
}

@article{dayal_alma_2022,
	title = {The {ALMA} {REBELS} survey: the dust content of \textit{z} ∼ 7 {Lyman} break galaxies},
	volume = {512},
	copyright = {https://creativecommons.org/licenses/by/4.0/},
	issn = {0035-8711, 1365-2966},
	shorttitle = {The {ALMA} {REBELS} survey},
	url = {https://academic.oup.com/mnras/article/512/1/989/6541865},
	doi = {10.1093/mnras/stac537},
	language = {en},
	number = {1},
	urldate = {2025-03-26},
	journal = {MNRAS},
	author = {Dayal, P and Ferrara, A and Sommovigo, L and Bouwens, R and Oesch, P A and Smit, R and Gonzalez, V and Schouws, S and Stefanon, M and Kobayashi, C and Bremer, J and Algera, H S B and Aravena, M and Bowler, R A A and da Cunha, E and Fudamoto, Y and Graziani, L and Hodge, J and Inami, H and De Looze, I and Pallottini, A and Riechers, D and Schneider, R and Stark, D and Endsley, R},
	month = mar,
	year = {2022},
	pages = {989--1002},
}

@article{salpeter_luminosity_1955,
	title = {The {Luminosity} {Function} and {Stellar} {Evolution}.},
	volume = {121},
	issn = {0004-637X, 1538-4357},
	url = {http://adsabs.harvard.edu/doi/10.1086/145971},
	doi = {10.1086/145971},
	language = {en},
	urldate = {2025-03-27},
	journal = {ApJ},
	author = {Salpeter, Edwin E.},
	month = jan,
	year = {1955},
	pages = {161},
}

@article{shaver_galactic_1983,
	title = {The galactic abundance gradient⋆},
	volume = {204},
	issn = {0035-8711},
	url = {https://doi.org/10.1093/mnras/204.1.53},
	doi = {10.1093/mnras/204.1.53},
	number = {1},
	urldate = {2025-03-28},
	journal = {MNRAS},
	author = {Shaver, P. A. and McGee, R. X. and Newton, Lynette M. and Danks, A. C. and Pottasch, S. R.},
	month = sep,
	year = {1983},
	pages = {53--112},
}

@article{adams_epochs_2024,
	title = {{EPOCHS}. {II}. {The} {Ultraviolet} {Luminosity} {Function} from 7.5 {\textless} z {\textless} 13.5 {Using} 180 arcmin2 of {Deep}, {Blank} {Fields} from the {PEARLS} {Survey} and {Public} {JWST} {Data}},
	volume = {965},
	issn = {0004-637X},
	url = {https://ui.adsabs.harvard.edu/abs/2024ApJ...965..169A},
	doi = {10.3847/1538-4357/ad2a7b},
	urldate = {2025-04-09},
	journal = {ApJ},
	author = {Adams, Nathan J. and Conselice, Christopher J. and Austin, Duncan and Harvey, Thomas and Ferreira, Leonardo and Trussler, James and Juodžbalis, Ignas and Li, Qiong and Windhorst, Rogier and Cohen, Seth H. and Jansen, Rolf A. and Summers, Jake and Tompkins, Scott and Driver, Simon P. and Robotham, Aaron and D'Silva, Jordan C. J. and Yan, Haojing and Coe, Dan and Frye, Brenda and Grogin, Norman A. and Koekemoer, Anton M. and Marshall, Madeline A. and Pirzkal, Nor and Ryan, Russell E. and Maksym, W. Peter and Rutkowski, Michael J. and Willmer, Christopher N. A. and Hammel, Heidi B. and Nonino, Mario and Bhatawdekar, Rachana and Wilkins, Stephen M. and Bradley, Larry D. and Broadhurst, Tom and Cheng, Cheng and Dole, Hervé and Hathi, Nimish P. and Zitrin, Adi},
	month = apr,
	year = {2024},
	note = {Publisher: IOP
ADS Bibcode: 2024ApJ...965..169A},
	pages = {169},
}

@article{gnedin_cosmological_2000,
	title = {Cosmological {Reionization} by {Stellar} {Sources}},
	volume = {535},
	issn = {0004-637X},
	url = {https://iopscience.iop.org/article/10.1086/308876/meta},
	doi = {10.1086/308876},
	language = {en},
	number = {2},
	urldate = {2025-04-15},
	journal = {ApJ},
	author = {Gnedin, Nickolay Y.},
	month = jun,
	year = {2000},
	note = {Publisher: IOP Publishing},
	pages = {530},
}

@article{kravtsov_tumultuous_2004,
	title = {The {Tumultuous} {Lives} of {Galactic} {Dwarfs} and the {Missing} {Satellites} {Problem}},
	volume = {609},
	issn = {0004-637X},
	url = {https://iopscience.iop.org/article/10.1086/421322/meta},
	doi = {10.1086/421322},
	language = {en},
	number = {2},
	urldate = {2025-04-18},
	journal = {ApJ},
	author = {Kravtsov, Andrey V. and Gnedin, Oleg Y. and Klypin, Anatoly A.},
	month = jul,
	year = {2004},
	note = {Publisher: IOP Publishing},
	pages = {482},
}

@article{gnedin_probing_1998,
	title = {Probing the {Universe} with the {Lyalpha} forest - {I}. {Hydrodynamics} of the low-density intergalactic medium},
	volume = {296},
	issn = {0035-8711},
	url = {https://ui.adsabs.harvard.edu/abs/1998MNRAS.296...44G},
	doi = {10.1046/j.1365-8711.1998.01249.x},
	urldate = {2025-04-23},
	journal = {MNRAS},
	author = {Gnedin, Nickolay Y. and Hui, Lam},
	month = may,
	year = {1998},
	note = {Publisher: OUP
ADS Bibcode: 1998MNRAS.296...44G},
	pages = {44--55},
}

@article{harikane_pure_2024,
	title = {Pure {Spectroscopic} {Constraints} on {UV} {Luminosity} {Functions} and {Cosmic} {Star} {Formation} {History} from 25 {Galaxies} at z spec = 8.61-13.20 {Confirmed} with {JWST}/{NIRSpec}},
	volume = {960},
	issn = {0004-637X},
	url = {https://ui.adsabs.harvard.edu/abs/2024ApJ...960...56H},
	doi = {10.3847/1538-4357/ad0b7e},
	urldate = {2025-04-25},
	journal = {ApJ},
	author = {Harikane, Yuichi and Nakajima, Kimihiko and Ouchi, Masami and Umeda, Hiroya and Isobe, Yuki and Ono, Yoshiaki and Xu, Yi and Zhang, Yechi},
	month = jan,
	year = {2024},
	note = {Publisher: IOP
ADS Bibcode: 2024ApJ...960...56H},
	pages = {56},
}

@article{hutter_astraeus_2021,
	title = {Astraeus {I}: the interplay between galaxy formation and reionization},
	volume = {503},
	issn = {0035-8711},
	shorttitle = {Astraeus {I}},
	url = {https://ui.adsabs.harvard.edu/abs/2021MNRAS.503.3698H},
	doi = {10.1093/mnras/stab602},
	urldate = {2025-04-03},
	journal = {MNRAS},
	author = {Hutter, Anne and Dayal, Pratika and Yepes, Gustavo and Gottlöber, Stefan and Legrand, Laurent and Ucci, Graziano},
	month = may,
	year = {2021},
	note = {Publisher: OUP
ADS Bibcode: 2021MNRAS.503.3698H},
	pages = {3698--3723},
}

@article{murray_hmfcalc_2013,
	title = {{HMFcalc}: {An} online tool for calculating dark matter halo mass functions},
	volume = {3},
	issn = {2213-1337},
	shorttitle = {{HMFcalc}},
	url = {https://ui.adsabs.harvard.edu/abs/2013A&C.....3...23M},
	doi = {10.1016/j.ascom.2013.11.001},
	urldate = {2025-05-27},
	journal = {A\&C},
	author = {Murray, S. G. and Power, C. and Robotham, A. S. G.},
	month = nov,
	year = {2013},
	note = {Publisher: Elsevier
ADS Bibcode: 2013A\&C.....3...23M},
	pages = {23},
}

@article{padovani_stellar_1993,
	title = {Stellar {Mass} {Loss} in {Elliptical} {Galaxies} and the {Fueling} of {Active} {Galactic} {Nuclei}},
	volume = {416},
	issn = {0004-637X},
	url = {https://ui.adsabs.harvard.edu/abs/1993ApJ...416...26P},
	doi = {10.1086/173212},
	urldate = {2025-05-27},
	journal = {ApJ},
	author = {Padovani, Paolo and Matteucci, Francesca},
	month = oct,
	year = {1993},
	note = {Publisher: IOP
ADS Bibcode: 1993ApJ...416...26P},
	pages = {26},
}

@article{bouwens_uv_2015,
	title = {{UV} {Luminosity} {Functions} at {Redshifts} z ∼ 4 to z ∼ 10: 10,000 {Galaxies} from {HST} {Legacy} {Fields}},
	volume = {803},
	issn = {0004-637X},
	shorttitle = {{UV} {Luminosity} {Functions} at {Redshifts} z ∼ 4 to z ∼ 10},
	url = {https://ui.adsabs.harvard.edu/abs/2015ApJ...803...34B},
	doi = {10.1088/0004-637X/803/1/34},
	urldate = {2023-05-01},
	journal = {ApJ},
	author = {Bouwens, R. J. and Illingworth, G. D. and Oesch, P. A. and Trenti, M. and Labbé, I. and Bradley, L. and Carollo, M. and van Dokkum, P. G. and Gonzalez, V. and Holwerda, B. and Franx, M. and Spitler, L. and Smit, R. and Magee, D.},
	month = apr,
	year = {2015},
	note = {ADS Bibcode: 2015ApJ...803...34B},
	pages = {34},
}

@article{bouwens_bright_2016,
	title = {The {Bright} {End} of the z ∼ 9 and z ∼ 10 {UV} {Luminosity} {Functions} {Using} {All} {Five} {CANDELS} {Fields}*},
	volume = {830},
	issn = {0004-637X},
	url = {https://ui.adsabs.harvard.edu/abs/2016ApJ...830...67B},
	doi = {10.3847/0004-637X/830/2/67},
	urldate = {2023-05-01},
	journal = {ApJ},
	author = {Bouwens, R. J. and Oesch, P. A. and Labbé, I. and Illingworth, G. D. and Fazio, G. G. and Coe, D. and Holwerda, B. and Smit, R. and Stefanon, M. and van Dokkum, P. G. and Trenti, M. and Ashby, M. L. N. and Huang, J. -S. and Spitler, L. and Straatman, C. and Bradley, L. and Magee, D.},
	month = oct,
	year = {2016},
	note = {ADS Bibcode: 2016ApJ...830...67B},
	pages = {67},
}

@article{calvi_bright_2016,
	title = {Bright galaxies at {Hubble}'s redshift detection frontier: {Preliminary} results and design from the redshift z{\textasciitilde}9-10 {BoRG} pure-parallel {HST} survey},
	volume = {817},
	issn = {1538-4357},
	shorttitle = {Bright galaxies at {Hubble}'s redshift detection frontier},
	url = {http://arxiv.org/abs/1512.05363},
	doi = {10.3847/0004-637X/817/2/120},
	number = {2},
	urldate = {2023-05-01},
	journal = {ApJ},
	author = {Calvi, V. and Trenti, M. and Stiavelli, M. and Oesch, P. and Bradley, L. D. and Schmidt, K. B. and Coe, D. and Brammer, G. and Bernard, S. and Bouwens, R. J. and Carrasco, D. and Carollo, C. M. and Holwerda, B. W. and MacKenty, J. W. and Mason, C. A. and Shull, J. M. and Treu, T.},
	month = jan,
	year = {2016},
	note = {arXiv:1512.05363 [astro-ph]},
	pages = {120},
}

@article{finkelstein_evolution_2015,
	title = {The {Evolution} of the {Galaxy} {Rest}-frame {Ultraviolet} {Luminosity} {Function} over the {First} {Two} {Billion} {Years}},
	volume = {810},
	issn = {0004-637X},
	url = {https://ui.adsabs.harvard.edu/abs/2015ApJ...810...71F},
	doi = {10.1088/0004-637X/810/1/71},
	urldate = {2023-05-01},
	journal = {ApJ},
	author = {Finkelstein, Steven L. and Ryan, Jr., Russell E. and Papovich, Casey and Dickinson, Mark and Song, Mimi and Somerville, Rachel S. and Ferguson, Henry C. and Salmon, Brett and Giavalisco, Mauro and Koekemoer, Anton M. and Ashby, Matthew L. N. and Behroozi, Peter and Castellano, Marco and Dunlop, James S. and Faber, Sandy M. and Fazio, Giovanni G. and Fontana, Adriano and Grogin, Norman A. and Hathi, Nimish and Jaacks, Jason and Kocevski, Dale D. and Livermore, Rachael and McLure, Ross J. and Merlin, Emiliano and Mobasher, Bahram and Newman, Jeffrey A. and Rafelski, Marc and Tilvi, Vithal and Willner, S. P.},
	month = sep,
	year = {2015},
	note = {ADS Bibcode: 2015ApJ...810...71F},
	pages = {71},
}

@article{atek_new_2015,
	title = {New {Constraints} on the {Faint} {End} of the {UV} {Luminosity} {Function} at z {\textasciitilde} 7-8 {Using} the {Gravitational} {Lensing} of the {Hubble} {Frontier} {Fields} {Cluster} {A2744}},
	volume = {800},
	issn = {0004-637X},
	url = {https://ui.adsabs.harvard.edu/abs/2015ApJ...800...18A},
	doi = {10.1088/0004-637X/800/1/18},
	urldate = {2023-05-18},
	journal = {ApJ},
	author = {Atek, Hakim and Richard, Johan and Kneib, Jean-Paul and Jauzac, Mathilde and Schaerer, Daniel and Clement, Benjamin and Limousin, Marceau and Jullo, Eric and Natarajan, Priyamvada and Egami, Eiichi and Ebeling, Harald},
	month = feb,
	year = {2015},
	note = {ADS Bibcode: 2015ApJ...800...18A},
	pages = {18},
}

@article{bouwens_new_2021,
	title = {New {Determinations} of the {UV} {Luminosity} {Functions} from z ∼ 9 to 2 {Show} a {Remarkable} {Consistency} with {Halo} {Growth} and a {Constant} {Star} {Formation} {Efficiency}},
	volume = {162},
	issn = {1538-3881},
	url = {https://dx.doi.org/10.3847/1538-3881/abf83e},
	doi = {10.3847/1538-3881/abf83e},
	language = {en},
	number = {2},
	urldate = {2023-05-12},
	journal = {AJ},
	author = {Bouwens, R. J. and Oesch, P. A. and Stefanon, M. and Illingworth, G. and Labbé, I. and Reddy, N. and Atek, H. and Montes, M. and Naidu, R. and Nanayakkara, T. and Nelson, E. and Wilkins, S.},
	month = jul,
	year = {2021},
	note = {Publisher: The American Astronomical Society},
	pages = {47},
}

@article{bouwens_uv_2023,
	title = {{UV} luminosity density results at z {\textgreater} 8 from the first {JWST}/{NIRCam} {Fields}: {Limitations} of early data sets and the need for spectroscopy},
	issn = {0035-8711},
	shorttitle = {{UV} luminosity density results at z {\textgreater} 8 from the first {JWST}/{NIRCam} {Fields}},
	url = {https://doi.org/10.1093/mnras/stad1014},
	doi = {10.1093/mnras/stad1014},
	urldate = {2023-05-12},
	journal = {MNRAS},
	author = {Bouwens, Rychard and Illingworth, Garth and Oesch, Pascal and Stefanon, Mauro and Naidu, Rohan and van Leeuwen, Ivana and Magee, Dan},
	month = apr,
	year = {2023},
	pages = {stad1014},
}

@article{bouwens_evolution_2023,
	title = {Evolution of the {UV} {LF} from z ∼ 15 to z ∼ 8 {Using} {New} {JWST} {NIRCam} {Medium}-{Band} {Observations} over the {HUDF}/{XDF}},
	issn = {0035-8711},
	url = {https://doi.org/10.1093/mnras/stad1145},
	doi = {10.1093/mnras/stad1145},
	urldate = {2023-05-12},
	journal = {MNRAS},
	author = {Bouwens, Rychard J and Stefanon, Mauro and Brammer, Gabriel and Oesch, Pascal A and Herard-Demanche, Thomas and Illingworth, Garth D and Matthee, Jorryt and Naidu, Rohan P and van Dokkum, Pieter G and van Leeuwen, Ivana F},
	month = apr,
	year = {2023},
	pages = {stad1145},
}

@article{donnan_evolution_2023,
	title = {The evolution of the galaxy {UV} luminosity function at redshifts z ≃ 8 - 15 from deep {JWST} and ground-based near-infrared imaging},
	volume = {518},
	issn = {0035-8711},
	url = {https://ui.adsabs.harvard.edu/abs/2023MNRAS.518.6011D},
	doi = {10.1093/mnras/stac3472},
	urldate = {2023-11-01},
	journal = {MNRAS},
	author = {Donnan, C. T. and McLeod, D. J. and Dunlop, J. S. and McLure, R. J. and Carnall, A. C. and Begley, R. and Cullen, F. and Hamadouche, M. L. and Bowler, R. A. A. and Magee, D. and McCracken, H. J. and Milvang-Jensen, B. and Moneti, A. and Targett, T.},
	month = feb,
	year = {2023},
	note = {ADS Bibcode: 2023MNRAS.518.6011D},
	pages = {6011--6040},
}

@article{ishigaki_full-data_2018,
	title = {Full-data {Results} of {Hubble} {Frontier} {Fields}: {UV} {Luminosity} {Functions} at z ∼ 6–10 and a {Consistent} {Picture} of {Cosmic} {Reionization}},
	volume = {854},
	issn = {0004-637X},
	shorttitle = {Full-data {Results} of {Hubble} {Frontier} {Fields}},
	url = {https://dx.doi.org/10.3847/1538-4357/aaa544},
	doi = {10.3847/1538-4357/aaa544},
	language = {en},
	number = {1},
	urldate = {2023-05-12},
	journal = {ApJ},
	author = {Ishigaki, Masafumi and Kawamata, Ryota and Ouchi, Masami and Oguri, Masamune and Shimasaku, Kazuhiro and Ono, Yoshiaki},
	month = feb,
	year = {2018},
	note = {Publisher: The American Astronomical Society},
	pages = {73},
}

@article{mclure_new_2013,
	title = {A new multifield determination of the galaxy luminosity function at z = 7-9 incorporating the 2012 {Hubble} {Ultra}-{Deep} {Field} imaging},
	volume = {432},
	issn = {0035-8711},
	url = {https://ui.adsabs.harvard.edu/abs/2013MNRAS.432.2696M},
	doi = {10.1093/mnras/stt627},
	urldate = {2023-05-18},
	journal = {MNRAS},
	author = {McLure, R. J. and Dunlop, J. S. and Bowler, R. A. A. and Curtis-Lake, E. and Schenker, M. and Ellis, R. S. and Robertson, B. E. and Koekemoer, A. M. and Rogers, A. B. and Ono, Y. and Ouchi, M. and Charlot, S. and Wild, V. and Stark, D. P. and Furlanetto, S. R. and Cirasuolo, M. and Targett, T. A.},
	month = jul,
	year = {2013},
	note = {ADS Bibcode: 2013MNRAS.432.2696M},
	pages = {2696--2716},
}

@article{schenker_uv_2013,
	title = {The {UV} {Luminosity} {Function} of {Star}-forming {Galaxies} via {Dropout} {Selection} at {Redshifts} z {\textasciitilde} 7 and 8 from the 2012 {Ultra} {Deep} {Field} {Campaign}},
	volume = {768},
	issn = {0004-637X},
	url = {https://ui.adsabs.harvard.edu/abs/2013ApJ...768..196S},
	doi = {10.1088/0004-637X/768/2/196},
	urldate = {2023-05-18},
	journal = {ApJ},
	author = {Schenker, Matthew A. and Robertson, Brant E. and Ellis, Richard S. and Ono, Yoshiaki and McLure, Ross J. and Dunlop, James S. and Koekemoer, Anton and Bowler, Rebecca A. A. and Ouchi, Masami and Curtis-Lake, Emma and Rogers, Alexander B. and Schneider, Evan and Charlot, Stephane and Stark, Daniel P. and Furlanetto, Steven R. and Cirasuolo, Michele},
	month = may,
	year = {2013},
	note = {ADS Bibcode: 2013ApJ...768..196S},
	pages = {196},
}

@article{willott_steep_2024,
	title = {A {Steep} {Decline} in the {Galaxy} {Space} {Density} beyond {Redshift} 9 in the {CANUCS} {UV} {Luminosity} {Function}},
	volume = {966},
	issn = {0004-637X},
	url = {https://dx.doi.org/10.3847/1538-4357/ad35bc},
	doi = {10.3847/1538-4357/ad35bc},
	language = {en},
	number = {1},
	urldate = {2025-07-09},
	journal = {ApJ},
	author = {Willott, Chris J. and Desprez, Guillaume and Asada, Yoshihisa and Sarrouh, Ghassan T. E. and Abraham, Roberto and Bradač, Maruša and Brammer, Gabe and Estrada-Carpenter, Vince and Iyer, Kartheik G. and Martis, Nicholas S. and Matharu, Jasleen and Mowla, Lamiya and Muzzin, Adam and Noirot, Gaël and Sawicki, Marcin and Strait, Victoria and Rihtaršič, Gregor and Withers, Sunna},
	month = apr,
	year = {2024},
	note = {Publisher: The American Astronomical Society},
	pages = {74},
}

@article{todini_dust_2001,
	title = {Dust formation in primordial {Type} {II} supernovae},
	volume = {325},
	issn = {0035-8711},
	url = {https://doi.org/10.1046/j.1365-8711.2001.04486.x},
	doi = {10.1046/j.1365-8711.2001.04486.x},
	number = {2},
	urldate = {2025-07-09},
	journal = {MNRAS},
	author = {Todini, Paolo and Ferrara, Andrea},
	month = aug,
	year = {2001},
	pages = {726--736},
}

@article{kobayashi_origin_2020,
	title = {The {Origin} of {Elements} from {Carbon} to {Uranium}},
	volume = {900},
	issn = {0004-637X},
	url = {https://dx.doi.org/10.3847/1538-4357/abae65},
	doi = {10.3847/1538-4357/abae65},
	language = {en},
	number = {2},
	urldate = {2025-07-10},
	journal = {ApJ},
	author = {Kobayashi, Chiaki and Karakas, Amanda I. and Lugaro, Maria},
	month = sep,
	year = {2020},
	note = {Publisher: The American Astronomical Society},
	pages = {179},
}

@article{efstathiou_suppressing_1992,
	title = {Suppressing the formation of dwarf galaxies via photoionization},
	volume = {256},
	issn = {0035-8711, 1365-2966},
	url = {https://academic.oup.com/mnras/article-lookup/doi/10.1093/mnras/256.1.43P},
	doi = {10.1093/mnras/256.1.43p},
	language = {en},
	number = {1},
	urldate = {2025-07-15},
	journal = {MNRAS},
	author = {Efstathiou, G.},
	month = may,
	year = {1992},
	note = {Publisher: Oxford University Press (OUP)},
	pages = {43P--47P},
}

@article{renaud_star_2018,
	title = {Star clusters in evolving galaxies},
	volume = {81},
	issn = {1387-6473},
	url = {https://www.sciencedirect.com/science/article/pii/S1387647318300010},
	doi = {10.1016/j.newar.2018.03.001},
	urldate = {2025-07-15},
	journal = {NewAR},
	author = {Renaud, Florent},
	month = apr,
	year = {2018},
	pages = {1--38},
}

@article{krumholz_big_2014,
	title = {The big problems in star formation: {The} star formation rate, stellar clustering, and the initial mass function},
	volume = {539},
	issn = {0370-1573},
	shorttitle = {The big problems in star formation},
	url = {https://ui.adsabs.harvard.edu/abs/2014PhR...539...49K},
	doi = {10.1016/j.physrep.2014.02.001},
	urldate = {2025-07-15},
	journal = {PhR},
	author = {Krumholz, Mark R.},
	month = jun,
	year = {2014},
	note = {Publisher: Elsevier
ADS Bibcode: 2014PhR...539...49K},
	pages = {49--134},
}

@article{balser_metallicity-electron_2024,
	title = {The {Metallicity}–{Electron} {Temperature} {Relationship} in {H} {II} {Regions}},
	volume = {964},
	issn = {0004-637X},
	url = {https://ui.adsabs.harvard.edu/abs/2024ApJ...964...47B},
	doi = {10.3847/1538-4357/ad2458},
	urldate = {2025-07-19},
	journal = {ApJ},
	author = {Balser, Dana S. and Wenger, Trey V.},
	month = mar,
	year = {2024},
	note = {Publisher: IOP
ADS Bibcode: 2024ApJ...964...47B},
	pages = {47},
}

@article{bate_statistical_2022,
	title = {The statistical properties of stars at redshift, z = 5, compared with the present epoch},
	volume = {519},
	issn = {0035-8711},
	url = {https://doi.org/10.1093/mnras/stac3481},
	doi = {10.1093/mnras/stac3481},
	number = {1},
	urldate = {2025-07-19},
	journal = {MNRAS},
	author = {Bate, Matthew R},
	month = feb,
	year = {2023},
	pages = {688--708},
}

@article{ferrara_super-early_2023,
	title = {Super-early {JWST} galaxies, outflows, and {Lyα} visibility in the {Epoch} of {Reionization}},
	volume = {684},
	copyright = {© The Authors 2024},
	issn = {0004-6361, 1432-0746},
	url = {https://www.aanda.org/articles/aa/abs/2024/04/aa48321-23/aa48321-23.html},
	doi = {10.1051/0004-6361/202348321},
	language = {en},
	urldate = {2025-07-19},
	journal = {A\&A},
	author = {Ferrara, A.},
	month = apr,
	year = {2024},
	note = {Publisher: EDP Sciences},
	pages = {A207},
}

@article{hutter_astraeus_2023,
	title = {astraeus – {VIII}. {A} new framework for {Lyman}-α emitters applied to different reionization scenarios},
	volume = {524},
	issn = {0035-8711},
	url = {https://doi.org/10.1093/mnras/stad2230},
	doi = {10.1093/mnras/stad2230},
	number = {4},
	urldate = {2025-07-19},
	journal = {MNRAS},
	author = {Hutter, Anne and Trebitsch, Maxime and Dayal, Pratika and Gottlöber, Stefan and Yepes, Gustavo and Legrand, Laurent},
	month = oct,
	year = {2023},
	pages = {6124--6148},
}

@article{menon_bursts_2024,
	title = {Bursts of {Star} {Formation} and {Radiation}-driven {Outflows} {Produce} {Efficient} {LyC} {Leakage} from {Dense} {Compact} {Star} {Clusters}},
	volume = {987},
	issn = {0004-637X},
	url = {https://dx.doi.org/10.3847/1538-4357/add2f9},
	doi = {10.3847/1538-4357/add2f9},
	language = {en},
	number = {1},
	urldate = {2025-07-19},
	journal = {ApJ},
	author = {Menon, Shyam H. and Burkhart, Blakesley and Somerville, Rachel S. and Thompson, Todd A. and Sternberg, Amiel},
	month = jun,
	year = {2025},
	note = {Publisher: The American Astronomical Society},
	pages = {12},
}

@article{menon_infrared_2022,
	title = {Infrared radiation feedback does not regulate star cluster formation},
	volume = {517},
	issn = {0035-8711},
	url = {https://doi.org/10.1093/mnras/stac2702},
	doi = {10.1093/mnras/stac2702},
	number = {1},
	urldate = {2025-07-19},
	journal = {MNRAS},
	author = {Menon, Shyam H and Federrath, Christoph and Krumholz, Mark R},
	month = nov,
	year = {2022},
	pages = {1313--1338},
}

@article{menon_outflows_2023,
	title = {Outflows driven by direct and reprocessed radiation pressure in massive star clusters},
	volume = {521},
	issn = {0035-8711},
	url = {https://doi.org/10.1093/mnras/stad856},
	doi = {10.1093/mnras/stad856},
	number = {4},
	urldate = {2025-07-19},
	journal = {MNRAS},
	author = {Menon, Shyam H and Federrath, Christoph and Krumholz, Mark R},
	month = jun,
	year = {2023},
	pages = {5160--5176},
}

@article{murray_thehalomod_2021,
	title = {{TheHaloMod}: {An} online calculator for the halo model},
	volume = {36},
	issn = {2213-1337},
	shorttitle = {{TheHaloMod}},
	url = {https://www.sciencedirect.com/science/article/pii/S221313372100041X},
	doi = {10.1016/j.ascom.2021.100487},
	urldate = {2025-07-19},
	journal = {A\&C},
	author = {Murray, S. G. and Diemer, B. and Chen, Z. and Neuhold, A. G. and Schnapp, M. A. and Peruzzi, T. and Blevins, D. and Engelman, T.},
	month = jul,
	year = {2021},
	pages = {100487},
}

@ARTICLE{Chon2024,
       author = {{Chon}, Sunmyon and {Hosokawa}, Takashi and {Omukai}, Kazuyuki and {Schneider}, Raffaella},
        title = "{Impact of radiative feedback on the initial mass function of metal-poor stars}",
      journal = {\mnras},
         year = 2024,
        month = may,
       volume = {530},
       number = {3},
        pages = {2453-2474},
          doi = {10.1093/mnras/stae1027},
archivePrefix = {arXiv},
       eprint = {2312.13339},
 primaryClass = {astro-ph.GA},
       adsurl = {https://ui.adsabs.harvard.edu/abs/2024MNRAS.530.2453C}
}

@ARTICLE{Fukushima2020,
       author = {{Fukushima}, Hajime and {Yajima}, Hidenobu and {Sugimura}, Kazuyuki and {Hosokawa}, Takashi and {Omukai}, Kazuyuki and {Matsumoto}, Tomoaki},
        title = "{Star cluster formation and cloud dispersal by radiative feedback: dependence on metallicity and compactness}",
      journal = {\mnras},
         year = 2020,
        month = sep,
       volume = {497},
       number = {3},
        pages = {3830-3845},
          doi = {10.1093/mnras/staa2062},
archivePrefix = {arXiv},
       eprint = {2005.13401},
 primaryClass = {astro-ph.GA},
       adsurl = {https://ui.adsabs.harvard.edu/abs/2020MNRAS.497.3830F}
}

@ARTICLE{Fukushima2021,
       author = {{Fukushima}, Hajime and {Yajima}, Hidenobu},
        title = "{Radiation hydrodynamics simulations of massive star cluster formation in giant molecular clouds}",
      journal = {\mnras},
         year = 2021,
        month = oct,
       volume = {506},
       number = {4},
        pages = {5512-5539},
          doi = {10.1093/mnras/stab2099},
archivePrefix = {arXiv},
       eprint = {2104.10892},
 primaryClass = {astro-ph.GA},
       adsurl = {https://ui.adsabs.harvard.edu/abs/2021MNRAS.506.5512F}
}

@ARTICLE{Fukushima2023,
       author = {{Fukushima}, Hajime and {Yajima}, Hidenobu},
        title = "{The formation of globular clusters with top-heavy initial mass functions}",
      journal = {\mnras},
         year = 2023,
        month = sep,
       volume = {524},
       number = {1},
        pages = {1422-1430},
          doi = {10.1093/mnras/stad1956},
archivePrefix = {arXiv},
       eprint = {2303.12405},
 primaryClass = {astro-ph.GA},
       adsurl = {https://ui.adsabs.harvard.edu/abs/2023MNRAS.524.1422F}
}

@ARTICLE{Garcia2023,
       author = {{Garcia}, Fred Angelo Batan and {Ricotti}, Massimo and {Sugimura}, Kazuyuki and {Park}, Jongwon},
        title = "{Star cluster formation and survival in the first galaxies}",
      journal = {\mnras},
         year = 2023,
        month = jun,
       volume = {522},
       number = {2},
        pages = {2495-2515},
          doi = {10.1093/mnras/stad1092},
archivePrefix = {arXiv},
       eprint = {2212.13946},
 primaryClass = {astro-ph.GA},
       adsurl = {https://ui.adsabs.harvard.edu/abs/2023MNRAS.522.2495G}
}

@ARTICLE{Geen2017,
       author = {{Geen}, Sam and {Soler}, Juan D. and {Hennebelle}, Patrick},
        title = "{Interpreting the star formation efficiency of nearby molecular clouds with ionizing radiation}",
      journal = {\mnras},
         year = 2017,
        month = nov,
       volume = {471},
       number = {4},
        pages = {4844-4855},
          doi = {10.1093/mnras/stx1765},
archivePrefix = {arXiv},
       eprint = {1703.10071},
 primaryClass = {astro-ph.GA},
       adsurl = {https://ui.adsabs.harvard.edu/abs/2017MNRAS.471.4844G}
}

@ARTICLE{He2019,
       author = {{He}, Chong-Chong and {Ricotti}, Massimo and {Geen}, Sam},
        title = "{Simulating star clusters across cosmic time - I. Initial mass function, star formation rates, and efficiencies}",
      journal = {\mnras},
         year = 2019,
        month = oct,
       volume = {489},
       number = {2},
        pages = {1880-1898},
          doi = {10.1093/mnras/stz2239},
archivePrefix = {arXiv},
       eprint = {1904.07889},
 primaryClass = {astro-ph.GA},
       adsurl = {https://ui.adsabs.harvard.edu/abs/2019MNRAS.489.1880H}
}

@ARTICLE{Kim2018,
       author = {{Kim}, Jeong-Gyu and {Kim}, Woong-Tae and {Ostriker}, Eve C.},
        title = "{Modeling UV Radiation Feedback from Massive Stars. II. Dispersal of Star-forming Giant Molecular Clouds by Photoionization and Radiation Pressure}",
      journal = {\apj},
         year = 2018,
        month = may,
       volume = {859},
       number = {1},
          eid = {68},
        pages = {68},
          doi = {10.3847/1538-4357/aabe27},
archivePrefix = {arXiv},
       eprint = {1804.04664},
 primaryClass = {astro-ph.GA},
       adsurl = {https://ui.adsabs.harvard.edu/abs/2018ApJ...859...68K}
}

@ARTICLE{Raskutti2016,
       author = {{Raskutti}, Sudhir and {Ostriker}, Eve C. and {Skinner}, M. Aaron},
        title = "{Numerical Simulations of Turbulent Molecular Clouds Regulated by Radiation Feedback Forces. I. Star Formation Rate and Efficiency}",
      journal = {\apj},
         year = 2016,
        month = oct,
       volume = {829},
       number = {2},
          eid = {130},
        pages = {130},
          doi = {10.3847/0004-637X/829/2/130},
archivePrefix = {arXiv},
       eprint = {1608.04469},
 primaryClass = {astro-ph.GA},
       adsurl = {https://ui.adsabs.harvard.edu/abs/2016ApJ...829..130R}
}

@ARTICLE{Reina-Campos2025,
       author = {{Reina-Campos}, Marta and {Gnedin}, Oleg Y. and {Sills}, Alison and {Li}, Hui},
        title = "{The Star Clusters as Links between Galaxy Evolution and Star Formation Project. I. Numerical Method}",
      journal = {\apj},
         year = 2025,
        month = jan,
       volume = {978},
       number = {1},
          eid = {15},
        pages = {15},
          doi = {10.3847/1538-4357/ad909f},
archivePrefix = {arXiv},
       eprint = {2408.04694},
 primaryClass = {astro-ph.GA},
       adsurl = {https://ui.adsabs.harvard.edu/abs/2025ApJ...978...15R}
}

@ARTICLE{Tanvir2024,
       author = {{Tanvir}, Tabassum S. and {Krumholz}, Mark R.},
        title = "{The metallicity dependence of the stellar initial mass function}",
      journal = {\mnras},
         year = 2024,
        month = jan,
       volume = {527},
       number = {3},
        pages = {7306-7316},
          doi = {10.1093/mnras/stad3581},
archivePrefix = {arXiv},
       eprint = {2305.20039},
 primaryClass = {astro-ph.GA},
       adsurl = {https://ui.adsabs.harvard.edu/abs/2024MNRAS.527.7306T}
}

@ARTICLE{Adamo2024,
       author = {{Adamo}, Angela and {Bradley}, Larry D. and {Vanzella}, Eros and {Claeyssens}, Ad{\'e}la{\"\i}de and {Welch}, Brian and {Diego}, Jose M. and {Mahler}, Guillaume and {Oguri}, Masamune and {Sharon}, Keren and {Abdurro'uf} and {Hsiao}, Tiger Yu-Yang and {Xu}, Xinfeng and {Messa}, Matteo and {Lassen}, Augusto E. and {Zackrisson}, Erik and {Brammer}, Gabriel and {Coe}, Dan and {Kokorev}, Vasily and {Ricotti}, Massimo and {Zitrin}, Adi and {Fujimoto}, Seiji and {Inoue}, Akio K. and {Resseguier}, Tom and {Rigby}, Jane R. and {Jim{\'e}nez-Teja}, Yolanda and {Windhorst}, Rogier A. and {Hashimoto}, Takuya and {Tamura}, Yoichi},
        title = "{Bound star clusters observed in a lensed galaxy 460 Myr after the Big Bang}",
      journal = {\nat},
         year = 2024,
        month = aug,
       volume = {632},
       number = {8025},
        pages = {513-516},
          doi = {10.1038/s41586-024-07703-7},
archivePrefix = {arXiv},
       eprint = {2401.03224},
 primaryClass = {astro-ph.GA},
       adsurl = {https://ui.adsabs.harvard.edu/abs/2024Natur.632..513A}
}

@ARTICLE{Topping2025,
       author = {{Topping}, Michael W. and {Stark}, Daniel P. and {Senchyna}, Peter and {Chen}, Zuyi and {Zitrin}, Adi and {Endsley}, Ryan and {Charlot}, St{\'e}phane and {Furtak}, Lukas J. and {Maseda}, Michael V. and {Plat}, Adele and {Smit}, Renske and {Mainali}, Ramesh and {Chevallard}, Jacopo and {Molyneux}, Stephen and {Rigby}, Jane R.},
        title = "{Deep Rest-UV JWST/NIRSpec Spectroscopy of Early Galaxies: The Demographics of C IV and N-emitters in the Reionization Era}",
      journal = {\apj},
         year = 2025,
        month = feb,
       volume = {980},
       number = {2},
          eid = {225},
        pages = {225},
          doi = {10.3847/1538-4357/ada95c},
archivePrefix = {arXiv},
       eprint = {2407.19009},
 primaryClass = {astro-ph.GA},
       adsurl = {https://ui.adsabs.harvard.edu/abs/2025ApJ...980..225T}
}

@ARTICLE{Narayanan2026,
       author = {{Narayanan}, Desika and {Torrey}, Paul and {Stark}, Daniel and {Chisholm}, John and {Finkelstein}, Steven and {Garcia}, Alex and {Kelley-Derzon}, Jessica and {Marinacci}, Federico and {Sales}, Laura and {Savitch}, Ethan and {Vogelsberger}, Mark and {Zimmerman}, Dhruv},
        title = "{The Growth of Dust in Galaxies in the First Billion Years with Applications to Blue Monsters}",
      journal = {The Open Journal of Astrophysics},
         year = 2026,
        month = apr,
       volume = {9},
        pages = {59986},
          doi = {10.33232/001c.159986},
archivePrefix = {arXiv},
       eprint = {2509.18266},
 primaryClass = {astro-ph.GA},
       adsurl = {https://ui.adsabs.harvard.edu/abs/2026OJAp....959986N}
}

@article{hunter_matplotlib_2007,
    title = {Matplotlib: {A} {2D} {Graphics} {Environment}},
    volume = {9},
    issn = {1558-366X},
    shorttitle = {Matplotlib},
    url = {https://ieeexplore.ieee.org/document/4160265},
    doi = {10.1109/MCSE.2007.55},
    number = {3},
    urldate = {2026-07-21},
    journal = {Computing in Science \& Engineering},
    author = {Hunter, John D.},
    month = may,
    year = {2007},
    pages = {90--95},
}

@article{van_der_walt_numpy_2011,
    title = {The {NumPy} {Array}: {A} {Structure} for {Efficient} {Numerical} {Computation}},
    volume = {13},
    issn = {1558-366X},
    shorttitle = {The {NumPy} {Array}},
    url = {https://ieeexplore.ieee.org/document/5725236},
    doi = {10.1109/MCSE.2011.37},
    number = {2},
    urldate = {2026-07-21},
    journal = {Computing in Science \& Engineering},
    author = {van der Walt, Stefan and Colbert, S. Chris and Varoquaux, Gael},
    month = mar,
    year = {2011},
    pages = {22--30},
}

@ARTICLE{Isobe2023,
       author = {{Isobe}, Yuki and {Ouchi}, Masami and {Tominaga}, Nozomu and {Watanabe}, Kuria and {Nakajima}, Kimihiko and {Umeda}, Hiroya and {Yajima}, Hidenobu and {Harikane}, Yuichi and {Fukushima}, Hajime and {Xu}, Yi and {Ono}, Yoshiaki and {Zhang}, Yechi},
        title = "{JWST Identification of Extremely Low C/N Galaxies with [N/O] {\ensuremath{\gtrsim}} 0.5 at z 6-10 Evidencing the Early CNO-cycle Enrichment and a Connection with Globular Cluster Formation}",
      journal = {\apj},
         year = 2023,
        month = dec,
       volume = {959},
       number = {2},
          eid = {100},
        pages = {100},
          doi = {10.3847/1538-4357/ad09be},
archivePrefix = {arXiv},
       eprint = {2307.00710},
 primaryClass = {astro-ph.GA},
       adsurl = {https://ui.adsabs.harvard.edu/abs/2023ApJ...959..100I}
}

@ARTICLE{Marques-Chaves2024,
       author = {{Marques-Chaves}, R. and {Schaerer}, D. and {Kuruvanthodi}, A. and {Korber}, D. and {Prantzos}, N. and {Charbonnel}, C. and {Weibel}, A. and {Izotov}, Y.~I. and {Messa}, M. and {Brammer}, G. and {Dessauges-Zavadsky}, M. and {Oesch}, P.},
        title = "{Extreme N-emitters at high redshift: Possible signatures of supermassive stars and globular cluster or black hole formation in action}",
      journal = {\aap},
         year = 2024,
        month = jan,
       volume = {681},
          eid = {A30},
        pages = {A30},
          doi = {10.1051/0004-6361/202347411},
archivePrefix = {arXiv},
       eprint = {2307.04234},
 primaryClass = {astro-ph.GA},
       adsurl = {https://ui.adsabs.harvard.edu/abs/2024A&A...681A..30M}
}

@ARTICLE{Senchyna2024,
       author = {{Senchyna}, Peter and {Plat}, Adele and {Stark}, Daniel P. and {Rudie}, Gwen C. and {Berg}, Danielle and {Charlot}, St{\'e}phane and {James}, Bethan L. and {Mingozzi}, Matilde},
        title = "{GN-z11 in Context: Possible Signatures of Globular Cluster Precursors at Redshift 10}",
      journal = {\apj},
         year = 2024,
        month = may,
       volume = {966},
       number = {1},
          eid = {92},
        pages = {92},
          doi = {10.3847/1538-4357/ad235e},
       adsurl = {https://ui.adsabs.harvard.edu/abs/2024ApJ...966...92S}
}

@ARTICLE{Watanabe2024,
       author = {{Watanabe}, Kuria and {Ouchi}, Masami and {Nakajima}, Kimihiko and {Isobe}, Yuki and {Tominaga}, Nozomu and {Suzuki}, Akihiro and {Ishigaki}, Miho N. and {Nomoto}, Ken'ichi and {Takahashi}, Koh and {Harikane}, Yuichi and {Hatano}, Shun and {Kusakabe}, Haruka and {Moriya}, Takashi J. and {Nishigaki}, Moka and {Ono}, Yoshiaki and {Onodera}, Masato and {Sugahara}, Yuma},
        title = "{EMPRESS. XIII. Chemical Enrichment of Young Galaxies Near and Far at z {\ensuremath{\sim}} 0 and 4{\textendash}10: Fe/O, Ar/O, S/O, and N/O Measurements with a Comparison of Chemical Evolution Models}",
      journal = {\apj},
         year = 2024,
        month = feb,
       volume = {962},
       number = {1},
          eid = {50},
        pages = {50},
          doi = {10.3847/1538-4357/ad13ff},
archivePrefix = {arXiv},
       eprint = {2305.02078},
 primaryClass = {astro-ph.GA},
       adsurl = {https://ui.adsabs.harvard.edu/abs/2024ApJ...962...50W}
}

@ARTICLE{Andalman2025,
       author = {{Andalman}, Zachary L. and {Teyssier}, Romain and {Dekel}, Avishai},
        title = "{On the origin of the high star formation efficiency in massive galaxies at Cosmic Dawn}",
      journal = {\mnras},
         year = 2025,
        month = jul,
       volume = {540},
       number = {4},
        pages = {3350-3383},
          doi = {10.1093/mnras/staf930},
archivePrefix = {arXiv},
       eprint = {2410.20530},
 primaryClass = {astro-ph.GA},
       adsurl = {https://ui.adsabs.harvard.edu/abs/2025MNRAS.540.3350A}
}

@ARTICLE{Ceverino2024,
       author = {{Ceverino}, D. and {Nakazato}, Y. and {Yoshida}, N. and {Klessen}, R.~S. and {Glover}, S.~C.~O.},
        title = "{Redshift-dependent galaxy formation efficiency at z = 5 ‑ 13 in the FirstLight Simulations}",
      journal = {\aap},
         year = 2024,
        month = sep,
       volume = {689},
          eid = {A244},
        pages = {A244},
          doi = {10.1051/0004-6361/202450224},
archivePrefix = {arXiv},
       eprint = {2404.02537},
 primaryClass = {astro-ph.GA},
       adsurl = {https://ui.adsabs.harvard.edu/abs/2024A&A...689A.244C}
}

@ARTICLE{Chen2026,
       author = {{Chen}, Hou-Zun and {Li}, Zhaozhou and {Dekel}, Avishai and {Yao}, Zhiyuan and {Mandelker}, Nir and {Kang}, Xi},
        title = "{Feedback-Free Star Formation in Clusters within a Galaxy Simulated at High Resolution in Cosmic Dawn}",
      journal = {arXiv e-prints},
         year = 2026,
        month = jun,
          eid = {arXiv:2606.12605},
        pages = {arXiv:2606.12605},
          doi = {10.48550/arXiv.2606.12605},
archivePrefix = {arXiv},
       eprint = {2606.12605},
 primaryClass = {astro-ph.GA},
       adsurl = {https://ui.adsabs.harvard.edu/abs/2026arXiv260612605C}
}

@ARTICLE{Chon2022,
       author = {{Chon}, Sunmyon and {Ono}, Haruka and {Omukai}, Kazuyuki and {Schneider}, Raffaella},
        title = "{Impact of the cosmic background radiation on the initial mass function of metal-poor stars}",
      journal = {\mnras},
         year = 2022,
        month = aug,
       volume = {514},
       number = {3},
        pages = {4639-4654},
          doi = {10.1093/mnras/stac1549},
archivePrefix = {arXiv},
       eprint = {2205.15328},
 primaryClass = {astro-ph.GA},
       adsurl = {https://ui.adsabs.harvard.edu/abs/2022MNRAS.514.4639C}
}

@ARTICLE{Dekel2023,
       author = {{Dekel}, Avishai and {Sarkar}, Kartick C. and {Birnboim}, Yuval and {Mandelker}, Nir and {Li}, Zhaozhou},
        title = "{Efficient formation of massive galaxies at cosmic dawn by feedback-free starbursts}",
      journal = {\mnras},
         year = 2023,
        month = aug,
       volume = {523},
       number = {3},
        pages = {3201-3218},
          doi = {10.1093/mnras/stad1557},
archivePrefix = {arXiv},
       eprint = {2303.04827},
 primaryClass = {astro-ph.GA},
       adsurl = {https://ui.adsabs.harvard.edu/abs/2023MNRAS.523.3201D}
}

@ARTICLE{Ferrara2023,
       author = {{Ferrara}, Andrea and {Pallottini}, Andrea and {Dayal}, Pratika},
        title = "{On the stunning abundance of super-early, luminous galaxies revealed by JWST}",
      journal = {\mnras},
         year = 2023,
        month = jul,
       volume = {522},
       number = {3},
        pages = {3986-3991},
          doi = {10.1093/mnras/stad1095},
archivePrefix = {arXiv},
       eprint = {2208.00720},
 primaryClass = {astro-ph.GA},
       adsurl = {https://ui.adsabs.harvard.edu/abs/2023MNRAS.522.3986F}
}

@ARTICLE{Fiore2023,
       author = {{Fiore}, Fabrizio and {Ferrara}, Andrea and {Bischetti}, Manuela and {Feruglio}, Chiara and {Travascio}, Andrea},
        title = "{Dusty-wind-clear JWST Super-early Galaxies}",
      journal = {\apjl},
         year = 2023,
        month = feb,
       volume = {943},
       number = {2},
          eid = {L27},
        pages = {L27},
          doi = {10.3847/2041-8213/acb5f2},
archivePrefix = {arXiv},
       eprint = {2211.08937},
 primaryClass = {astro-ph.GA},
       adsurl = {https://ui.adsabs.harvard.edu/abs/2023ApJ...943L..27F}
}

@ARTICLE{Garaldi2026,
       author = {{Garaldi}, Enrico and {Popovic}, Filip and {Kannan}, Rahul and {Smith}, Aaron and {Puchwein}, Ewald and {Yoshida}, Naoki and {Nagamine}, Kentaro and {Peroux}, Celine and {Keating}, Laura and {Vogelsberger}, Mark and {McClymont}, William and {Shen}, Xuejian and {Tacchella}, Sandro and {Hernquist}, Lars},
        title = "{The Thesan-Zoom project: bursty star formation is incompatible with prolonged dust survival}",
      journal = {arXiv e-prints},
         year = 2026,
        month = jul,
          eid = {arXiv:2607.08824},
        pages = {arXiv:2607.08824},
archivePrefix = {arXiv},
       eprint = {2607.08824},
 primaryClass = {astro-ph.GA},
       adsurl = {https://ui.adsabs.harvard.edu/abs/2026arXiv260708824G}
}

@ARTICLE{Gelli2024,
       author = {{Gelli}, Viola and {Mason}, Charlotte and {Hayward}, Christopher C.},
        title = "{The Impact of Mass-dependent Stochasticity at Cosmic Dawn}",
      journal = {\apj},
         year = 2024,
        month = nov,
       volume = {975},
       number = {2},
          eid = {192},
        pages = {192},
          doi = {10.3847/1538-4357/ad7b36},
archivePrefix = {arXiv},
       eprint = {2405.13108},
 primaryClass = {astro-ph.GA},
       adsurl = {https://ui.adsabs.harvard.edu/abs/2024ApJ...975..192G}
}

@ARTICLE{Jeong2025,
       author = {{Jeong}, Tae Bong and {Jeon}, Myoungwon and {Song}, Hyunmi and {Bromm}, Volker},
        title = "{Simulating High-redshift Galaxies: Enhancing UV Luminosity with Star Formation Efficiency and a Top-heavy IMF}",
      journal = {\apj},
         year = 2025,
        month = feb,
       volume = {980},
       number = {1},
          eid = {10},
        pages = {10},
          doi = {10.3847/1538-4357/ada27d},
archivePrefix = {arXiv},
       eprint = {2411.17007},
 primaryClass = {astro-ph.GA},
       adsurl = {https://ui.adsabs.harvard.edu/abs/2025ApJ...980...10J}
}

@ARTICLE{Katz2024,
       author = {{Katz}, Harley and {Ji}, Alexander P and {Telford}, Grace and {Senchyna}, Peter},
        title = "{Early Bright Galaxies from Helium Enhancements in High-Redshift Star Clusters}",
      journal = {The Open Journal of Astrophysics},
         year = 2024,
        month = nov,
       volume = {7},
          eid = {106},
        pages = {106},
          doi = {10.33232/001c.126253},
archivePrefix = {arXiv},
       eprint = {2410.14846},
 primaryClass = {astro-ph.GA},
       adsurl = {https://ui.adsabs.harvard.edu/abs/2024OJAp....7E.106K}
}

@ARTICLE{Katz2025,
       author = {{Katz}, Harley and {Cameron}, Alex J. and {Saxena}, Aayush and {Barrufet}, Laia and {Choustikov}, Nichloas and {Cleri}, Nikko J. and {de Graff}, Anna and {Ellis}, Richard S. and {Fosbury}, Robert A.~E. and {Heintz}, Kasper E. and {Maseda}, Michael and {Matthee}, Jorryt and {McConachie}, Ian and {Oesch}, Pascal A.},
        title = "{21 Balmer Jump Street: The Nebular Continuum at High Redshift and Implications for the Bright Galaxy Problem, UV Continuum Slopes, and Early Stellar Populations}",
      journal = {The Open Journal of Astrophysics},
         year = 2025,
        month = jul,
       volume = {8},
          eid = {104},
        pages = {104},
          doi = {10.33232/001c.142570},
archivePrefix = {arXiv},
       eprint = {2408.03189},
 primaryClass = {astro-ph.GA},
       adsurl = {https://ui.adsabs.harvard.edu/abs/2025OJAp....8E.104K}
}

@ARTICLE{Li2024,
       author = {{Li}, Zhaozhou and {Dekel}, Avishai and {Sarkar}, Kartick C. and {Aung}, Han and {Giavalisco}, Mauro and {Mandelker}, Nir and {Tacchella}, Sandro},
        title = "{Feedback-free starbursts at cosmic dawn: Observable predictions for JWST}",
      journal = {\aap},
         year = 2024,
        month = oct,
       volume = {690},
          eid = {A108},
        pages = {A108},
          doi = {10.1051/0004-6361/202348727},
archivePrefix = {arXiv},
       eprint = {2311.14662},
 primaryClass = {astro-ph.GA},
       adsurl = {https://ui.adsabs.harvard.edu/abs/2024A&A...690A.108L}
}

@ARTICLE{Lu2025,
       author = {{Lu}, Shengdong and {Frenk}, Carlos S. and {Bose}, Sownak and {Lacey}, Cedric G. and {Cole}, Shaun and {Baugh}, Carlton M. and {Helly}, John C.},
        title = "{A comparison of pre-existing {\ensuremath{\Lambda}}CDM predictions with the abundance of JWST galaxies at high redshift}",
      journal = {\mnras},
         year = 2025,
        month = jan,
       volume = {536},
       number = {1},
        pages = {1018-1034},
          doi = {10.1093/mnras/stae2646},
archivePrefix = {arXiv},
       eprint = {2406.02672},
 primaryClass = {astro-ph.GA},
       adsurl = {https://ui.adsabs.harvard.edu/abs/2025MNRAS.536.1018L}
}

@ARTICLE{Mason2023,
       author = {{Mason}, Charlotte A. and {Trenti}, Michele and {Treu}, Tommaso},
        title = "{The brightest galaxies at cosmic dawn}",
      journal = {\mnras},
         year = 2023,
        month = may,
       volume = {521},
       number = {1},
        pages = {497-503},
          doi = {10.1093/mnras/stad035},
archivePrefix = {arXiv},
       eprint = {2207.14808},
 primaryClass = {astro-ph.GA},
       adsurl = {https://ui.adsabs.harvard.edu/abs/2023MNRAS.521..497M}
}

@ARTICLE{Mayer2025,
       author = {{Mayer}, Lucio and {van Donkelaar}, Floor and {Messa}, Matteo and {Capelo}, Pedro R. and {Adamo}, Angela},
        title = "{In Situ Formation of Star Clusters at z > 7 via Galactic Disk Fragmentation: Shedding Light on Ultracompact Clusters and Overmassive Black Holes Seen by JWST}",
      journal = {\apjl},
         year = 2025,
        month = mar,
       volume = {981},
       number = {2},
          eid = {L28},
        pages = {L28},
          doi = {10.3847/2041-8213/adadfe},
archivePrefix = {arXiv},
       eprint = {2411.00670},
 primaryClass = {astro-ph.GA},
       adsurl = {https://ui.adsabs.harvard.edu/abs/2025ApJ...981L..28M}
}

@ARTICLE{Menci2024,
       author = {{Menci}, N. and {Sen}, A.~A. and {Castellano}, M.},
        title = "{The Excess of JWST Bright Galaxies: A Possible Origin in the Ground State of Dynamical Dark Energy in the Light of DESI 2024 Data}",
      journal = {\apj},
         year = 2024,
        month = dec,
       volume = {976},
       number = {2},
          eid = {227},
        pages = {227},
          doi = {10.3847/1538-4357/ad8d5b},
archivePrefix = {arXiv},
       eprint = {2410.22940},
 primaryClass = {astro-ph.CO},
       adsurl = {https://ui.adsabs.harvard.edu/abs/2024ApJ...976..227M}
}

@ARTICLE{Menon2024,
       author = {{Menon}, Shyam H. and {Lancaster}, Lachlan and {Burkhart}, Blakesley and {Somerville}, Rachel S. and {Dekel}, Avishai and {Krumholz}, Mark R.},
        title = "{The Interplay between the Initial Mass Function and Star Formation Efficiency through Radiative Feedback at High Stellar Surface Densities}",
      journal = {\apjl},
         year = 2024,
        month = jun,
       volume = {967},
       number = {2},
          eid = {L28},
        pages = {L28},
          doi = {10.3847/2041-8213/ad462d},
archivePrefix = {arXiv},
       eprint = {2405.00813},
 primaryClass = {astro-ph.GA},
       adsurl = {https://ui.adsabs.harvard.edu/abs/2024ApJ...967L..28M}
}

@ARTICLE{Pacucci2022,
       author = {{Pacucci}, Fabio and {Dayal}, Pratika and {Harikane}, Yuichi and {Inoue}, Akio K. and {Loeb}, Abraham},
        title = "{Are the newly-discovered z   13 drop-out sources starburst galaxies or quasars?}",
      journal = {\mnras},
         year = 2022,
        month = jul,
       volume = {514},
       number = {1},
        pages = {L6-L10},
          doi = {10.1093/mnrasl/slac035},
archivePrefix = {arXiv},
       eprint = {2201.00823},
 primaryClass = {astro-ph.GA},
       adsurl = {https://ui.adsabs.harvard.edu/abs/2022MNRAS.514L...6P}
}

@ARTICLE{Shen2023,
       author = {{Shen}, Xuejian and {Vogelsberger}, Mark and {Boylan-Kolchin}, Michael and {Tacchella}, Sandro and {Kannan}, Rahul},
        title = "{The impact of UV variability on the abundance of bright galaxies at z {\ensuremath{\geq}} 9}",
      journal = {\mnras},
         year = 2023,
        month = nov,
       volume = {525},
       number = {3},
        pages = {3254-3261},
          doi = {10.1093/mnras/stad2508},
archivePrefix = {arXiv},
       eprint = {2305.05679},
 primaryClass = {astro-ph.GA},
       adsurl = {https://ui.adsabs.harvard.edu/abs/2023MNRAS.525.3254S}
}

@ARTICLE{Somerville2025,
       author = {{Somerville}, Rachel S. and {Yung}, L.~Y. Aaron and {Lancaster}, Lachlan and {Menon}, Shyam and {Sommovigo}, Laura and {Finkelstein}, Steven L.},
        title = "{Density-modulated star formation efficiency: implications for the observed abundance of ultraviolet luminous galaxies at z > 10}",
      journal = {\mnras},
         year = 2025,
        month = dec,
       volume = {544},
       number = {4},
        pages = {3774-3798},
          doi = {10.1093/mnras/staf1824},
archivePrefix = {arXiv},
       eprint = {2505.05442},
 primaryClass = {astro-ph.GA},
       adsurl = {https://ui.adsabs.harvard.edu/abs/2025MNRAS.544.3774S}
}

@ARTICLE{Sun2023,
       author = {{Sun}, Guochao and {Faucher-Gigu{\`e}re}, Claude-Andr{\'e} and {Hayward}, Christopher C. and {Shen}, Xuejian and {Wetzel}, Andrew and {Cochrane}, Rachel K.},
        title = "{Bursty Star Formation Naturally Explains the Abundance of Bright Galaxies at Cosmic Dawn}",
      journal = {\apjl},
         year = 2023,
        month = oct,
       volume = {955},
       number = {2},
          eid = {L35},
        pages = {L35},
          doi = {10.3847/2041-8213/acf85a},
archivePrefix = {arXiv},
       eprint = {2307.15305},
 primaryClass = {astro-ph.GA},
       adsurl = {https://ui.adsabs.harvard.edu/abs/2023ApJ...955L..35S}
}

@ARTICLE{Toyouchi2025,
       author = {{Toyouchi}, Daisuke and {Yajima}, Hidenobu and {Ferrara}, Andrea and {Nagamine}, Kentaro},
        title = "{Bridging Theory and Observations: Insights into Star Formation Efficiency and Dust Attenuation in z > 5 Galaxies}",
      journal = {\mnras},
         year = 2025,
        month = jul,
       volume = {541},
       number = {4},
        pages = {3606-3626},
          doi = {10.1093/mnras/staf1182},
archivePrefix = {arXiv},
       eprint = {2502.12538},
 primaryClass = {astro-ph.GA},
       adsurl = {https://ui.adsabs.harvard.edu/abs/2025MNRAS.541.3606T}
}

@ARTICLE{Trinca2024,
       author = {{Trinca}, Alessandro and {Schneider}, Raffaella and {Valiante}, Rosa and {Graziani}, Luca and {Ferrotti}, Arianna and {Omukai}, Kazuyuki and {Chon}, Sunmyon},
        title = "{Exploring the nature of UV-bright z {\ensuremath{\gtrsim}} 10 galaxies detected by JWST: star formation, black hole accretion, or a non-universal IMF?}",
      journal = {\mnras},
         year = 2024,
        month = apr,
       volume = {529},
       number = {4},
        pages = {3563-3581},
          doi = {10.1093/mnras/stae651},
archivePrefix = {arXiv},
       eprint = {2305.04944},
 primaryClass = {astro-ph.GA},
       adsurl = {https://ui.adsabs.harvard.edu/abs/2024MNRAS.529.3563T}
}

@ARTICLE{Ziparo2023,
       author = {{Ziparo}, Francesco and {Ferrara}, Andrea and {Sommovigo}, Laura and {Kohandel}, Mahsa},
        title = "{Blue monsters. Why are JWST super-early, massive galaxies so blue?}",
      journal = {\mnras},
         year = 2023,
        month = apr,
       volume = {520},
       number = {2},
        pages = {2445-2450},
          doi = {10.1093/mnras/stad125},
archivePrefix = {arXiv},
       eprint = {2209.06840},
 primaryClass = {astro-ph.GA},
       adsurl = {https://ui.adsabs.harvard.edu/abs/2023MNRAS.520.2445Z}
}

@ARTICLE{Abdurrouf2025,
       author = {{Abdurro'uf} and {Coe}, Dan and {Resseguier}, Tom and {Murphy}, Calla and {Xu}, Xinfeng and {Adamo}, Angela and {Roy}, Namrata and {Henry}, Alaina and {Kokorev}, Vasily and {Brammer}, Gabriel and {Fujimoto}, Seiji and {Ferguson}, Henry C. and {Pagul}, Amanda and {Windhorst}, Rogier A. and {Heckman}, Timothy and {Diego}, Jose M. and {Akins}, Hollis B. and {Allingham}, Joseph and {Amor{\'\i}n}, Ricardo O. and {Berg}, Danielle A. and {Brada{\v{c}}}, Maru{\v{s}}a and {Bradley}, Larry D. and {Chen}, Wenlei and {Chisholm}, John and {Conselice}, Christopher J. and {Dayal}, Pratika and {Dessauges-Zavadsky}, Miroslava and {Faisst}, Andreas L. and {Finkelstein}, Steven L. and {Fudamoto}, Yoshinobu and {Furtak}, Lukas J. and {Harikane}, Yuichi and {Hsiao}, Tiger Yu-Yang and {Jimenez-Teja}, Yolanda and {Koekemoer}, Anton M. and {Larson}, Rebecca L. and {Lucas}, Ray A. and {Messa}, Matteo and {Mowla}, Lamiya and {Nakane}, Minami and {Noirot}, Ga{\"e}l and {Pan}, Richard and {Pascale}, Massimo and {Richard}, Johan and {Ricotti}, Massimo and {Robbins}, Luke and {Schaerer}, Daniel and {Sun}, Fengwu and {Vanzella}, Eros and {Welch}, Brian and {Willott}, Chris and {Zitrin}, Adi},
        title = "{Spatially Resolved Physical Properties of Young Star Clusters and Star-forming Clumps in the Brightest z>6 Galaxy, the Strongly Lensed Cosmic Spear at z=6.2}",
      journal = {arXiv e-prints},
         year = 2025,
        month = dec,
          eid = {arXiv:2512.08054},
        pages = {arXiv:2512.08054},
          doi = {10.48550/arXiv.2512.08054},
archivePrefix = {arXiv},
       eprint = {2512.08054},
 primaryClass = {astro-ph.GA},
       adsurl = {https://ui.adsabs.harvard.edu/abs/2025arXiv251208054A}
}

@ARTICLE{Bradac2025,
       author = {{Brada{\v{c}}}, Maru{\v{s}}a and {Jude{\v{z}}}, Jon and {Willott}, Chris and {Rihtar{\v{s}}ic}, Gregor and {Martis}, Nicholas S. and {Harshan}, Anishya and {Felicioni}, Giordano and {Asada}, Yoshihisa and {Desprez}, Guillaume and {Clowe}, Douglas and {Gonzalez}, Anthony H. and {Jones}, Christine and {Lemaux}, Brian C. and {Markevitch}, Maxim and {Markov}, Vladan and {Mowla}, Lamiya and {Noirot}, Ga{\"e}l and {Peter}, Annika H.~G. and {Robertson}, Andrew and {Sarrouh}, Ghassan T.~E. and {Sawicki}, Marcin and {Schrabback}, Tim and {Tripodi}, Roberta},
        title = "{Star Formation under a Cosmic Microscope: Highly Magnified z = 11 Galaxy behind the Bullet Cluster}",
      journal = {\apjl},
         year = 2025,
        month = dec,
       volume = {995},
       number = {2},
          eid = {L74},
        pages = {L74},
          doi = {10.3847/2041-8213/ae27d2},
archivePrefix = {arXiv},
       eprint = {2509.20446},
 primaryClass = {astro-ph.GA},
       adsurl = {https://ui.adsabs.harvard.edu/abs/2025ApJ...995L..74B}
}

@ARTICLE{Bunker2023,
       author = {{Bunker}, Andrew J. and {Saxena}, Aayush and {Cameron}, Alex J. and {Willott}, Chris J. and {Curtis-Lake}, Emma and {Jakobsen}, Peter and {Carniani}, Stefano and {Smit}, Renske and {Maiolino}, Roberto and {Witstok}, Joris and {Curti}, Mirko and {D'Eugenio}, Francesco and {Jones}, Gareth C. and {Ferruit}, Pierre and {Arribas}, Santiago and {Charlot}, Stephane and {Chevallard}, Jacopo and {Giardino}, Giovanna and {de Graaff}, Anna and {Looser}, Tobias J. and {L{\"u}tzgendorf}, Nora and {Maseda}, Michael V. and {Rawle}, Tim and {Rix}, Hans-Walter and {Del Pino}, Bruno Rodr{\'\i}guez and {Alberts}, Stacey and {Egami}, Eiichi and {Eisenstein}, Daniel J. and {Endsley}, Ryan and {Hainline}, Kevin and {Hausen}, Ryan and {Johnson}, Benjamin D. and {Rieke}, George and {Rieke}, Marcia and {Robertson}, Brant E. and {Shivaei}, Irene and {Stark}, Daniel P. and {Sun}, Fengwu and {Tacchella}, Sandro and {Tang}, Mengtao and {Williams}, Christina C. and {Willmer}, Christopher N.~A. and {Baker}, William M. and {Baum}, Stefi and {Bhatawdekar}, Rachana and {Bowler}, Rebecca and {Boyett}, Kristan and {Chen}, Zuyi and {Circosta}, Chiara and {Helton}, Jakob M. and {Ji}, Zhiyuan and {Kumari}, Nimisha and {Lyu}, Jianwei and {Nelson}, Erica and {Parlanti}, Eleonora and {Perna}, Michele and {Sandles}, Lester and {Scholtz}, Jan and {Suess}, Katherine A. and {Topping}, Michael W. and {{\"U}bler}, Hannah and {Wallace}, Imaan E.~B. and {Whitler}, Lily},
        title = "{JADES NIRSpec Spectroscopy of GN-z11: Lyman-{\ensuremath{\alpha}} emission and possible enhanced nitrogen abundance in a z = 10.60 luminous galaxy}",
      journal = {\aap},
         year = 2023,
        month = sep,
       volume = {677},
          eid = {A88},
        pages = {A88},
          doi = {10.1051/0004-6361/202346159},
archivePrefix = {arXiv},
       eprint = {2302.07256},
 primaryClass = {astro-ph.GA},
       adsurl = {https://ui.adsabs.harvard.edu/abs/2023A&A...677A..88B}
}

@ARTICLE{Cameron2024,
       author = {{Cameron}, Alex J. and {Katz}, Harley and {Witten}, Callum and {Saxena}, Aayush and {Laporte}, Nicolas and {Bunker}, Andrew J.},
        title = "{Nebular dominated galaxies: insights into the stellar initial mass function at high redshift}",
      journal = {\mnras},
         year = 2024,
        month = oct,
       volume = {534},
       number = {1},
        pages = {523-543},
          doi = {10.1093/mnras/stae1547},
archivePrefix = {arXiv},
       eprint = {2311.02051},
 primaryClass = {astro-ph.GA},
       adsurl = {https://ui.adsabs.harvard.edu/abs/2024MNRAS.534..523C}
}

@ARTICLE{Cameron2026,
       author = {{Cameron}, Alex J. and {Carreira}, Courtney and {Simmonds}, Charlotte and {Bunker}, Andrew J. and {Saxena}, Aayush and {Carniani}, Stefano and {Charlot}, St{\'e}phane and {Chevallard}, Jacopo and {Curtis-Lake}, Emma and {Hainline}, Kevin and {Hausen}, Ryan and {Ji}, Xihan and {Ji}, Zhiyuan and {Johnson}, Benjamin D. and {Rinaldi}, Pierluigi and {Robertson}, Brant and {Scholtz}, Jan and {Silcock}, Maddie S. and {Tacchella}, Sandro and {Trussler}, James A.~A. and {{\"U}bler}, Hannah and {Williams}, Christina C. and {Willmer}, Christopher N.~A. and {Willott}, Chris and {Witstok}, Joris},
        title = "{JADES: Evolution of nitrogen abundances in star-forming galaxies from z \raisebox{-0.5ex}\textasciitilde 1.5-7}",
      journal = {arXiv e-prints},
         year = 2026,
        month = jan,
          eid = {arXiv:2601.15964},
        pages = {arXiv:2601.15964},
          doi = {10.48550/arXiv.2601.15964},
archivePrefix = {arXiv},
       eprint = {2601.15964},
 primaryClass = {astro-ph.GA},
       adsurl = {https://ui.adsabs.harvard.edu/abs/2026arXiv260115964C}
}

@ARTICLE{Cullen2025,
       author = {{Cullen}, F. and {Carnall}, A.~C. and {Scholte}, D. and {McLeod}, D.~J. and {McLure}, R.~J. and {Arellano-C{\'o}rdova}, K.~Z. and {Stanton}, T.~M. and {Donnan}, C.~T. and {Dunlop}, J.~S. and {Shapley}, A.~E. and {Barrufet}, L. and {Begley}, R. and {Bondestam}, C. and {Cirasuolo}, M. and {Leung}, H.-H. and {Pollock}, C.~L. and {Stevenson}, S.},
        title = "{The JWST EXCELS survey: an extremely metal-poor galaxy at z = 8.271 hosting an unusual population of massive stars}",
      journal = {\mnras},
         year = 2025,
        month = jul,
       volume = {540},
       number = {3},
        pages = {2176-2194},
          doi = {10.1093/mnras/staf838},
archivePrefix = {arXiv},
       eprint = {2501.11099},
 primaryClass = {astro-ph.GA},
       adsurl = {https://ui.adsabs.harvard.edu/abs/2025MNRAS.540.2176C}
}

@ARTICLE{Labbe2023,
       author = {{Labb{\'e}}, Ivo and {van Dokkum}, Pieter and {Nelson}, Erica and {Bezanson}, Rachel and {Suess}, Katherine A. and {Leja}, Joel and {Brammer}, Gabriel and {Whitaker}, Katherine and {Mathews}, Elijah and {Stefanon}, Mauro and {Wang}, Bingjie},
        title = "{A population of red candidate massive galaxies  600 Myr after the Big Bang}",
      journal = {\nat},
         year = 2023,
        month = apr,
       volume = {616},
       number = {7956},
        pages = {266-269},
          doi = {10.1038/s41586-023-05786-2},
archivePrefix = {arXiv},
       eprint = {2207.12446},
 primaryClass = {astro-ph.GA},
       adsurl = {https://ui.adsabs.harvard.edu/abs/2023Natur.616..266L}
}

@ARTICLE{Mowla2024,
       author = {{Mowla}, Lamiya and {Iyer}, Kartheik and {Asada}, Yoshihisa and {Desprez}, Guillaume and {Tan}, Vivian Yun Yan and {Martis}, Nicholas and {Sarrouh}, Ghassan and {Strait}, Victoria and {Abraham}, Roberto and {Brada{\v{c}}}, Maru{\v{s}}a and {Brammer}, Gabriel and {Muzzin}, Adam and {Pacifici}, Camilla and {Ravindranath}, Swara and {Sawicki}, Marcin and {Willott}, Chris and {Estrada-Carpenter}, Vince and {Jahan}, Nusrath and {Noirot}, Ga{\"e}l and {Matharu}, Jasleen and {Rihtar{\v{s}}i{\v{c}}}, Gregor and {Zabl}, Johannes},
        title = "{Formation of a low-mass galaxy from star clusters in a 600-million-year-old Universe}",
      journal = {\nat},
         year = 2024,
        month = dec,
       volume = {636},
       number = {8042},
        pages = {332-336},
          doi = {10.1038/s41586-024-08293-0},
archivePrefix = {arXiv},
       eprint = {2402.08696},
 primaryClass = {astro-ph.GA},
       adsurl = {https://ui.adsabs.harvard.edu/abs/2024Natur.636..332M}
}

@ARTICLE{Nakane2025,
       author = {{Nakane}, Minami and {Kokorev}, Vasily and {Fujimoto}, Seiji and {Ouchi}, Masami and {McLeod}, Derek J. and {Golubchik}, Miriam and {Oguri}, Masamune and {Zitrin}, Adi and {Bondestam}, Cecilia and {Donnan}, Callum T. and {Brammer}, Gabriel and {Finkelstein}, Steven L. and {Willott}, Chris and {Rihtarsic}, Gregor and {Desprez}, Guillaume and {Adamo}, Angela and {Vanzella}, Eros and {Brada{\v{c}}}, Maru{\v{s}}a and {Messa}, Matteo and {Yanagisawa}, Hiroto and {Sun}, Fengwu and {Ferguson}, Henry C. and {Lucas}, Ray A. and {Coe}, Dan and {Richard}, Johan and {Abdurro'uf} and {Akins}, Hollis B. and {Allingham}, Joseph F.~V. and {Amor{\'\i}n}, Ricardo O. and {Asada}, Yoshihisa and {Atek}, Hakim and {Bezanson}, Rachel and {Bradley}, Larry D. and {Chisholm}, John and {Conselice}, Christopher J. and {Dayal}, Pratika and {Dessauges-Zavadsky}, Miroslava and {Diego}, Jose M. and {Faisst}, Andreas L. and {Fei}, Qinyue and {Frye}, Brenda L. and {Fudamoto}, Yoshinobu and {Furtak}, Lukas J. and {Harikane}, Yuichi and {Hsiao}, Tiger Yu-Yang and {Jim{\'e}nez-Teja}, Yolanda and {Kartaltepe}, Jeyhan S. and {Kiyota}, Tomokazu and {Koekemoer}, Anton M. and {Lagos}, Claudia del P. and {Magdis}, Georgios E. and {Meena}, Ashish Kumar and {Mowla}, Lamiya and {Noirot}, Ga{\"e}l and {Oesch}, Pascal A. and {Ono}, Yoshiaki and {Ortiz}, III, Rafael and {Pan}, Richard and {Papovich}, Casey and {Pierel}, Justin D. and {Ricotti}, Massimo and {Robbins}, Luke and {Schaerer}, Daniel and {Schneider}, Raffaella and {Treu}, Tommaso and {Valentino}, Francesco and {Windhorst}, Rogier A. and {Bauer}, Franz E. and {Bromm}, Volker and {Egami}, Eiichi and {Gonz{\'a}lez-Otero}, Mauro and {Kohno}, Kotaro and {Labbe}, Ivo and {Matthee}, Jorryt and {Mun}, Marcie and {Naidu}, Rohan P. and {Tripodi}, Roberta},
        title = "{VENUS: A Strongly Lensed Clumpy Galaxy at $z\sim11-12$ behind the Galaxy Cluster MACS J0257.1-2325}",
      journal = {arXiv e-prints},
         year = 2025,
        month = nov,
          eid = {arXiv:2511.14483},
        pages = {arXiv:2511.14483},
          doi = {10.48550/arXiv.2511.14483},
archivePrefix = {arXiv},
       eprint = {2511.14483},
 primaryClass = {astro-ph.GA},
       adsurl = {https://ui.adsabs.harvard.edu/abs/2025arXiv251114483N}
}

@ARTICLE{Vanzella2023a,
       author = {{Vanzella}, Eros and {Claeyssens}, Ad{\'e}la{\"\i}de and {Welch}, Brian and {Adamo}, Angela and {Coe}, Dan and {Diego}, Jose M. and {Mahler}, Guillaume and {Khullar}, Gourav and {Kokorev}, Vasily and {Oguri}, Masamune and {Ravindranath}, Swara and {Furtak}, Lukas J. and {Hsiao}, Tiger Yu-Yang and {Abdurro'uf} and {Mandelker}, Nir and {Brammer}, Gabriel and {Bradley}, Larry D. and {Brada{\v{c}}}, Maru{\v{s}}a and {Conselice}, Christopher J. and {Dayal}, Pratika and {Nonino}, Mario and {Andrade-Santos}, Felipe and {Windhorst}, Rogier A. and {Pirzkal}, Nor and {Sharon}, Keren and {de Mink}, S.~E. and {Fujimoto}, Seiji and {Zitrin}, Adi and {Eldridge}, Jan J. and {Norman}, Colin},
        title = "{JWST/NIRCam Probes Young Star Clusters in the Reionization Era Sunrise Arc}",
      journal = {\apj},
         year = 2023,
        month = mar,
       volume = {945},
       number = {1},
          eid = {53},
        pages = {53},
          doi = {10.3847/1538-4357/acb59a},
archivePrefix = {arXiv},
       eprint = {2211.09839},
 primaryClass = {astro-ph.GA},
       adsurl = {https://ui.adsabs.harvard.edu/abs/2023ApJ...945...53V}
}

@ARTICLE{ArrabalHaro2023,
       author = {{Arrabal Haro}, Pablo and {Dickinson}, Mark and {Finkelstein}, Steven L. and {Kartaltepe}, Jeyhan S. and {Donnan}, Callum T. and {Burgarella}, Denis and {Carnall}, Adam C. and {Cullen}, Fergus and {Dunlop}, James S. and {Fern{\'a}ndez}, Vital and {Fujimoto}, Seiji and {Jung}, Intae and {Krips}, Melanie and {Larson}, Rebecca L. and {Papovich}, Casey and {P{\'e}rez-Gonz{\'a}lez}, Pablo G. and {Amor{\'\i}n}, Ricardo O. and {Bagley}, Micaela B. and {Buat}, V{\'e}ronique and {Casey}, Caitlin M. and {Chworowsky}, Katherine and {Cohen}, Seth H. and {Ferguson}, Henry C. and {Giavalisco}, Mauro and {Huertas-Company}, Marc and {Hutchison}, Taylor A. and {Kocevski}, Dale D. and {Koekemoer}, Anton M. and {Lucas}, Ray A. and {McLeod}, Derek J. and {McLure}, Ross J. and {Pirzkal}, Norbert and {Seill{\'e}}, Lise-Marie and {Trump}, Jonathan R. and {Weiner}, Benjamin J. and {Wilkins}, Stephen M. and {Zavala}, Jorge A.},
        title = "{Confirmation and refutation of very luminous galaxies in the early Universe}",
      journal = {\nat},
         year = 2023,
        month = oct,
       volume = {622},
       number = {7984},
        pages = {707-711},
          doi = {10.1038/s41586-023-06521-7},
archivePrefix = {arXiv},
       eprint = {2303.15431},
 primaryClass = {astro-ph.GA},
       adsurl = {https://ui.adsabs.harvard.edu/abs/2023Natur.622..707A}
}

@ARTICLE{Bouwens2023b,
       author = {{Bouwens}, Rychard J. and {Stefanon}, Mauro and {Brammer}, Gabriel and {Oesch}, Pascal A. and {Herard-Demanche}, Thomas and {Illingworth}, Garth D. and {Matthee}, Jorryt and {Naidu}, Rohan P. and {van Dokkum}, Pieter G. and {van Leeuwen}, Ivana F.},
        title = "{Evolution of the UV LF from z   15 to z   8 using new JWST NIRCam medium-band observations over the HUDF/XDF}",
      journal = {\mnras},
         year = 2023,
        month = jul,
       volume = {523},
       number = {1},
        pages = {1036-1055},
          doi = {10.1093/mnras/stad1145},
archivePrefix = {arXiv},
       eprint = {2211.02607},
 primaryClass = {astro-ph.GA},
       adsurl = {https://ui.adsabs.harvard.edu/abs/2023MNRAS.523.1036B}
}

@ARTICLE{Donnan2024,
       author = {{Donnan}, C.~T. and {McLure}, R.~J. and {Dunlop}, J.~S. and {McLeod}, D.~J. and {Magee}, D. and {Arellano-C{\'o}rdova}, K.~Z. and {Barrufet}, L. and {Begley}, R. and {Bowler}, R.~A.~A. and {Carnall}, A.~C. and {Cullen}, F. and {Ellis}, R.~S. and {Fontana}, A. and {Illingworth}, G.~D. and {Grogin}, N.~A. and {Hamadouche}, M.~L. and {Koekemoer}, A.~M. and {Liu}, F. -Y. and {Mason}, C. and {Santini}, P. and {Stanton}, T.~M.},
        title = "{JWST PRIMER: A new multi-field determination of the evolving galaxy UV luminosity function at redshifts $\mathbf{z \simeq 9-15}$}",
      journal = {arXiv e-prints},
         year = 2024,
        month = mar,
          eid = {arXiv:2403.03171},
        pages = {arXiv:2403.03171},
          doi = {10.48550/arXiv.2403.03171},
archivePrefix = {arXiv},
       eprint = {2403.03171},
 primaryClass = {astro-ph.GA},
       adsurl = {https://ui.adsabs.harvard.edu/abs/2024arXiv240303171D}
}

@ARTICLE{Finkelstein2023,
       author = {{Finkelstein}, Steven L. and {Bagley}, Micaela B. and {Ferguson}, Henry C. and {Wilkins}, Stephen M. and {Kartaltepe}, Jeyhan S. and {Papovich}, Casey and {Yung}, L.~Y. Aaron and {Arrabal Haro}, Pablo and {Behroozi}, Peter and {Dickinson}, Mark and {Kocevski}, Dale D. and {Koekemoer}, Anton M. and {Larson}, Rebecca L. and {Le Bail}, Aur{\'e}lien and {Morales}, Alexa M. and {P{\'e}rez-Gonz{\'a}lez}, Pablo G. and {Burgarella}, Denis and {Dav{\'e}}, Romeel and {Hirschmann}, Michaela and {Somerville}, Rachel S. and {Wuyts}, Stijn and {Bromm}, Volker and {Casey}, Caitlin M. and {Fontana}, Adriano and {Fujimoto}, Seiji and {Gardner}, Jonathan P. and {Giavalisco}, Mauro and {Grazian}, Andrea and {Grogin}, Norman A. and {Hathi}, Nimish P. and {Hutchison}, Taylor A. and {Jha}, Saurabh W. and {Jogee}, Shardha and {Kewley}, Lisa J. and {Kirkpatrick}, Allison and {Long}, Arianna S. and {Lotz}, Jennifer M. and {Pentericci}, Laura and {Pierel}, Justin D.~R. and {Pirzkal}, Nor and {Ravindranath}, Swara and {Ryan}, Russell E. and {Trump}, Jonathan R. and {Yang}, Guang and {Bhatawdekar}, Rachana and {Bisigello}, Laura and {Buat}, V{\'e}ronique and {Calabr{\`o}}, Antonello and {Castellano}, Marco and {Cleri}, Nikko J. and {Cooper}, M.~C. and {Croton}, Darren and {Daddi}, Emanuele and {Dekel}, Avishai and {Elbaz}, David and {Franco}, Maximilien and {Gawiser}, Eric and {Holwerda}, Benne W. and {Huertas-Company}, Marc and {Jaskot}, Anne E. and {Leung}, Gene C.~K. and {Lucas}, Ray A. and {Mobasher}, Bahram and {Pandya}, Viraj and {Tacchella}, Sandro and {Weiner}, Benjamin J. and {Zavala}, Jorge A.},
        title = "{CEERS Key Paper. I. An Early Look into the First 500 Myr of Galaxy Formation with JWST}",
      journal = {\apjl},
         year = 2023,
        month = mar,
       volume = {946},
       number = {1},
          eid = {L13},
        pages = {L13},
          doi = {10.3847/2041-8213/acade4},
archivePrefix = {arXiv},
       eprint = {2211.05792},
 primaryClass = {astro-ph.GA},
       adsurl = {https://ui.adsabs.harvard.edu/abs/2023ApJ...946L..13F}
}

@ARTICLE{Harikane2025,
       author = {{Harikane}, Yuichi and {Inoue}, Akio K. and {Ellis}, Richard S. and {Ouchi}, Masami and {Nakazato}, Yurina and {Yoshida}, Naoki and {Ono}, Yoshiaki and {Sun}, Fengwu and {Sato}, Riku A. and {Ferrami}, Giovanni and {Fujimoto}, Seiji and {Kashikawa}, Nobunari and {McLeod}, Derek J. and {P{\'e}rez-Gonz{\'a}lez}, Pablo G. and {Sawicki}, Marcin and {Sugahara}, Yuma and {Xu}, Yi and {Yamanaka}, Satoshi and {Carnall}, Adam C. and {Cullen}, Fergus and {Dunlop}, James S. and {Egami}, Eiichi and {Grogin}, Norman and {Isobe}, Yuki and {Koekemoer}, Anton M. and {Laporte}, Nicolas and {Lee}, Chien-Hsiu and {Magee}, Dan and {Matsuo}, Hiroshi and {Matsuoka}, Yoshiki and {Mawatari}, Ken and {Nakajima}, Kimihiko and {Nakane}, Minami and {Tamura}, Yoichi and {Umeda}, Hiroya and {Yanagisawa}, Hiroto},
        title = "{JWST, ALMA, and Keck Spectroscopic Constraints on the UV Luminosity Functions at z {\ensuremath{\sim}} 7─14: Clumpiness and Compactness of the Brightest Galaxies in the Early Universe}",
      journal = {\apj},
         year = 2025,
        month = feb,
       volume = {980},
       number = {1},
          eid = {138},
        pages = {138},
          doi = {10.3847/1538-4357/ad9b2c},
archivePrefix = {arXiv},
       eprint = {2406.18352},
 primaryClass = {astro-ph.GA},
       adsurl = {https://ui.adsabs.harvard.edu/abs/2025ApJ...980..138H}
}

@ARTICLE{Livermore2017,
       author = {{Livermore}, R.~C. and {Finkelstein}, S.~L. and {Lotz}, J.~M.},
        title = "{Directly Observing the Galaxies Likely Responsible for Reionization}",
      journal = {\apj},
         year = 2017,
        month = feb,
       volume = {835},
       number = {2},
          eid = {113},
        pages = {113},
          doi = {10.3847/1538-4357/835/2/113},
archivePrefix = {arXiv},
       eprint = {1604.06799},
 primaryClass = {astro-ph.GA},
       adsurl = {https://ui.adsabs.harvard.edu/abs/2017ApJ...835..113L}
}

@ARTICLE{McLeod2026,
       author = {{McLeod}, D.~J. and {Dunlop}, J.~S. and {McLure}, R.~J. and {Donnan}, C.~T. and {Begley}, R. and {Antonogiannaki}, S. and {Magee}, D. and {Illingworth}, G.~D. and {Arrabal Haro}, P. and {Bondestam}, C. and {Carnall}, A.~C. and {Cullen}, F. and {Dickinson}, M. and {Ellis}, R.~S. and {Frye}, B.~L. and {Golawska}, H. and {Grogin}, N.~A. and {Holst}, I.~J.~B. and {Kamieneski}, P.~S. and {Leung}, H.-H. and {Liu}, F.-Y. and {Stanton}, T.~M. and {Tittley}, E.~R.},
        title = "{A search for the first galaxies across $>0.6$ deg$^2$ of JWST imaging: new evidence for a rapid decline in star-formation activity at $z>12$}",
      journal = {arXiv e-prints},
         year = 2026,
        month = apr,
          eid = {arXiv:2604.16666},
        pages = {arXiv:2604.16666},
          doi = {10.48550/arXiv.2604.16666},
archivePrefix = {arXiv},
       eprint = {2604.16666},
 primaryClass = {astro-ph.GA},
       adsurl = {https://ui.adsabs.harvard.edu/abs/2026arXiv260416666M}
}

@ARTICLE{PerezGonzalez2023,
       author = {{P{\'e}rez-Gonz{\'a}lez}, Pablo G. and {Costantin}, Luca and {Langeroodi}, Danial and {Rinaldi}, Pierluigi and {Annunziatella}, Marianna and {Ilbert}, Olivier and {Colina}, Luis and {N{\o}rgaard-Nielsen}, Hans Ulrik and {Greve}, Thomas R. and {{\"O}stlin}, G{\"o}ran and {Wright}, Gillian and {Alonso-Herrero}, Almudena and {{\'A}lvarez-M{\'a}rquez}, Javier and {Caputi}, Karina I. and {Eckart}, Andreas and {Le F{\`e}vre}, Olivier and {Labiano}, {\'A}lvaro and {Garc{\'\i}a-Mar{\'\i}n}, Macarena and {Hjorth}, Jens and {Kendrew}, Sarah and {Pye}, John P. and {Tikkanen}, Tuomo and {van der Werf}, Paul and {Walter}, Fabian and {Ward}, Martin and {Bik}, Arjan and {Boogaard}, Leindert and {Bosman}, Sarah E.~I. and {G{\'o}mez}, Alejandro Crespo and {Gillman}, Steven and {Iani}, Edoardo and {Jermann}, Iris and {Melinder}, Jens and {Meyer}, Romain A. and {Moutard}, Thibaud and {van Dishoek}, Ewine and {Henning}, Thomas and {Lagage}, Pierre-Olivier and {Guedel}, Manuel and {Peissker}, Florian and {Ray}, Tom and {Vandenbussche}, Bart and {Garc{\'\i}a-Argum{\'a}nez}, {\'A}ngela and {Mar{\'\i}a M{\'e}rida}, Rosa},
        title = "{Life beyond 30: Probing the -20 < M $_{UV}$ < -17 Luminosity Function at 8 < z < 13 with the NIRCam Parallel Field of the MIRI Deep Survey}",
      journal = {\apjl},
         year = 2023,
        month = jul,
       volume = {951},
       number = {1},
          eid = {L1},
        pages = {L1},
          doi = {10.3847/2041-8213/acd9d0},
archivePrefix = {arXiv},
       eprint = {2302.02429},
 primaryClass = {astro-ph.GA},
       adsurl = {https://ui.adsabs.harvard.edu/abs/2023ApJ...951L...1P}
}

@ARTICLE{Whitler2025,
       author = {{Whitler}, Lily and {Stark}, Daniel P. and {Topping}, Michael W. and {Robertson}, Brant and {Rieke}, Marcia and {Hainline}, Kevin N. and {Endsley}, Ryan and {Chen}, Zuyi and {Baker}, William M. and {Bhatawdekar}, Rachana and {Bunker}, Andrew J. and {Carniani}, Stefano and {Charlot}, St{\'e}phane and {Chevallard}, Jacopo and {Curtis-Lake}, Emma and {Egami}, Eiichi and {Eisenstein}, Daniel J. and {Helton}, Jakob M. and {Ji}, Zhiyuan and {Johnson}, Benjamin D. and {P{\'e}rez-Gonz{\'a}lez}, Pablo G. and {Rinaldi}, Pierluigi and {Tacchella}, Sandro and {Williams}, Christina C. and {Willmer}, Christopher N.~A. and {Willott}, Chris and {Witstok}, Joris},
        title = "{The z {\ensuremath{\gtrsim}} 9 Galaxy UV Luminosity Function from the JWST Advanced Deep Extragalactic Survey: Insights into Early Galaxy Evolution and Reionization}",
      journal = {\apj},
         year = 2025,
        month = oct,
       volume = {992},
       number = {1},
          eid = {63},
        pages = {63},
          doi = {10.3847/1538-4357/adfddc},
archivePrefix = {arXiv},
       eprint = {2501.00984},
 primaryClass = {astro-ph.GA},
       adsurl = {https://ui.adsabs.harvard.edu/abs/2025ApJ...992...63W}
}

% Alternatively you could enter them by hand, like this:
% This method is tedious and prone to error if you have lots of references
%\begin{thebibliography}{99}
%\bibitem[\protect\citeauthoryear{Author}{2012}]{Author2012}
%Author A.~N., 2013, Journal of Improbable Astronomy, 1, 1
%\bibitem[\protect\citeauthoryear{Others}{2013}]{Others2013}
%Others S., 2012, Journal of Interesting Stuff, 17, 198
%\end{thebibliography}

%%%%%%%%%%%%%%%%%%%%%%%%%%%%%%%%%%%%%%%%%%%%%%%%%%

%%%%%%%%%%%%%%%%% APPENDICES %%%%%%%%%%%%%%%%%%%%%

\appendix

\section{Time evolution of the specific UV luminosity and ionisation production rate} \label{app:SSP_lum_fits}

The fitted specific UV luminosity per unit stellar mass formed is
{\footnotesize
\begin{align}\label{eq:luv_instant}
\log_{10}&\frac{L_{\rm UV}}{\rm erg,s^{-1}Å^{-1}} = 0.777\log_{10}\1 \mlim - 6.65\2 + 31.29 \\\notag
&+ \gamma_N\3\1\mathrm{max}\3f_{\rm massive},f_{\rm min}\4 + \delta_N\2^{\beta_N}
- \1f_{\rm min} + \delta_N\2^{\beta_N}\4\\\notag
&+\gamma_1\1\mathrm{min}\3\mathrm{max}\3\tau,\tau_0\4,\tau_1\4-\tau_0\2\\\notag
&+\gamma_2\1\mathrm{max}\3\tau,\tau_1\4-\tau_1\2 ,
\end{align}
}
with
{\footnotesize
\begin{align}
\delta_N &= 0.002\sqrt{\mlim -10},\quad
\beta_N = 18 \1\mlim\2^{-0.778},\\
\gamma_N &= 1 + 0.035\1\sqrt{\mlim}-5\2,\\
\tau_0 &= -0.276\log_{10}\1\mlim-6.49\2+9.83,\\
\gamma_1 &=0.289\log_{10}\1\mlim+8.66\2-0.118,\\
\gamma_2 &= -1.25
- \mathrm{max}\3f_{\rm massive}-f_{\rm min},0\4^{1.31}
\11-0.016\3\sqrt{\mlim}+1.4\4\2,\\
\tau_1 &= \log_{10}\1377\1\mlim\2^{-1.2}+1.3\2+6,\quad
\tau = t/{\rm yr},
\end{align}
}
and
{\footnotesize
\begin{align}
f_{\rm min} =
\frac{\1100-\mlim\2\1\mlim\2^{-1.35}}
{\1100-\mlim\2\1\mlim\2^{-1.35}
-\dfrac{\1\mlim\2^{-0.35}-0.1^{-0.35}}{0.35}}.
\end{align}
}
Here, $\mlim \equiv m_{\max}/\msun$ is the maximum stellar mass in solar‑mass units, $\tau$ is the SSP age in years, and $f_{\rm massive}$ is the IMF mass fraction in the high‑mass log‑flat segment. The quantity $f_{\rm min}$ is the value of $f_{\rm massive}$ at $m_c=\mlim$, corresponding to the limit in which the log-flat segment vanishes. Equation~\eqref{eq:luv_instant} reduces to the cloud-mass-limited Salpeter IMF calibration for $f_{\rm massive}=0$ and to the global Evolving/Salpeter IMF calibration for $\mlim=100\,\msun$.
Applying the same procedure to the ionising photon production rate yields
{\footnotesize
\begin{align}\label{eq:nion_instant}
\log_{10}&\frac{\dot{Q}}{\rm s^{-1}}
= 1.889\log_{10}\1\mlim-8.99\2+43.01 \\\notag
&+\1\frac{\mathrm{max}\3f_{\rm massive}-f_{\rm min},0\4}
{1-f_{\rm min}}\2^\delta
\11+0.025\sqrt{\mlim}\2\\\notag
&+\gamma\1\mathrm{max}\3\tau,\tau_0\4-\tau_0\2,
\end{align}
}
with
{\footnotesize
\begin{align}
\delta &= 2.4-\log_{10}\mlim,\\
\gamma &= -3.9
- \1\mathrm{min}\3f_{\rm massive},0.905\4^{\beta}
- f_{\rm min}^{\beta}\2\alpha,\\
\beta &= 7-2.75\log_{10}\1\mlim+5\2,\\
\alpha &= 1-0.07\log_{10}\1\mlim\2,\\
\tau_0 &= \log_{10}\1 1013\1\mlim\2^{-1.8}+2.26\2+6.
\end{align}
}

\section{Cloud Mass Limited Initial Mass Functions}\label{app:CMLIMFs}
% -------------------------------------------------------------------------
\subsection{Cloud mass limited Salpeter IMF}\label{app:salpeterCloudIMF}
% -------------------------------------------------------------------------

We first find the expected number of stars per unit stellar mass formed. We normalise the IMF, and find the mean stellar mass. The mean number of stars per unit stellar mass is then the inverse of this:

\begin{align}\label{eq:dndmSal}
    \frac{\intd N}{\intd m} &= cm^{\alpha}\\
    N=1 &= \int_{\mstarmin}^{\mstarmax} cm^\alpha\intd m\\
    &= c\dfrac{1}{\alpha +1}\1{\mstarmax}^{\alpha+1}-{\mstarmin}^{\alpha+1}\2\\
    \implies c&= \dfrac{\alpha+1}{{\mstarmax}^{\alpha+1}-{\mstarmin}^{\alpha+1}}\label{eq:cSal}\\
    \mu_m &= \int_{\mstarmin}^{\mstarmax} mcm^\alpha\intd m\\
    &= \dfrac{\alpha + 1}{\alpha+ 2}\dfrac{{\mstarmax}^{\alpha+2}-{\mstarmin}^{\alpha+2}}{{\mstarmax}^{\alpha+1}-{\mstarmin}^{\alpha+1}}\\
    N_\star &= \dfrac{\alpha + 2}{\alpha + 1}\dfrac{{\mstarmax}^{\alpha + 1}-{\mstarmin}^{\alpha + 1}}{{\mstarmax}^{\alpha + 2}-{\mstarmin}^{\alpha + 2}}
\end{align}
where $\mstarmax$ and $\mstarmin$ are the upper and lower stellar mass limits for the IMF, $\alpha = -2.35$ is the powerlaw exponent for the Salpeter IMF, and $N_\star$ is the mean number of stars per unit stellar mass formed. We then find the expected mass of the most massive star. We do this by finding the expected number of stars with a mass greater than $\mlim$, as a function of the number of stars formed. We set this equal to 1, and isolate $\mlim$ to find the maximum expected stellar mass as a function of the number of stars:
\begin{align}
    N\1m\geq \mlim\2 &= N_{\rm tot}\int_{\mlim}^{\mstarmax} c m^\alpha\intd m\\
    &= N_{\rm tot} \dfrac{\mstarmax^{\alpha+1}-\1\mlim\2^{\alpha+1}}{\mstarmax^{\alpha+1}-\mstarmin^{\alpha+1}}=1\\
    \mlim &= \1\dfrac{\mstarmax^{1+\alpha}(N_{\rm tot}-1) + \mstarmin^{1+\alpha}}{N_{\rm tot}}\2^{1/(1+\alpha)}
\end{align}
where $N_{\rm tot} = N_\star m$ is the total number of stars to be formed. With the new upper mass limit, we must renormalise the IMF, which we do in the same way as in Eqs. \ref{eq:dndmSal}-\ref{eq:cSal}, but integrating over $m\frac{\intd N}{\intd m} $ instead of just $\frac{\intd N}{\intd m} $ to normalise the IMF to 1 solar mass instead of 1 star. We find
\begin{equation}\label{eq:mlimSal}
    \frac{\intd N}{\intd m} =\dfrac{\alpha+2}{ \1\mlim\2^{\alpha+2}-\mstarmin^{\alpha+2}}m^\alpha
\end{equation}

as the new normalised Salpeter IMF with an upper mass limit of $\mlim$.

% -------------------------------------------------------------------------
\subsection{Cloud mass limited Evolving IMF}\label{app:evolcingCloudIMF}
% -------------------------------------------------------------------------

We derive the cloud mass limited evolving IMF in much the same way as above. The full evolving IMF is defined as
\begin{align}\label{eq:evoIMF}
    \frac{\intd N}{\intd m} &=\begin{cases}
        (1-f_{\rm massive})c_1m^\alpha,&\mstarmin<m\leq m_c\\
        f_{\rm massive}c_2m^{-1},&m_c<m\leq \mstarmax
    \end{cases}
\end{align}
with
\begin{align}
    (1-f_{\rm massive})c_1m_c^{\alpha} &= f_{\rm massive}c_2m_c^{-1}\\
    c_2 &= \frac{1-f_{\rm massive}}{f_{\rm massive}}m_c^{\alpha+1}c_1
\end{align}
We normalise the IMF:
{\footnotesize
\begin{align}
    N=1&=c_1(1-f_{\rm massive})\\\notag&\times\1\int_{\mstarmin}^{m_c} m^{\alpha}\intd m + \int_{m_c}^{\mstarmax} m_c^{\alpha+1}m^{-1}\intd m\2\\
    &= c_1(1-f_{\rm massive})\1\dfrac{m_c^{\alpha+1}-\mstarmin^{\alpha+1}}{\alpha+1}+m_c^{\alpha+1}\ln\frac{\mstarmax}{m_c}\2\\
    \imp\,c_1 &= \dfrac{1+\alpha}{1-f_{\rm massive}}\dfrac{m_c^{-(1+\alpha)}}{1-\1\frac{\mstarmin}{m_c}\2^{\alpha+1}+(1+\alpha)\ln\frac{\mstarmax}{m_c}}\label{eq:c1EvoIMF}\\
    c&= \dfrac{(1+\alpha)m_c^{-(1+\alpha)}}{1-\1\frac{\mstarmin}{m_c}\2^{\alpha+1}+(1+\alpha)\ln\frac{\mstarmax}{m_c}}
\end{align}
}
Then we can find $N_\star$ as
\begin{align}
    \mu_m &= c\1\int_{\mstarmin}^{m_c}m^{\alpha+1}\intd m+m_c^{\alpha+1}\int_{m_c}^{\mstarmax}\intd m\2\\
    &= c\1\dfrac{m_c^{\alpha+2}-\mstarmin^{\alpha+2}}{\alpha+1}+m_c^{\alpha+1}(\mstarmax-m_c)\2\\
    N_\star &= \dfrac{1-\1\frac{\mstarmin}{m_c}\2^{\alpha+1}+(1+\alpha)\ln\frac{\mstarmax}{m_c}}{\frac{1+\alpha}{2+\alpha}\11-\1\frac{\mstarmin}{m_c}\2^{\alpha+2}\2+(1+\alpha)(\mstarmax-m_c)}
\end{align}
For the upper mass limit we must consider two different cases, one where $\mlim\geq m_c$, and one where $\mlim<m_c$. For the first case we find:
\begin{align}
    1&=N\1m\geq\mlim\geq m_c\2\\
    &=N_{\rm tot} \int_{\mlim} ^{\mstarmax} cm_c^{\alpha+1}m^{-1}\intd m\\
    &=N_{\rm tot}cm_c^{\alpha+1}\ln\frac{\mstarmax}{\mlim} \\
    \imp \mlim&=\mstarmax \exp\3{\frac{-1}{N_{\rm tot}cm_c^{\alpha+1}}}\4\\
    &= \mstarmax \exp\3{\frac{-1}{N_{\rm tot}}\1\frac{1-\1\frac{\mstarmin}{m_c}\2^{\alpha+1}}{1+\alpha}+\ln\frac{\mstarmax}{m_c}\2}\4\\
    \mlim&= \mstarmax \1\dfrac{\mstarmax}{m_c}\2^{-1/N_{\rm tot}}\exp\3{\dfrac{\1\frac{\mstarmin}{m_c}\2^{\alpha+1}-1}{N_{\rm tot}\1\alpha+1\2}}\4
\end{align}
For the second case, with $\mlim<m_c$, we find:
\begin{align}
    1&=N(m\geq \mlim,m_c>\mlim)\\
    &=N_{\rm tot}c\1\int_{\mlim}^{m_c}m^{\alpha}\intd m+\int_{m_c}^{\mstarmax} m_c^{\alpha+1}m^{-1}\intd m\2\\
    &=N_{\rm tot}cm_c^{\alpha+1}\3\dfrac{1}{\alpha+1}\11-\1\dfrac{\mlim}{m_c}\2^{\alpha+1}\2+\ln\frac{\mstarmax}{m_c}\4
\end{align}
\begin{equation}
    \imp1-\1\dfrac{\mlim}{m_c}\2^{\alpha+1}=\dfrac{1+\alpha}{N_{\rm tot}cm_c^{\alpha+1}}-(\alpha+1)\ln\frac{\mstarmax}{m_c}\\
\end{equation}
{\footnotesize
\begin{align}
    \mlim&=m_c\31-\dfrac{1+\alpha}{N_{\rm tot}cm_c^{\alpha+1}}+(\alpha+1)\ln\frac{\mstarmax}{m_c}\4^{1/(\alpha+1)}\\
    \mlim&=m_c\3\dfrac{(N_{\rm tot}-1)\11+(\alpha+1)\ln\frac{\mstarmax}{m_c}\2-\1\frac{\mstarmin}{m_c}\2^{\alpha+1}}{N_{\rm tot}}\4^{1/(\alpha+1)}
\end{align}
}
This allows us to now renormalise the IMF with the new upper mass limit. For the case where $\mlim<m_c$, we find that the IMF reduces to a Salpeter IMF, and we can simply use Eq. \ref{eq:mlimSal}. For the case where $m_c\leq\mlim$ we can repeat the steps in Eqs. \ref{eq:evoIMF}-\ref{eq:c1EvoIMF}, replacing $\mstarmax$ with $\mlim$, and integrating over $m\frac{\intd N}{\intd m} $ instead of just $\frac{\intd N}{\intd m} $, to normalise the IMF to number of stars per stellar mass formed. Doing this we find
{\small
\begin{align}
    \frac{\intd N}{\intd m} =\begin{cases}
        c_0m^\alpha,&\mstarmin\leq m\leq m_c\\
        c_0m_c^{\alpha+1}m^{-1},&m_c< m\leq \mlim
    \end{cases}\\
    c_0 = \dfrac{(\alpha+2)}{m_c^{\alpha+1}(\mlim-m_c)(\alpha+2)+m_c^{\alpha+2}-\mstarmin^{\alpha+2}}
\end{align}
}
\newcommand{\id}{\mathbf{1}_{\left(N=n\right)}}
\newcommand{\idm}{\mathbf{1}_{\left(N=m\right)}}
\section{Derivation of the variance on number of clouds to reach a given mass}\label{app:variance}
We need to find the variance on the number of clouds $N$, drawn randomly from the cloud mass distribution, in order for the total mass of the clouds to reach $M_{\rm total}=\sum_{i=1}^NM_i$.

Given the cloud mass distribution with powerlaw index $\alpha=-2$, and minimum and maximum cloud masses $\mmin$ and $\mmax$, we can find the variance and expectation value of $\mcl$ as
\begin{align}
    E[\mcl] &= \mu
    = \frac{\alpha+1}{\mmax{}^{\alpha+1}-\mmin{}^{\alpha+1}}\log\1\dfrac{\mmax}{\mmin}\2\\
    Var[\mcl] &= \sigma^2= \mmax\mmin-\mu^2\\
    \intertext{We Can find the expected value of the total mass of $n$ clouds, $E[S_n]$}
    E[S_n] &= E\left[\sum_{i=1}^nM_i\right] = n\mu\\
    %&= \sum_{i=1}^n\left[M_i\right]\\
    %&= \sum_{i=1}^n\mu\\
    %&= n\mu\\
    \intertext{We can similarly find the variance $Var[S_n]$ as}
    Var[S_n] &= Var\left[\sum_{i=1}^nM_i\right]\\
    \intertext{By Bienaymé's identity this is}
    &= \sum_{i=1}^nVar[M_i] = n\sigma^2\\
    %&= \sum_{i=1}^n\sigma^2\\
    %&= n\sigma^2\\
    \intertext{From this we can find $E[S_n^2]$}
    E[S_n^2] &= Var[S_n] + E[S_n]^2\\
    &= n\sigma^2+n^2\mu^2
    \intertext{We can now find the expectation value of $S_N$, the sum of $N$ cloud masses, where $N$ is a random variable as well:}
    E[S_N]&= E\left[\sum_{n=1}^\infty S_n\id\right]\\
    &= \sum_{n=1}^\infty E\3S_n\id\4\\
    &= \sum_{n=1}^\infty E\3S_n\4P(N=n)\\
    &= \sum_{n=1}^\infty \mu n P(N=n)\\
    %&= \mu \sum_{n=1}^\infty n P(N=n)\\
    &= \mu E[N]\\
    \intertext{where $\id=1$ when $N=n$ and $0$ otherwise. In the same way we can find the expectation value of $N_N^2$}
    E[S_N^2] &= E\3\1\sum_{n=1}^\infty S_n\id\2^2\4\\
    &= E\3\sum_{n=1}^\infty\sum_{m=1}^\infty S_nS_m\id\idm\4\\
    &= E\3\sum_{n=1}^\infty S_n^2\id\4\\
    &= \sum_{n=1}^\infty E[S_n^2]P(N=n)\\
    &= \sum_{n=1}^\infty (n\sigma^2+n^2\mu^2)P(N=n)\\
    &= \sigma^2\1\sum_{n=1}^\infty nP(N=n)\2 + \mu^2\1\sum_{n=1}^\infty n^2P(N=n)\2\\
    &= \sigma^2 E[N] + \mu^2E[N^2]\\
    \intertext{This now lets ud find the variance on the sum $S_N$ as}
    Var[S_N] &= E[S_N^2]-E[S_N]^2\\
    &= \sigma^2E[N] + \mu^2E[N^2]-\mu^2E[N^2]\\
    &= \sigma^2 E[N] - \mu^2Var[N]
    \intertext{Finally we can find the variance of N as}
    Var[N] &= \dfrac{\sigma^2E[N]}{\mu^2} - \dfrac{Var[S_N]}{\mu^2}\\
    \intertext{The expected number of clouds to reach the total mass $M$ is}
    E[N] &= \dfrac{M}{\mu}
    \intertext{Finally, since we know the total mass, $S_N=M$, then the variance of $S_N$ becomes zero, and thus}
    Var[N] &= \dfrac{M\sigma^2}{\mu^3}
\end{align}

% Don't change these lines
\bsp	% typesetting comment
\label{lastpage}
\end{document}